\documentclass[times, twocolumn]{aastex62}
\usepackage{graphicx,amsmath,natbib}
\usepackage{color}

\newcommand{\teff}{$T_{\rm eff}$}

\newcommand{\logg}{log($g$)}

\shorttitle{The JWST ERS Program for Exoplanet Direct Imaging \& Spectroscopy}
\shortauthors{Hinkley et al.}

\begin{document}

\title{The JWST Early Release Science Program for the Direct Imaging \& Spectroscopy of Exoplanetary Systems}

\author[0000-0001-8074-2562]{Sasha Hinkley}\affiliation{University of Exeter, Astrophysics Group, Physics Building, Stocker Road, Exeter, EX4 4QL, UK.}
\author[0000-0001-5365-4815]{Aarynn L.~Carter}\affiliation{Department of Astronomy \& Astrophysics, University of California, Santa Cruz, CA 95064, USA}
\author[0000-0003-2259-3911]{Shrishmoy Ray}\affiliation{University of Exeter, Astrophysics Group, Physics Building, Stocker Road, Exeter, EX4 4QL, UK.}
\author{Andrew Skemer}\affiliation{Department of Astronomy \& Astrophysics, University of California, Santa Cruz, CA 95064, USA}
\author{Beth Biller}\affiliation{SUPA, Institute for Astronomy, The University of Edinburgh, Royal Observatory, Blackford Hill, Edinburgh, EH9 3HJ, UK}
\author{Elodie Choquet}\affiliation{Aix Marseille Univ., CNRS, CNES, LAM, Marseille, France}
\author[0000-0001-6205-9233]{Maxwell A. Millar-Blanchaer}\affiliation{Department of Physics, University of California, Santa Barbara, Santa Barbara, CA, USA}
\author{Stephanie Sallum}\affiliation{Department of Physics and Astronomy, University of California, Irvine, Irvine, CA, USA}
\author{Brittany Miles}\affiliation{Department of Astronomy \& Astrophysics, University of California, Santa Cruz, CA 95064, USA}
\author{Niall Whiteford}\affiliation{Department of Astrophysics, American Museum of Natural History, Central Park West at 79th Street, NY 10024, USA}
\author[0000-0001-8718-3732]{Polychronis Patapis}\affiliation{Institute for Particle Physics \& Astrophysics, ETH Zurich, 8092 Zurich, Switzerland}
\author[0000-0002-3191-8151]{Marshall Perrin}\affiliation{Space Telescope Science Institute, 3700 San Martin Drive, Baltimore, MD 21218, USA}
\author{Laurent Pueyo}\affiliation{Space Telescope Science Institute, 3700 San Martin Drive, Baltimore, MD 21218, USA}
\author{Glenn Schneider}\affiliation{Steward Observatory, University of Arizona, 933 N. Cherry Ave, Tucson, AZ 85721-0065 USA}
\author[0000-0002-2805-7338]{Karl Stapelfeldt}\affiliation{Jet Propulsion Laboratory, California Institute of Technology, M/S 321-100, 4800 Oak Grove Drive, Pasadena, CA 91109, USA}
\author[0000-0003-0774-6502]{Jason Wang}\affiliation{Department of Astronomy, California Institute of Technology, Pasadena, CA 91125, USA}\affiliation{Center for Interdisciplinary Exploration and Research in Astrophysics (CIERA) and Department of Physics and Astronomy, Northwestern University, Evanston, IL 60208, USA}
\author[0000-0002-4479-8291]{Kimberly Ward-Duong}\affiliation{Department of Astronomy, Smith College, Northampton MA 01063, USA}\affiliation{Space Telescope Science Institute, 3700 San Martin Drive, Baltimore, MD 21218, USA}
\author{Brendan P.~Bowler}\affiliation{Department of Astronomy, The University of Texas at Austin, 2515 Speedway Boulevard Stop C1400, Austin, TX 78712, USA}
\author[0000-0001-9353-2724]{Anthony Boccaletti}\affiliation{LESIA, Observatoire de Paris, Universit\'e PSL, CNRS, Sorbonne Universit\'e, Universit\'e Paris Cit\'e, 5 place Jules Janssen, 92195 Meudon, France}
\author[0000-0001-8627-0404]{Julien H.~Girard}\affiliation{Space Telescope Science Institute, 3700 San Martin Drive, Baltimore, MD 21218, USA}
\author{Dean Hines}\affiliation{Space Telescope Science Institute, 3700 San Martin Drive, Baltimore, MD 21218, USA}
\author[0000-0002-6221-5360]{Paul Kalas}\affiliation{Department of Astronomy, University of California at Berkeley, CA 94720, USA}
\author[0000-0003-2769-0438]{Jens Kammerer}\affiliation{Space Telescope Science Institute, 3700 San Martin Drive, Baltimore, MD 21218, USA}
\author[0000-0003-0626-1749]{Pierre Kervella}\affiliation{LESIA, Observatoire de Paris, Universit\'e PSL, CNRS, Sorbonne Universit\'e, Universit\'e Paris Cit\'e, 5 place Jules Janssen, 92195 Meudon, France}
\author[0000-0002-0834-6140]{Jarron Leisenring}\affiliation{Steward Observatory, University of Arizona, 933 N. Cherry Ave, Tucson, AZ 85721-0065 USA}
\author{Eric Pantin}\affiliation{IRFU/DAp D\'epartement D'Astrophysique CE Saclay, Gif-sur-Yvette, France}
\author[0000-0003-2969-6040]{Yifan Zhou}\affiliation{Department of Astronomy, The University of Texas at Austin, 2515 Speedway Boulevard Stop C1400, Austin, TX 78712, USA}
\author{Michael Meyer}\affiliation{Department of Astronomy, University of Michigan, Ann Arbor, MI 48109, USA}
\author[0000-0003-2232-7664]{Michael C. Liu}\affiliation{Institute of Astronomy, University of Hawaii, 2860 Woodlawn Drive, Honolulu, HI 96822, USA}
\author{Mickael Bonnefoy}\affiliation{Universit\'e Grenoble Alpes / CNRS, Institut de Plan\'etologie et d’Astrophysique de Grenoble, 38000 Grenoble, France}
\author{Thayne Currie}\affiliation{NASA-Ames Research Center, Moffett Field, California, USA}
\author[0000-0003-0241-8956]{Michael McElwain}\affiliation{NASA-Goddard Space Flight Center, Greenbelt, MD, USA}
\author{Stanimir Metchev}\affiliation{Department of Physics and Astronomy, Centre for Planetary Science and Exploration, The University of Western Ontario, London, ON N6A 3K7, Canada}
\author{Mark Wyatt}\affiliation{Institute of Astronomy, University of Cambridge, Madingley Road, Cambridge CB3 0HA, UK}
\author[0000-0002-4006-6237]{Olivier Absil}\affiliation{Space Sciences, Technologies, and Astrophysics Research (STAR) Institute, University of Li\`ege, B-4000 Li\`ege Belgium}
\author{Jea Adams}\affiliation{Center for Astrophysics $\vert$ Harvard \& Smithsonian, 60 Garden St, Cambridge, MA 02138, USA}
\author{Travis Barman}\affiliation{Lunar \& Planetary Laboratory, University of Arizona, Tucson, AZ 85721, USA}
\author{Isabelle Baraffe}\affiliation{University of Exeter, Astrophysics Group, Physics Building, Stocker Road, Exeter, EX4 4QL, UK.}\affiliation{\'Ecole Normale Sup\'erieure, Lyon, CRAL (UMR CNRS 5574), Universit\'e de Lyon, France}
\author[0000-0002-7520-8389]{Mariangela Bonavita}\affiliation{School of Physical Sciences, The Open University, Walton Hall, Milton Keynes, MK7 6AA}
\author[0000-0001-8568-6336]{Mark Booth}\affiliation{Astrophysikalisches Institut und Universit\"atssternwarte, Friedrich-Schiller-Universit\"at Jena, Schillerg\"asschen 2-3, D-07745 Jena, Germany}
\author{Marta Bryan}\affiliation{Department of Astronomy, University of California at Berkeley, CA 94720, USA}
\author{Gael Chauvin}\affiliation{Universit\'e Grenoble Alpes / CNRS, Institut de Plan\'etologie et d’Astrophysique de Grenoble, 38000 Grenoble, France}
\author[0000-0002-8382-0447]{Christine Chen}\affiliation{Space Telescope Science Institute, 3700 San Martin Drive, Baltimore, MD 21218, USA}\affiliation{Department of Physics and Astronomy, The Johns Hopkins University, 3400 N. Charles Street, Baltimore, MD 21218, US}
\author[0000-0002-3729-2663]{Camilla Danielski}\affiliation{Instituto de Astrofísica de Andalucía, CSIC, Glorieta de la Astronomía s/n, 18008, Granada, Spain}\affiliation{AIM, CEA, CNRS, Universit\`e Paris-Saclay, Universit\`e Paris Diderot, Sorbonne Paris Cit\`e, F-91191 Gif-sur-Yvette, France}
\author[0000-0003-1863-4960]{Matthew De Furio}\affiliation{Department of Astronomy, University of Michigan, Ann Arbor, MI 48109, USA}
\author[0000-0002-8332-8516]{Samuel M.~Factor}\affiliation{Department of Astronomy, The University of Texas at Austin, 2515 Speedway Boulevard Stop C1400, Austin, TX 78712, USA}
\author{Michael P. Fitzgerald}\affiliation{Department of Physics \& Astronomy, University of California, Los Angeles, CA. 90095 USA}
\author[0000-0002-9843-4354]{Jonathan J.~Fortney}\affiliation{Department of Astronomy \& Astrophysics, University of California, Santa Cruz, CA 95064, USA}
\author{Carol Grady}\affiliation{Eureka Scientific, 2452 Delmer St., Suite 100, Oakland CA 96402-3017, USA.}
\author[0000-0002-7162-8036]{Alexandra Greenbaum}\affiliation{IPAC, Mail Code 100-22, Caltech, 1200 E. California Blvd., Pasadena, CA 91125, USA}
\author{Thomas Henning}\affiliation{Max-Planck-Institut f\"ur Astronomie, K\"onigstuhl 17, 69117 Heidelberg, Germany}
\author{Kielan K.~W.~Hoch}\affiliation{Space Telescope Science Institute, 3700 San Martin Drive, Baltimore, MD 21218, USA}
\author[0000-0001-8345-593X]{Markus Janson}\affiliation{Department of Astronomy, Stockholm University, 10691, Stockholm, Sweden}
\author{Grant Kennedy}\affiliation{Department of Physics, University of Warwick, Coventry CV4 7AL, UK}
\author[0000-0002-7064-8270]{Matthew Kenworthy}\affiliation{Leiden Observatory, Leiden University, Postbus 9513, 2300 RA Leiden, The Netherlands}
\author{Adam Kraus}\affiliation{Department of Astronomy, The University of Texas at Austin, 2515 Speedway Boulevard Stop C1400, Austin, TX 78712, USA}
\author[0000-0002-4677-9182]{Masayuki Kuzuhara}\affiliation{Astrobiology Center, 2-21-1, Osawa, Mitaka, Tokyo, 181-8588, Japan}\affiliation{National Astronomical Observatory of Japan, 2-21-1 Osawa, Mitaka, Tokyo 181-8588, Japan}
\author{Pierre-Olivier Lagage}\affiliation{AIM, CEA, CNRS, Universit\`e Paris-Saclay, Universit\`e Paris Diderot, Sorbonne Paris Cit\`e, F-91191 Gif-sur-Yvette, France}
\author{Anne-Marie Lagrange}\affiliation{Universit\'e Grenoble Alpes / CNRS, Institut de Plan\'etologie et d’Astrophysique de Grenoble, 38000 Grenoble, France}
\author{Ralf Launhardt}\affiliation{Max-Planck-Institut f\"ur Astronomie, K\"onigstuhl 17, 69117 Heidelberg, Germany}
\author{Cecilia Lazzoni}\affiliation{University of Exeter, Astrophysics Group, Physics Building, Stocker Road, Exeter, EX4 4QL, UK.}\affiliation{INAF--Osservatorio Astronomico di Padova, Vicolo dell'Osservatorio 5, I-35122, Padova, Italy}
\author{James Lloyd}\affiliation{Department of Astronomy, Cornell University, Ithaca NY, USA}
\author{Sebastian Marino}\affiliation{Jesus College, University of Cambridge, Jesus Lane, Cambridge CB5 8BL, UK}\affiliation{Institute of Astronomy, University of Cambridge, Madingley Road, Cambridge CB3 0HA, UK}
\author{Mark Marley} \affiliation{Lunar \& Planetary Laboratory, University of Arizona, Tucson, AZ 85721, USA}
\author[0000-0001-6301-896X]{Raquel Martinez}\affiliation{Department of Physics and Astronomy, University of California, Irvine, Irvine, CA, USA}
\author[0000-0002-4164-4182]{Christian Marois}\affiliation{Herzberg Astronomy \& Astrophysics Research Centre, National Research Council of Canada, 5071 West Saanich Road, Victoria, BC V9E 2E7, Canada}\affiliation{Department of Physics \& Astronomy, University of Victoria, 3800 Finnerty Road, Victoria, BC V8P 5C2, Canada}
\author[0000-0003-3017-9577]{Brenda Matthews}\affiliation{Herzberg Astronomy \& Astrophysics Research Centre, National Research Council of Canada, 5071 West Saanich Road, Victoria, BC V9E 2E7, Canada}\affiliation{Department of Physics \& Astronomy, University of Victoria, 3800 Finnerty Road, Victoria, BC V8P 5C2, Canada}
\author[0000-0003-0593-1560]{Elisabeth C.~Matthews}\affiliation{Observatoire Astronomique de l’Universit\'e de Gen\`eve, 51 Ch.~Pegasi, 1290 Versoix, Switzerland}
\author{Dimitri Mawet}\affiliation{Department of Astronomy, California Institute of Technology, Pasadena, CA 91125, USA}\affiliation{Jet Propulsion Laboratory, California Institute of Technology, M/S 321-100, 4800 Oak Grove Drive, Pasadena, CA 91109, USA}
\author{Johan Mazoyer}\affiliation{LESIA, Observatoire de Paris, Universit\'e PSL, CNRS, Sorbonne Universit\'e, Universit\'e Paris Cit\'e, 5 place Jules Janssen, 92195 Meudon, France}
\author{Mark Phillips}\affiliation{Institute of Astronomy, University of Hawaii, 2860 Woodlawn Drive, Honolulu, HI 96822, USA}
\author{Simon Petrus}\affiliation{Universit\'e Grenoble Alpes / CNRS, Institut de Plan\'etologie et d’Astrophysique de Grenoble, 38000 Grenoble, France}\affiliation{N\'ucleo Milenio Formaci\'on Planetaria - NPF, Universidad de Valpara\'iso, Av. Gran Breta\~na 1111, Valpara\'iso, Chile}
\author[0000-0003-3829-7412]{Sascha P.~Quanz}\affiliation{Institute for Particle Physics \& Astrophysics, ETH Zurich, 8092 Zurich, Switzerland}
\author{Andreas Quirrenbach}\affiliation{Landessternwarte,Zentrum f\"ur Astronomie der Universit\"at Heidelberg, K\"onigstuhl 12, 69117 Heidelberg, Germany}
\author{Julien Rameau}\affiliation{Universit\'e Grenoble Alpes / CNRS, Institut de Plan\'etologie et d’Astrophysique de Grenoble, 38000 Grenoble, France}
\author[0000-0002-4388-6417]{Isabel Rebollido}\affiliation{Space Telescope Science Institute, 3700 San Martin Drive, Baltimore, MD 21218, USA}
\author[0000-0003-4203-9715]{Emily Rickman}\affiliation{European Space Agency (ESA), ESA Office, Space Telescope Science Institute, 3700 San Martin Drive, Baltimore, MD 21218, USA}
\author[0000-0001-9992-4067]{Matthias Samland}\affiliation{Max-Planck-Institut f\"ur Astronomie, K\"onigstuhl 17, 69117 Heidelberg, Germany}
\author[0000-0001-9855-8261]{B.~Sargent}\affiliation{Space Telescope Science Institute, 3700 San Martin Drive, Baltimore, MD 21218, USA}\affiliation{Center for Astrophysical Sciences, The William H. Miller III Department of Physics and Astronomy, Johns Hopkins University, Baltimore, MD 21218, USA}
\author{Joshua E.~Schlieder}\affiliation{Exoplanets \& Stellar Astrophysics Laboratory, NASA Goddard Space Flight Center, Greenbelt, MD, USA }
\author[0000-0003-1251-4124]{Anand Sivaramakrishnan}\affiliation{Space Telescope Science Institute, 3700 San Martin Drive, Baltimore, MD 21218, USA}
\author[0000-0003-0454-3718]{Jordan M.~Stone}\affiliation{US Naval Research Laboratory, Remote Sensing Division, 4555 Overlook Ave SW, Washington, DC 20375}
\author[0000-0002-6510-0681]{Motohide Tamura}\affiliation{Department of Astronomy, The University of Tokyo, 7-3-1, Hongo, Bunkyo-ku, Tokyo 113-0033, Japan}\affiliation{Astrobiology Center, 2-21-1, Osawa, Mitaka, Tokyo, 181-8588, Japan}\affiliation{National Astronomical Observatory of Japan, 2-21-1 Osawa, Mitaka, Tokyo 181-8588, Japan}
\author{Pascal Tremblin}\affiliation{Maison de la Simulation, CEA, CNRS, Univ. Paris-Sud, UVSQ, Universit\`e Paris-Saclay, F-91191 Gif-sur-Yvette, France}
\author{Taichi Uyama}\affiliation{Infrared Processing and Analysis Center, California Institute of Technology, 1200 E. California Blvd., Pasadena, CA 91125, USA}\affiliation{NASA Exoplanet Science Institute, Pasadena, CA 91125, USA}
\author{Malavika Vasist}\affiliation{Space Sciences, Technologies, and Astrophysics Research (STAR) Institute, University of Li\`ege, B-4000 Li\`ege Belgium}
\author[0000-0002-5902-7828]{Arthur Vigan}\affiliation{Aix Marseille Univ., CNRS, CNES, LAM, Marseille, France}
\author{Kevin Wagner}\affiliation{Steward Observatory, University of Arizona, 933 N. Cherry Ave, Tucson, AZ 85721-0065 USA}\affiliation{NASA Hubble Fellowship Program - Sagan Fellow}
\author{Marie Ygouf}\affiliation{Jet Propulsion Laboratory, California Institute of Technology, M/S 321-100, 4800 Oak Grove Drive, Pasadena, CA 91109, USA}

\begin{abstract}
The {\it direct} characterization of exoplanetary systems with high contrast imaging is among the highest priorities for the broader exoplanet community.  As large space missions will be necessary for detecting and characterizing exo-Earth twins, developing the techniques and technology for direct imaging of exoplanets is a driving focus for the community. For the first time, \textit{JWST} will directly observe extrasolar planets at mid-infrared wavelengths beyond 5\,$\mu$m, deliver detailed spectroscopy revealing much more precise chemical abundances and atmospheric conditions, and provide sensitivity to analogs of our solar system ice-giant planets at wide orbital separations, an entirely new class of exoplanet.  However, in order to maximise the scientific output over the lifetime of the mission, an exquisite understanding of the instrumental performance of \textit{JWST} is needed as early in the mission as possible.  In this paper, we describe our 55-hour Early Release Science Program that will utilize all four \textit{JWST} instruments to extend the characterisation of planetary mass companions to $\sim$15\,$\mu$m as well as image a circumstellar disk in the mid-infrared with unprecedented sensitivity. Our program will also assess the performance of the observatory in the key modes expected to be commonly used for exoplanet direct imaging and spectroscopy, optimize data calibration and processing, and generate representative datasets that will enable a broad user base to effectively plan for general observing programs in future cycles. 
\end{abstract}


\section{Introduction}
In the relatively short span of a quarter century, astronomers have transitioned from speculating about the prevalence of exoplanetary systems to discovering thousands, and it is now clear that most stars host planetary systems \citep[e.g.,][]{ckb12, dc13}.  The vast majority of these, however, have been identified only indirectly via the transit and radial velocity  detection methods, and  lie at close orbital separations from their host stars.

As the indirect transit and Doppler detection methods are inherently much less sensitive to wide-separation planets with long orbital periods, the direct imaging technique \citep[e.g.,][]{b16} will be the only approach to fully define the outermost architectures of planetary systems ($\sim$\,10 to hundreds of AU), and provide a more complete understanding of the true frequency of planetary mass companions to nearby stars \citep[e.g.,][]{ndm19,vfm21}.  In the last decade, imaging observations mostly at wavelengths of $\lesssim$2\,$\mu$m have produced numerous scattered light images of circumstellar disks \citep{ekf20}, and directly revealed  $\sim$\,10-20, young ($\lesssim$\,50 Myr), massive ($\gtrsim$1\,M$_\mathrm{Jup}$) planets \citep[e.g.,][]{mmb08,lbc10, rameau95086, cdl17, mgb15, bkg20,jgr21}.  Direct Imaging is also the only technique that will be capable of \textit{characterising} exoplanets at orbital radii $\gtrsim$0.5\,AU, as transit transmission spectroscopy \citep[e.g.,][]{sfn16} requires multiple transits to achieve a strong signal for Earth-mass planets on Earth-like orbits around Sun-like stars \citep{mkr17} resulting in prohibitively long time baselines. It is also the only technique projected to provide the in-depth characterization of such exo-Earths \citep[e.g.,][]{t19, qab21}.

By spatially separating the light of the host star and the extremely faint planet, the direct imaging technique is also naturally suited to {\it direct} spectroscopy of planets themselves, allowing detailed characterization \citep[e.g.,][]{bld10, mgb15,drp16,cdl17,cbu18}. In addition to providing information on atmospheric properties and compositions \citep[e.g.,][]{hbv15,kls21}, the direct imaging technique can provide powerful estimations of fundamental parameters, e.g. luminosity, effective temperature, and orbital properties \citep{gln19}.  The characterization power of the direct imaging method becomes even more pronounced when combined with other exoplanet detection techniques, such as precise radial velocity monitoring or astrometry \citep[e.g.,][]{nll20, lrn20, wvl21, lwr21}, to more fully constrain parameters (e.g.~planet mass) that are difficult to ascertain with one technique alone.

Going forward, direct imaging will ultimately provide direct, high-resolution (R$\sim$30,000-100,000) spectra of exoplanet atmospheres \citep[e.g.,][]{sdb15,mbd18,ovm21,wrm21}.  Obtaining photons directly from the atmosphere of the exoplanet itself will allow us to apply to exoplanetary atmospheres all of the spectroscopic techniques that have been applied to stars and brown dwarfs over the last century. Indeed, all of the detailed information that can be retrieved using high-resolution spectroscopy of stars and brown dwarfs (e.g.~chemical abundances, compositions, thermodynamic conditions, Doppler tomography) will also be directly obtained for exoplanetary atmospheres, allowing much more precise interpretation of the spectra \citep[][]{kbm13,bkm15}.  

At the same time, the direct imaging technique has been especially prolific at imaging both very young primordial (protoplanetary) disks as well as dusty circumstellar debris disks. Direct images of these disk structures uniquely allow the study of the dynamical interactions between circumstellar disks, and the planets that are dynamically sculpting them \citep{cpc16,mhv17,aqg18, ekf20}.  Upcoming space-based missions, the advent of the 30-40m telescopes, as well as updates to existing ground-based high contrast imaging platforms \citep[e.g.~MagAOX, GPI 2.0, SPHERE+, SCExAO;][]{mcm18, ckd20} will vastly improve exoplanet direct detection capabilities.  Consequently, the development of the direct imaging technique will be a major priority for the broader exoplanet community going forward, and is a key mode for current and future space missions including \textit{JWST} \citep{gmc06}.

\vspace{0.2in}
\subsection{JWST}
\textit{JWST} will be transformative for characterizing exoplanet atmospheres at  mid-infrared and longer wavelengths (3-28\,$\mu$m).  In addition to a handful of spectra of free-floating brown dwarfs from \textit{The Spitzer Space Telescope} and \textit{Akari} missions \citep{crm06,sy12}, direct images of exoplanets from 3 to 5\,$\mu$m exist \citep[e.g.,][]{gmm11, smh14, smz16}, but the extremely high telluric background has imposed a strong sensitivity barrier. Transit transmission spectroscopy of transiting ``Hot Jupiter'' exoplanets using the \textit{Spitzer Space Telescope} exist \citep[e.g.,][]{gcb07,gbc08}, but this class of planet has very different properties than young planets studied via direct imaging, or even solar system giant planets.  Even further, despite some tantalizing early detections \citep[e.g.,][]{wbp21}, exoplanets have never been \textit{directly} observed at wavelengths $\gtrsim$5\,$\mu$m.   

\begin{deluxetable*}{c|c|c|c|c|c|c}
\tabletypesize{\scriptsize}
\tablecaption{A Table showing the primary science targets to be observed within this ERS program}
\tablewidth{0pt}
\tablehead{ 
\colhead{Target} & 
\colhead{Host Star SpT} & 
\colhead{Distance (pc)} &
\colhead{Age (Myr)} & 
\colhead{Angular Separation ($^{\prime\prime}$)} &
\colhead{Mass (M$_\mathrm{Jup}$)} &
\colhead{References}
}
\startdata
HIP\,65426b   &    A2V       & 109.2$\pm$0.8        & 14$\pm$4 &  0.83           &   6-12     & \citet{cdl17, csb19}      \\
VHS\,1256\,b  & M7.5$\pm$0.5 & 22.2$^{+1.1}_{-1.2}$ & 150-300  &  8.06 $\pm$0.03 &  19$\pm$5  & \citet{gbp15, dlm20} \\
HD\,141569A    &    A2        & 110.6$\pm$0.5        & 5$\pm$3  &       -         &      -     &  \citet{wrb00,cka03}  \\
\enddata
\label{targtable}
\end{deluxetable*}

To date, much of the instrumentation dedicated to the purpose of high contrast imaging \citep[][]{hoz11,mcb18,bvm19} has operated at wavelengths $\lesssim$\,3$\mu$m. However, in addition to observing planets closer to the peak of their thermal emission, obtaining exoplanet photometry and spectroscopy at 3-5\,$\mu$m is a key probe of atmospheric compositions, and dramatically improves differentiation between equilibrium and disequilibrium atmospheric chemistry \citep{she12, ptb20, msm20}.  As well as clear markers of composition, many of the key atmospheric processes that shape the spectra of planetary mass companions, e.g.~vertical atmospheric mixing \citep{hb07,bmk11}, cloud physics, and the presence of patchy cloud structures \citep{cbi11} have signatures at the wavelengths observable by \textit{JWST}.  These sets of constraints cannot be entirely accessed through observations at 1-2\,$\mu$m alone.  Incorporating data at long wavelengths can dramatically help to break the persistent degeneracies \citep[see e.g.,][]{cbs07} in the interpretation of exoplanet spectra that might be in fact driven by enhanced elemental abundances, or various physical processes such as vertical atmospheric mixing \citep{kbm13}.  While \textit{JWST} will not have the angular resolution, nor the inner working angles, of 8-10m ground-based telescopes operating in the near-infrared \citep{ppv18}, the extremely precise flux measurements afforded by the stability and low thermal background at the second Lagrange point will allow the differentiation of dominant exoplanet atmospheric chemistries, as well as derive circumstellar disk compositions.

The long wavelength coverage offered by \textit{JWST} will provide abundances of the expected molecules in exoplanet atmospheres (CH$_4$, CO, CO$_2$, H$_2$O and NH$_3$), using multiple features and a broad wavelength range to distinguish between clouds and molecular absorption.  Calculating abundances for the major oxygen and carbon-bearing species will provide measurements of an object's C/O ratio.  Whether or not the C/O ratio of a companion matches that of its star may provide clues regarding the companion formation mechanism, and may help to distinguish zones of formation within the protoplanetary disk \citep[][]{omb11, cpa16,cev19, mmb22}. However, the utility and ease of a single metric such as the C/O ratio to capture the complex physics \citep[e.g.,][]{fmn13,cpa16,mbj17} inherent in the planet formation process has been debated.  For example, some works \citep[e.g.,][]{mvm16} have demonstrated how limitations in our understanding of disk chemistry, the planet formation process and migration can limit the utility of this metric.

Differences in the near-infrared Spectral Energy Distributions (SEDs) of exoplanets and field brown dwarfs \citep{cld04,mmb08} suggest that their atmospheres are not the same, perhaps due to differences in surface gravity \citep{msc12}, stellar insolation \citep{zmm16}, or cloud features.  This difference is readily apparent in color-magnitude diagrams, where planetary-mass objects tend to be much redder than the field brown dwarf population \citep{frc16, lda16}.  Silicate and iron clouds are typically used to explain the colors of red (L-type) brown dwarfs, but the colors can also be explained by thermo-chemical instabilities \citep{tam15, tac16} or the viewing angle \citep{vab17}. Although there have been tentative detections of silicates through an 8-12\,$\mu$m absorption feature in field brown dwarfs with \textit{Spitzer} \citep{crm06, lkc08}, extremely red exoplanets might have stronger features that point towards the cloud hypothesis.  Moving forward, it will be essential to gather empirical constraints on these SEDs in order to effectively model the sensitivity of \textit{JWST} to these substellar objects in future surveys.

\begin{figure*}
  \centering
  \vspace{-0.7in}
  \includegraphics[width=1.0\textwidth]{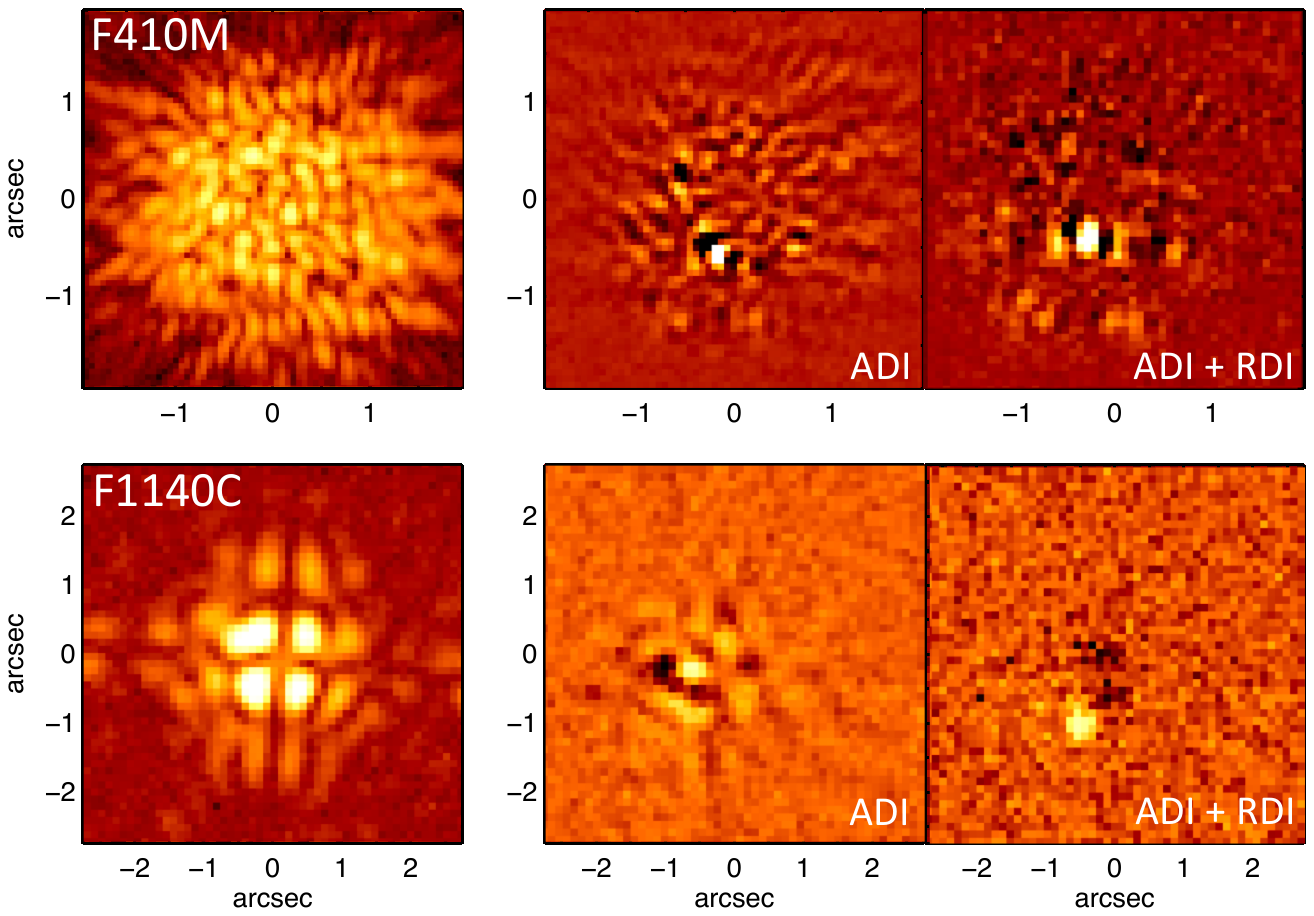}
   \vspace{-0.9in}
  \caption{Simulated coronagraphic images of the directly imaged exoplanet HIP\,65426b \citep{cdl17} in the 4.1$\mu$m NIRCam filter ({\it F410M, top row}), and the 11.4$\mu$m MIRI filter ({\it F1140C, bottom row}) using the simulation methods described in \citet{csd21}. In each set of figures, the left image shows the coronagraphic point spread function for each bandpass, while the middle and right panels show subsequent image post-processing employing Angular Differential Imaging (ADI) or a combination of ADI with Reference Differential Imaging (labelled ``ADI+RDI'') that utilizes a small library of images obtained using the small grid dither strategy \citep{lsp16}.} 
  \label{fig:hip65426_sim}
\end{figure*}

\textit{JWST} utilizes four science instruments operating in the infrared, with three of these operating at wavelengths shorter than 5\,$\mu$m. NIRCam \citep{rkh05, bre12} and NIRSpec \citep{bkf07, jfa22} are the primary near-infrared imager and spectrograph, respectively, but only NIRCam and MIRI offer a coronagraphic mode.  NIRCam offers a series of coronagraphic occulting masks \citep{gbb05,kbt07} and preflight estimates of the coronagraphic performance suggest that contrasts of $\sim$10$^{-4}-10^{-5}$ within $\sim$1$^{\prime\prime}$ will be achieved. NIRSpec will have numerous capabilities  for spectroscopically characterizing the atmospheres of extrasolar planets \cite[e.g.,][]{bfg22}, including obtaining direct spectroscopy for widely separated companions.  NIRISS \citep{dhb12} is equipped with an Aperture Masking Interferometer mode \citep[``AMI,''][]{slf12, asg14,sst20} with the goal of obtaining moderate contrast (10$^{-3} - 10^{-4}$) at separations comparable to (and within) the \textit{JWST} diffraction limit of 130-150\,mas at $\sim$4-5\,$\mu$m.  MIRI \citep{rwb15,wwg15} is the only instrument capable of observations at wavelengths longer than 5\,$\mu$m, and is equipped with four-quadrant phase mask (FQPM) coronagraphs \citep{rrb00, bbr06} operating at 10.65, 11.60, and 15.5\,$\mu$m, as well as a classical Lyot coronagraph operating at 23\,$\mu$m \citep{bbb05, cab08, blb15}.  In addition to offering very good contrast performance at small inner working angles of ${\sim}\lambda/D$ at wavelengths of 10-16\,$\mu$m, the 4QPM coronagraph design was chosen during the MIRI design phase as it was one of the few coronagraphs that had been validated both in laboratory testing as well as on sky. Below we describe some of the potentially transformative science that will be done with \textit{JWST} coronagraphy for identifying wide-separation giant planets, as well as characterizing circumstellar debris disks.

\subsection{A New Class of Planets: Wide-Separation Saturn \& Neptune Analogues}

The exquisite thermal infrared sensitivity afforded by \textit{JWST} means that it will be possible for the first time to directly image sub-Jupiter mass planets, an entirely new class of directly imaged planet \citep{bkt10,chb21}.  Previous studies \citep[e.g.,][]{jqc15} have demonstrated the gain in sensitivity when using space-based imaging at infrared wavelengths, similar to those to be used with the \textit{JWST} NIRCam instrument.  In the background-limited regime within the field of view, \textit{JWST} will have 50 times the sensitivity of the Gemini Planet Imager \citep{mgb15} operating at 2$\mu$m, and 500 times the sensitivity of the Keck/NIRC2 instrument at 4$\mu$m \citep{ppv18}. 
Analogues to our own solar system ice-giant planets should have very cold temperatures, and thus the peak wavelength of their emission will be shifted out of the near- and mid-infrared (1 to $\sim$5\,$\mu$m) into to the thermal infrared ($\gtrsim$10\,$\mu$m).  The very high telluric background from the ground imposes a stark sensitivity limit, even with new technologies (e.g. adaptive secondary mirrors). 
Thus, in addition to potentially directly detecting a handful of planets detected by the radial velocity technique \citep[e.g.,][]{bbs20}, the superior thermal infrared sensitivity of \textit{JWST} will allow imaging for the first time of wide, young Saturn analogues in many cases down to masses of 0.2 M$_\mathrm{Jup}$.  In the most favorable cases, \textit{JWST} will be capable of directly imaging wide young Neptune analogues \citep{bkt10,sbm16, chb21}.  

Early results from direct imaging surveys \citep[e.g.,][]{ndm19, wak19, vfm21} suggest that an abundant population of lower mass planetary companions exists at wide orbital separations (tens to hundreds of AU).  Extrapolating the mass distribution power laws derived by the GPI campaign \citep{ndm19} indicates several 0.1-1.0\,M$_\mathrm{Jup}$ planets \textit{per star} reside at separations of 10-100\,AU when considering the broadest range of host star masses (0.2-5 M$_{\odot}$). Along the same lines, \citet{fmp19} predict a factor of $\sim$2$-$4 increase in occurrence rate between 10$-$100\,AU for 0.1$-$13\,$M_\mathrm{Jup}$ planets compared to 1$-$13\,$M_\mathrm{Jup}$ planets.  Recent statistical arguments from microlensing efforts \citep[e.g.,][]{psm21} are consistent with this, suggesting $\sim$1.5 ice giant planets ($\lesssim$1\,M$_\mathrm{Jup}$) per star reside between 5-15\,AU. 
The sensitivity of \textit{JWST} to such a wide range of masses (several M$_\mathrm{Jup}$ down to $\lesssim$\,1 M$_\mathrm{Jup}$) has the potential to associate sub-populations of planets to various planet formation mechanisms.  For example, it is expected that the gravitational instability model \citep{kmy10, fr13} will preferentially form planets with masses $\gtrsim$3\,M$_\mathrm{Jup}$, while lower mass planets on such wide orbits are challenging to form via this model, and instead suggest a mechanism more akin to a model based on accretion of solids in a protoplanetary disk \citep[e.g.,][]{phb96}. Therefore, measuring the mass and separation functions of wide-orbit, low-mass companions is also a powerful constraint on the specifics of a formation mechanism based on core accretion.

\subsection{Studies of Circumstellar Disks with JWST}
Exoplanets and circumstellar debris disks go hand-in-hand:~the gravitational influence of giant planets in these systems shapes the belts of debris, generated by colliding remnant planetesimals \citep{w08, hdm18}, can exist in stable orbits around a star, and sculpts the dust distribution through scattering, secular interactions and resonances \citep{ckk09, w03, mm15}.  
Indeed, the structure of debris disks reveal dynamical relations between planets and their circumstellar environment, and any evaluation of \textit{JWST's} ability to characterize planetary systems must also include its ability to characterize diffuse, extended emission from circumstellar debris disks.

\begin{figure*}
  \centering
  \includegraphics[width=.95\textwidth]{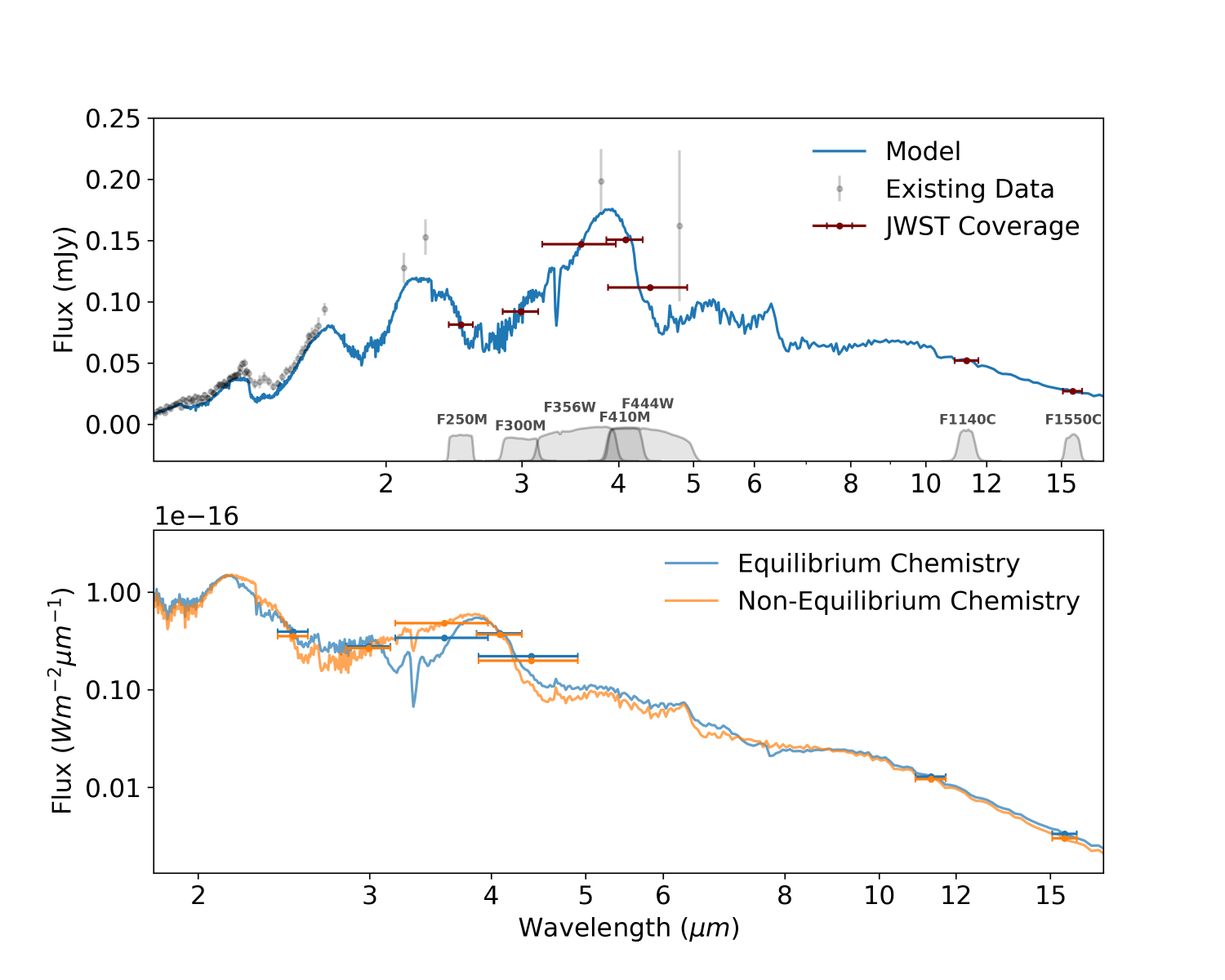}
  \caption{\textit{Top panel:} The spectral coverage of HIP\,65426b we will achieve using NIRCam and MIRI Coronagraphy (red points), compared to the existing VLT-SPHERE and NACO photometric points \citep{csb19} shown in black/grey, and overlaid with a model (\teff=1340\,K, \logg=3.95) computed using the tools described in \citet{tam15}.  
  The simulated \textit{JWST} data points reflect the integral of the flux density over the filter bandpass, divided by the width of the bandpass. 
  \textit{Bottom panel:} Two example synthetic atmosphere models (assuming the same physical parameters for each model) taken from \cite{ptb20} illustrating the effects of non-equilibrium chemistry (orange curve) compared to equilibrium chemistry (blue). Also shown are the photometric points from NIRCam and MIRI we will obtain, which will easily have the precision to differentiate between the two models, particularly at 3.56\,$\mu$m.  
  }
  \label{fig:hip65426_photometry}
\end{figure*}

With nearly an order of magnitude greater angular resolution than \textit{Spitzer},  \textit{JWST} will allow resolved imaging of asteroid belts in other systems for the first time.
The very large, contiguous fields-of-view of NIRCam and MIRI, coupled with their high sensitivity, will be critically important for studies of debris disks, providing access to regions at high contrast beyond the control zones of ground-based Adaptive Optics systems \citep[e.g., $\sim$0.5-1.0$^{\prime\prime}$,][]{mhv18,hml21}.  
H$_2$O ice features are strong in Saturn's rings and Kuiper belt objects \citep[e.g.,][]{fcc14} and are expected in debris disks, but sufficiently sensitive observations have been impossible from the ground.  In addition to constraining the dust scattering properties, NIRCam observations of debris disks may potentially directly detect the planets responsible for driving the ring structures, or provide very sensitive upper limits on their presence.    Due to very high CO$_2$ opacity in Earth's atmosphere, debris disks have {\it never} been imaged specifically at 15\,$\mu$m---on the steeply rising Wien side of the disk thermal emission. Spatially resolved 15\,$\mu$m debris disk observations will constrain the disk radial temperature profiles (and thus dust grain sizes), and disentangle the unresolved excess emission close to the star from the extended emission in the outer, bright debris structures.  \textit{JWST} is well poised to address these issues, but the best observing practices need to be established quickly.

\vspace{0.2in}
In this paper, we describe our upcoming 55-hour Early Release Science (ERS) Program dedicated to the direct imaging and spectroscopy of extrasolar planetary systems. In \S 2 we provide our rationale for an ERS program dedicated to this task, and in \S 3 we provide a detailed overview of the observational strategy of our program. In \S4 we describe our plans for rapid data processing as well as the Science Enabling Products (``SEPs'') our team will deliver shortly after the program is executed, and in \S5 we provide a summary and conclude.

\begin{deluxetable*}{cccccc}
\tabletypesize{\scriptsize}
\tablecaption{A Table showing the observing configuration for each of the major tasks in our program: coronagraphy of an exoplanet and a circumstellar disk, spectroscopy of a planetary mass companion, and aperture masking interferometry.  In addition to the instruments used, targets to be observed (both primary and calibration targets), wavelengths/resolution employed, and observing modes, the table also shows in parentheses the partitions of time dedicated purely to science observations (denoted ``$t_{\rm sci}$''), and the time charged to to the observatory (``$t_{\rm obs}$''), which includes observatory overheads. Our program was also awarded an additional $\sim$12 hrs to collect background observations using MIRI, bringing our true time allocation to $\sim$68 hrs. These observations were needed to calibrate some stray light effects discovered during MIRI commissioning. However, since such a time commitment is not expected to be typical for future observing programs, we have chosen not to list it in this table.}
\tablewidth{0pt}
\tablehead{ 
\colhead{\textbf{Task (Instruments)}} & 
\colhead{\textbf{Targets}     ($t_{\rm sci}$, $t_{\rm obs}$)} &
\colhead{\textbf{$\lambda$ coverage, $\lambda$/$\Delta\lambda$}} & 
\colhead{\textbf{Mode: Filters/Gratings}} &
\colhead{\textbf{$t_{\rm sci}$ (hr)}} &
\colhead{\textbf{$t_{\rm obs}$ (hr)}}
}
\startdata
\textbf{Exoplanet Coronagraphy}        &  Primary:~HIP\,65426\,b (5.0, 10.4)  &                                & NIRCam\,MASK335R:                                     &               &      \\ 
(NIRCam, MIRI)                                  &  Calibrator: HIP\,68245 (2.9, 6.0)      & 2-15\,$\mu$m         & F250M,\,F300M,\,F410M,\,F356W,\,F444W    &     7.9     &   16.4 \\ 
                                                           &                                                            &                                &  MIRI\,FQPM:~F1140C,\,F1550C                    &               &      \\ \hline 
\textbf{Disk Coronagraphy}                &  Primary:~HD\,141569A (8.1, 14.2)    & 3-15\,$\mu$m         & NIRCam\,MASK335R:~F300M,\,F360M         &     13.5   &   23.8 \\  
(NIRCam, MIRI)                                 &  Calibrator: HD\,140986 (5.4,9.6)       &                                 &  MIRI\,FQPM:~F1065C,\,F1140C,\,F1550C    &               &      \\ \hline
\textbf{Spectroscopy of a Planetary}  &                                                            & 1-28\,$\mu$m,         & NIRSpec\,IFS:~G140H, G235H,  G395H        &                &        \\
\textbf{Mass Companion}                   &   VHS\,1256\,b                                    & R$\sim$1500-3000  & MIRI MRS: All channels                                  & 2.7          & 6.3  \\ 
(NIRSpec, MIRI)                                 &                                                            &                                  &                                                                         &                &       \\  \hline
\textbf{Aperture Masking}              & Primary:~HIP\,65426 (3.0, 4.7)        &                   &                                        &       &      \\
\textbf{Interferometry}                & Calibrator:~HD\,115842 (0.6, 1.8)    &  3.8\,$\mu$m      &     F380M                              & 4.6   & 8.8  \\
(NIRISS)                               & Calibrator:~HD\,116084 (1.0, 2.3)    &                   &                                        &       &      \\   \hline 
                                       &                                      &                   &                                        &\textbf{Total: 28.7} & \textbf{Total: 55.3} 
\enddata
\label{tab:obstable}
\end{deluxetable*}

\vspace{0.25in}
\section{Rationale for a JWST Early Release Science Program}
\textit{JWST} will be a transformative observatory for directly characterizing both exoplanets and their circumstellar environments.  It will enable very sensitive, high-fidelity imaging and spectrophotometry of exoplanetary systems in the near/mid-infrared with NIRCam, NIRSpec, and NIRISS, and with MIRI at wavelengths $\gtrsim$5\,$\mu$m {\it for the first time}.  However, imaging young extrasolar giant planets and disks at very small angular separations from a host star is an extreme technical challenge: at  young ages ($\sim$1-100\,Myr) even the most massive planets are typically $\sim$10$^3$-10$^6$ times fainter than the host star, and buried in the halo of the instrumental Point Spread Function (PSF).  At such a high level of contrast, an exquisite understanding is required of the instrument response, PSF stability, and PSF subtraction techniques during post-processing.  To this end, successfully obtaining images of planets and circumstellar disks requires the correct choice of observation mode and extremely careful post-processing.

As described in \cite{ppv18}, the last $\sim$25 years of highly successful coronagraphy and high contrast imaging with \textit{The Hubble Space Telescope (HST)} provides a very useful set of lessons for the best path forward with \textit{JWST}.  
One clear lesson from the \textit{HST} legacy is that, particularly for non-optimal coronagraphs (e.g., those on board \textit{HST} or even \textit{JWST}), more of the contrast gains come from state-of-the-art starlight suppression work in the image post-processing stage than from suppression at the hardware level \citep[][]{p16, cpc16,ppv18, zbw21,szb22}.  
Indeed, exquisite calibration of the PSF is key, and typically this is carried out by using some variation of the Angular Differential Imaging technique \citep[``ADI,''][]{mld06} to utilize the rotation of the observatory to disentangle bona fide astrophysical sources (e.g., planets) from the residual scattered starlight. For each individual science image, the task is to use other images in the sequence that do not contain a signal of the planet to construct an optimized, synthetic model of the PSF, typically using least-squares \citep[``LS,''][]{lmd07}, or principal component analysis methods \citep[``PCA,''][]{spl12,aq12} to generate  synthetic PSF calibrations that are optimally matched to each science exposure.  However, \textit{JWST} will typically only achieve $\sim$5$^\circ$ of rotation when the spacecraft is rolled slightly, meaning that the sensitivity of \textit{JWST} will be limited at close separations from the star.  However, it has now been well demonstrated, especially for space-based imaging such as \textit{HST/NICMOS} \citep{shp11} with PSF morphologies that are highly stable in time, that using a vast suite of reference PSF images from other epochs/targets can alleviate this problem.  Indeed, over its lifetime, \textit{JWST} will provide such an extensive library of images, but not necessarily in the early phases of the mission.

Coronagraphic imaging with \textit{HST} required several cycles to optimize the observing strategy and PSF calibration methods, and the methods are still being refined \cite[e.g.,][]{sgd17,zbw21}.  Further, applying techniques originally developed for the ground to \textit{HST} archival coronagraphic data \citep{lmd09, shp11} has revealed vastly improved sensitivity to imaging planets and disks including numerous new discoveries from archival data \citep[][]{spp14,cpc16, szb22}.  Similar methods and algorithms may be applicable to \textit{JWST} images, provided the correct observing strategy and data processing methods are identified as early as possible, and used consistently going forward.  Although the \textit{JWST} instruments have been characterized during instrument testing, neither the in-flight performance of each instrument, nor the optimal strategy for obtaining data in flight, or even the optimal ways to post-process the data are well understood. 

The finite lifetime of the \textit{JWST} mission means that it is essential to correctly optimize and implement the observing strategies as early as possible for gaining spectrophotometry and images of exoplanets and circumstellar disks.  In anticipation of this need, in 2009 the Space Telescope Science Institute (STScI) appointed the \textit{JWST} Advisory Committee (JSTAC) to provide a recommendation on how best to maximize the scientific return of \textit{JWST}. One of the recommendations was to create a clear pathway ``to enable the community to understand the performance of \textit{JWST} prior to the submission of the first post-launch Cycle 2 proposals that will be submitted just months after the end of commissioning''\footnote{The full description of the recommendations of the JSTAC can be found at: http://www.stsci.edu/jwst/about/history/jwst-advisory-committee-jstac}. In response, the STScI Director created the \textit{JWST} Early Release Science (ERS) programs, comprised of $\sim$500 hours of Director's Discretionary Time (DDT).  This program was to be characterized by open community access to substantial, representative datasets, in key instrument modes.  Along with the SEPs described in \S4,  this effort will support the broader community in the preparation of future \textit{JWST} proposals, and engage a broad cross-section of the community to familiarize themselves with \textit{JWST} data and its scientific capabilities.\footnote{More information on the ERS programs can be found at: http://www.stsci.edu/jwst/science-planning/calls-for-proposals-and-policy}

From 2016-2017 numerous members of the community engaged in research related to the direct imaging and spectroscopy of exoplanets and circumstellar disks self-organized to formulate a program designed to address the key questions about \textit{JWST} performance that will inform future proposal cycles. 
A total of 106 proposals were submitted, and 13 proposals were ultimately selected spanning a range of science disciplines including extrasolar planets, solar system, stellar populations and stellar physics, Galaxies and intergalactic medium, and host galaxies to massive black holes.

\section{Overview of ERS Program 1386}
Our Program ``High Contrast Imaging of Exoplanets and Exoplanetary Systems with JWST'' (Program 1386) was ultimately awarded 55 hours of DDT to utilize all four \textit{JWST} instruments, and assess the performance of the observatory in representative modes that are highly applicable to our community going forward.  In addition, our team has been tasked to:~1) optimize data calibration and analysis methods; 2) make clear recommendations to our community about the best practices for \textit{JWST} observing; and 3) generate a set of SEPs, tools that will be essential for the broader community to plan for Cycle 2 proposals (\S4). The ultimate outcome of this program will be to rapidly establish the optimal strategies for imaging and spectroscopy of exoplanetary systems going forward, and provide the community with a set of tools to help prepare the strongest proposals for Cycle 2 and beyond.  

At the same time, our program will showcase the transformative science expected from \textit{JWST} related to the direct characterization of planetary systems such as highly sensitive coronagraphy, direct spectroscopy into the thermal infrared, and interferometry.  Table 1 provides an overview of the targets we will observe and their basic properties.  The exoplanet HIP\,65426b \citep{cdl17} and substellar object  VHS\,J125601.92-125723.9b \citep[][hereafter ``VHS\,1256\,b'']{gbp15, msb18} are both young companions to their host stars at wide orbital separations, and HD\,141569A, is a young (5$\pm$3\,Myr) circumstellar disk \citep{cka03} still potentially in the phase of forming a planetary system.

Our program will carry out:~1) coronagraphic imaging of HIP\,65426b and 2) HD\,141569A at wavelengths extending to 15.5\,$\mu$m; 3) spectroscopy of VHS\,1256b at resolution $R$ $\sim$\,1500-3000 out to 28\,$\mu$m; and 4) aperture masking interferometry of a bright star (HIP\,65426) at 3.8\,$\mu$m. Table 2 lists these four primary science tasks and the instruments they will employ, the primary targets for each of these tasks, as well as the objects selected for calibration.  Table 2 also lists the wavelength coverage and spectral resolution, as well as details of the chosen observing mode, e.g.~the NIRCam ``MASK335R'' or MIRI FQPM coronagraphy, NIRSpec Integral Field Spectrograph (IFS) or the MIRI Medium Resolution Spectrograph (MRS).  Table 2 also lists the pure ``science time'' (denoted ``$t_{\rm sci}$'') dedicated to each of these three tasks, as well as the overall time charged to the observatory (denoted ``$t_{\rm obs}$'') reflecting spacecraft, observatory, and instrument overheads.  In total, our program will utilize 55.3 hours of observing time, with 28.7 of these hours dedicated to pure science observations. A majority of this time will be dedicated to coronagraphic observations of HIP\,65426b and HD\,141569A ($\sim$40 hours observatory time, $\sim$21 hours of science). Below we describe each of these components of the program in greater detail.

\begin{figure*}
  \centering
  \includegraphics[width=1.0\textwidth]{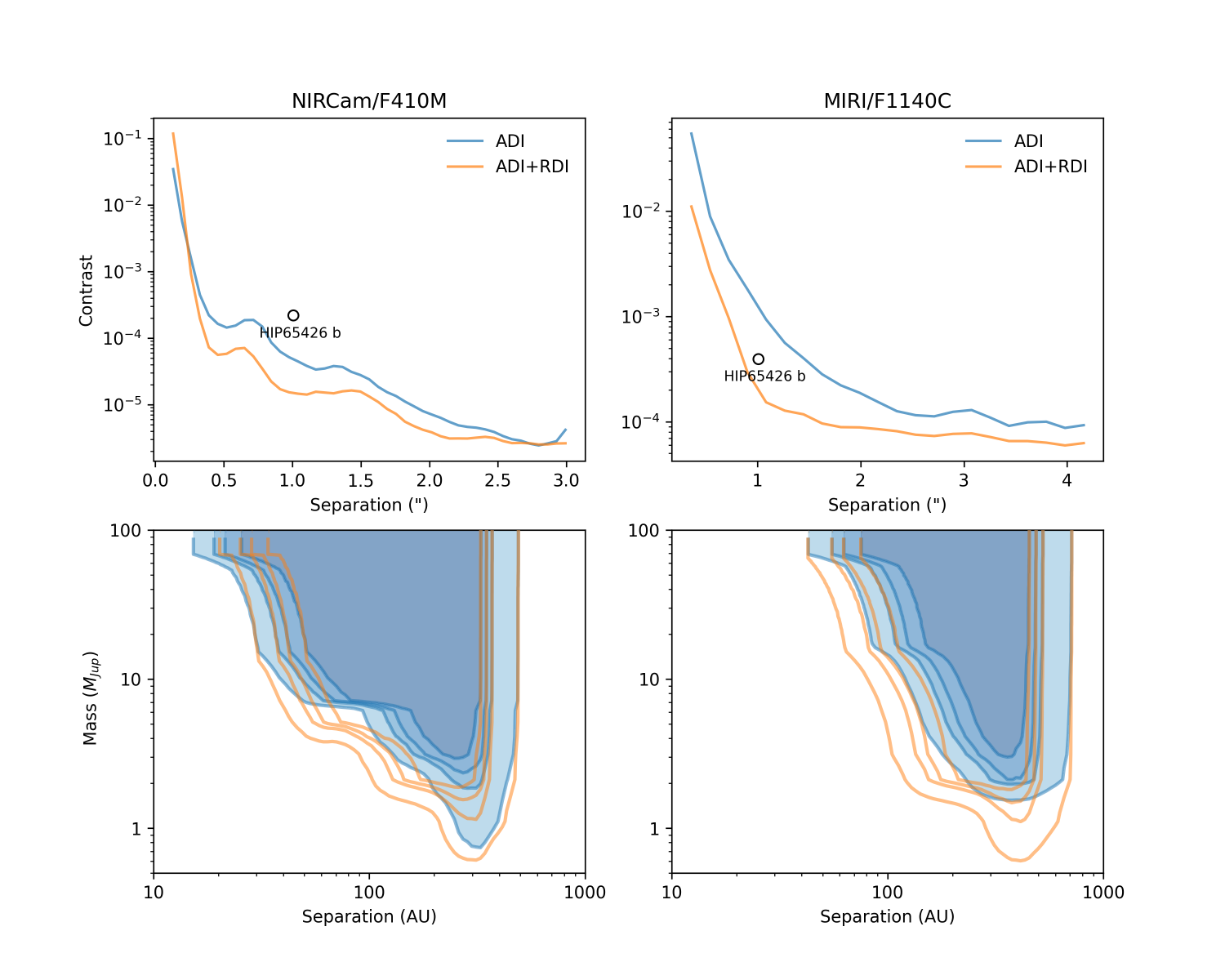}
  \vspace{-0.25in}
  \caption{
  \textit{Top row:} Simulated contrast curves using the updated version of the PanCAKE tool \citep{csd21} showing the predicted contrast performance in the NIRCam F410M (4.1\,$\mu$m) and MIRI F1140C (11.4\,$\mu$m) \textit{JWST} filters.  The two sets of curves illustrate the coronagraphic performance using PSF subtraction based both on Angular Differential Imaging (ADI) as well as a combination of ADI and Reference Differential Imaging (RDI), corresponding to the images shown in Figure~\ref{fig:hip65426_sim}. The theoretical contrast of the HIP\,65426b exoplanet is shown by the circular point. \textit{Bottom row:}~the corresponding detection probability maps \citep{b20} based on these contrast curves with the 30, 60, 70, and 80\% detection probability contours highlighted. 
  }
  \label{fig:detection_prob}
\end{figure*}


\subsection{Coronagraphy of an Extrasolar Planet}
HIP\,65426b is a 6-12\,M$_\mathrm{Jup}$ planetary mass companion \citep{cdl17} with a wide projected orbital separation of 92 AU, and a young age (14$\pm$4 Myr) based on its host star's high probability of membership to the Lower Centaurus Crux association \citep{dhd99,rir11}. This was the first major discovery by the SPHERE ``SHINE'' GTO survey \citep{dcb21, lgl21, vfm21}. Initial photometric measurements from 1 to 5$\mu$m \citep{csb19, sqt20} and medium resolution spectroscopy \citep{pbc21} of this object are consistent with a dusty, low surface gravity atmosphere with mid/late-L spectral type.  The angular separation of 830\,mas and contrast of $\sim$\,$10^{-4}$ relative to the host star make this object an ideal early target to be observed with the NIRCam and MIRI coronagraphs.  Figure~\ref{fig:hip65426_sim} shows synthetic coronagraphic images calculated using an updated version \citep[described in greater detail in][]{csd21} of the Pandeia Coronagraphy Advanced Kit for Extractions, \citep[PanCAKE,][]{ppv18, gbb18}.  PanCAKE is a Python-based package which extends the capabilities of the primary \textit{JWST} exposure time calculator (ETC) Pandeia \citep{ppl16} to be applicable to the coronagraphic observing modes of \textit{JWST}. 

Figure~\ref{fig:hip65426_sim} also highlights the effect of various image post-processing strategies on the significance of the detection of the underlying planet HIP\,65426b.  The central panels of Figure~\ref{fig:hip65426_sim} show the detection using a simple framework based on ADI, taking advantage of the fact that HIP\,65426 will be observed in two configurations separated by a physical ``roll'' of the observatory of a few degrees.  This well-tested methodology is based on the heritage of ``roll deconvolution'' methods developed for \textit{HST} \citep{ ksb12,sgh14}. Figure~\ref{fig:hip65426_sim} also shows the improvement in contrast that can be gained by tapping into a larger pool of PSF reference images to perform Reference Differential Imaging \citep[``RDI,''][]{shp11,rnm19, szb22}, and labelled ``ADI+RDI'' in the figure. This method takes advantage of the ``small-grid dither'' strategy in which repeated observations of a calibration star are obtained at six or nine positions separated by sub-pixel spacings, with the goal of more broadly sampling PSF variations due to varying placement behind the corongraphic masks. 

Numerous studies have demonstrated the improved coronagraphic suppression at small inner working angles by tapping into a much larger ``library'' of reference images \citep[e.g.,][]{shp11, spp14, cpc16, rnm19, wmr21, szb22}.  This potential improvement in inner working angle is even more crucial for \textit{JWST} which typically has an angular resolution of $\sim$100-140\,mas at 3-4\,$\mu$m, compared to the $\sim$30-50\,mas resolution of 8-10m ground-based telescopes in the near-infrared.  This improvement is most pronounced with coronagraphic datasets that a) have a large number of PSF reference images; and b) are highly stable in time, such as is expected to be the case for \textit{JWST}. Indeed, the pool of reference images will inevitably grow over the lifetime of \textit{JWST}, and our ability to suppress starlight at small inner working angles will improve over time as our library of reference images grows.  Thus, it is of paramount importance to determine the correct observing strategy early in the mission, so that the community may tap into the most uniform pool of reference images possible.  The coronagraphic datasets obtained as part of this ERS program will mark the first step in this process.


Figure~\ref{fig:hip65426_photometry} (top panel) shows the planned photometric coverage for the coronagraphic observations of HIP\,65426\,b. As listed in Table 2, the total wavelength coverage for this portion of the program spans from 2.50\,$\mu$m to 15.5\,$\mu$m using the NIRCam ``MASK335R'' (335 round coronagraphic mask) F250M, F300M, F410M, F356W, and F444W filters, as well as the F1140C and F1550C MIRI filters coupled to the FQPM.  The transmission profiles of each filter are shown in the top panel of Figure~\ref{fig:hip65426_photometry}, and the horizontal error bars on each point indicate the filter widths (in $\mu$m).  The continuous curve in the top panel of Figure~\ref{fig:hip65426_photometry} shows a model for a \teff=1340\,K and \logg=3.95 object, and is based on the results presented in \cite{csb19}.  The photometric points in the figure were calculated by integrating the model over the bandpass of each of the filters.  

The bottom panel of Figure~\ref{fig:hip65426_photometry} shows the power of the longer wavelength data that will be available in this ERS program to characterize the atmospheres of substellar objects. The figure shows two synthetic atmospheric models taken from \cite{ptb20} which presents a set of theoretical atmosphere and evolutionary models for very cool brown dwarfs and self-luminous giant exoplanets generated using the one-dimensional radiative-convective equilibrium code ATMO \citep{tac16}.   The figure shows an atmosphere model based on equilibrium atmospheric chemistry (blue curve), as well as a model characterized by disequilibrium chemistry (orange curve).  Evidence for exoplanet atmospheres characterized by disequilibrium chemistry already exist \citep[e.g.,][]{she12,kbm13, msm20}, and atmospheres can be driven out of equilibrium due to processes that happen on timescales faster than the net chemical reaction timescales that would bring the atmosphere back to equilibrium \citep[e.g., rapid vertical atmospheric mixing driven by gravity waves or convective overshooting;][]{fls96, kzm18}. The models plotted rely on the values of 1618K and log($g$) = 3.78 for the HIP\,65426\,b system from \citep{csb19} which derived these best fit parameters using a combination of 1-2.2\,$\mu$m spectroscopy and photometry from VLT-SPHERE, as well as $L^{\prime}$ and $M^{\prime}$ photometry from VLT-NACO. When only considering the band-averaged brightnesses, the two models are virtually identical at 2.50 $\mu$m, but are very clearly differentiated at 3.56\,$\mu$m.  
Of course, other atmospheric properties besides the presence of disequilibrium chemistry can affect the flux near 3.56\,$\mu$m. For example, a higher effective temperature ($T_{\rm eff}$) or a lower overall atmospheric metallicity leading to decreased CH$_4$ would change the strength of the 3.56\,$\mu$m feature. However these degeneracies can be broken via other parameters accessed by our datasets, e.g.~constraints on the effective temperature afforded by the long wavelength coverage, and contraints on the atmospheric metallicity from host star abundance measurements.  
Thus, the very high expected photometric precision of NIRCam, combined with long wavelength coverage and host star abundance constraints means that our data will be sensitive to the effects of disequilibrium chemistry, which has the largest impact near 3.56\,$\mu$m given the broad CH$_4$ feature.

\begin{figure*}
  \centering
  \includegraphics[width=0.9\textwidth]{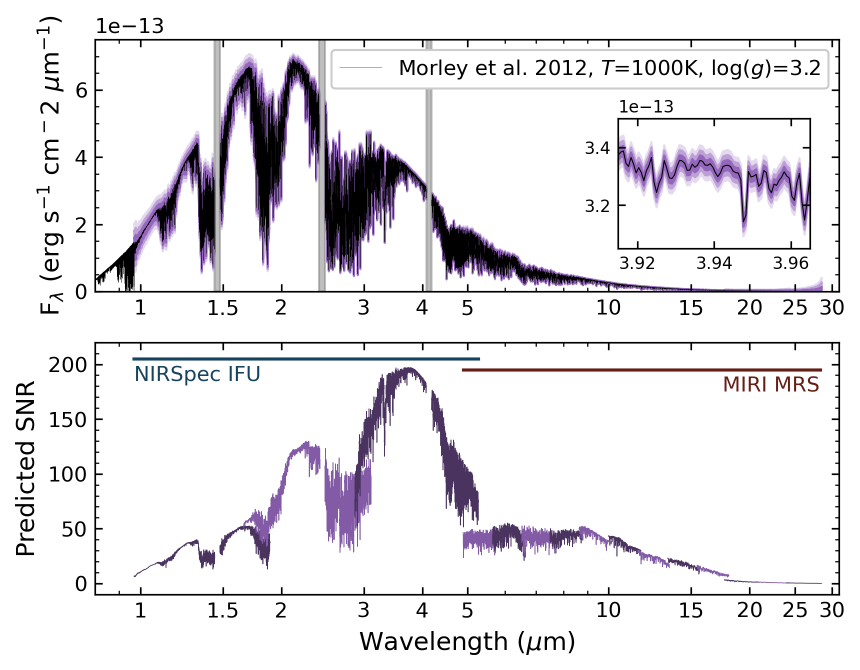}
  \caption{\textit{Top panel:} A synthetic spectrum of the VHS\,1256b planetary mass companion we will obtain using NIRSpec and MIRI, reflecting the actual spectral resolution, as well as a subset of the spectral range that will be covered (inset).  The purple regions signify the 1, 2, 3$\sigma$ confidence intervals in the spectrum, while the grey regions indicate portions of the spectrum where no data will be gathered due to detector gaps.  The input model spectrum is based on a synthetic model described in \cite{mfm12}, which has been used in conjunction with the estimated signal-to-noise ratio defined in the ETC, to determine the confidence intervals shown in the plot. \textit{Bottom Panel:} The predicted SNR for the VHS\,1256b spectrum, as determined from the \textit{JWST} ETC.}  
  \label{fig:vhs1256}
\end{figure*}

The significant photometric uncertainties indicated by the error bars on the existing ground-based data shown in Figure~\ref{fig:hip65426_photometry} (top panel) highlights that it is challenging for these data to provide precise constraints on the theoretical atmospheric models of HIP65426b.  However, the vastly higher precision NIRCam and MIRI photometry beyond $\sim$2\,$\mu$m, will allow much tighter constraints on the allowable atmospheric parameters.
Along with the vastly increased photometric precision, the longer wavelength coverage provided by our observations will provide a much greater source of leverage for performing an atmospheric retrieval analysis, similar to that described in \S\ref{sec:retrievals}. 
Specifically, we will carry out forward model fitting to extract the best fit model parameters reflecting the atmospheric compositions as well as a retrieval analysis to gather constraints on molecular abundances (e.g., CO, H$_2$O, CH$_4$), disequilibrium chemisty, and the presents of clouds.   
With high precision NIRCam and MIRI data beyond $\sim$2\,$\mu$m, particularly out to $\sim$15\,$\mu$m on the tail of the SED, our dataset will also provide a much better constraint on the broadband emission of HIP65426b.  

To demonstrate the sensitivity of coronagraphic observations using both NIRCam and MIRI, Figure~\ref{fig:detection_prob} shows the expected achievable contrast corresponding to the various stages of image post-processing at 4.1\,$\mu$m and 11.4\,$\mu$m shown in Figure~\ref{fig:hip65426_sim}.  The contrast curves were calculated using a modified version of PanCAKE described in \cite{csd21} for the F410M and F1140C filters that we will utilise in this program. Rather than the 5$\sigma$ contrast curves used historically \citep[e.g., ][]{hos07}, the curves are based on a $3\times10^{-7}$ false positive fraction, taking into account the statistical correction due to small angular separations \citep{mmw14}.
The contrast curves presented in Figure~\ref{fig:detection_prob} are also calibrated for the two-dimensional throughput of the coronagraphs as well as the inherent throughput of the Karhunen-Lo\`eve Image Projection subtraction routine \citep[``KLIP,''][]{spl12} which uses a principal component analysis approach to build a synthetic reference image for each science image from a library of reference images.  Importantly, these contrast curves also take into account dynamical changes in the wavefront error due to the varying thermal state of the optical telescope element (OTE) due to telescope slews, small variations in the tension of the physical structure of the telescope, or vibrations within the OTE (e.g.~due to the OTE stray light baffle/insulation closeouts).  
For each wavelength, Figure~\ref{fig:detection_prob} displays a curve assuming only a simple roll subtraction (termed ``ADI''). Further, Figure~\ref{fig:detection_prob} also shows the improvement in contrast by a factor of a $\sim$few to ten that can be gained by taking advantage of the small-grid dither strategy (labelled ``ADI+RDI'' in the figure).  As a calibration strategy for our observations of HIP\,65426\,b, we will observe the B2IV star HIP\,68245, which is close on the sky to HIP\,65426, but has $\sim$8 times the brightness in the mid-infrared (K=4.49 versus K=6.77 for HIP\,65426). The enhanced brightness of HIP68245 allows us to obtain a nine-point dithered observation of this calibrator without a prohibitively large amount of observing time. Table 2 shows that this calibrator star will be observed in roughly half the charged time as our observations of HIP\,65426 (6.0 hours versus 10.4 hours), but with nine times the data volume, enhancing the overall calibration process.  

\begin{figure}
  \centering
  \includegraphics[width=0.45\textwidth]{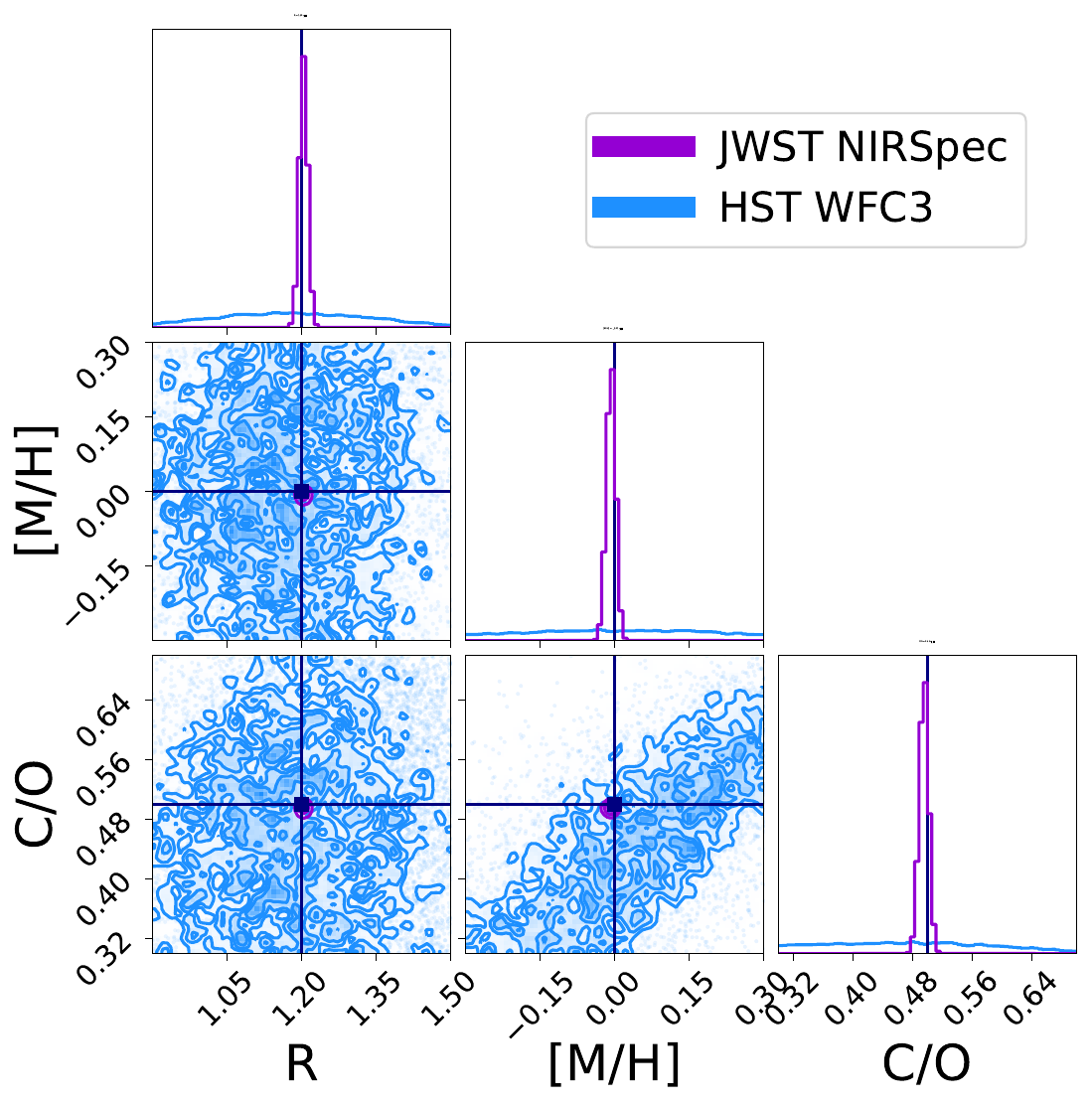}
  \caption{A demonstration of a simulated atmospheric retrieval analysis of VHS\,1256b using simulated \textit{HST WFC3} and \textit{JWST} NIRSpec data.  The dataset for NIRSpec was simulated at an SNR of 50 - 250 using the \textit{JWST} ETC, and a uniform SNR=50 for WFC3.  Retrieval analyses were performed to derive the model parameters, demonstrating the power \textit{JWST} can offer for precisely and accurately constraining crucial model parameters within retrievals.  
  }
  \label{fig:niall_retrieval}
\end{figure}


To calculate a representative estimation of the sensitivity to low mass companions (in Jupiter masses) as a function of projected orbital separation (in AU) of these observations, the lower panels of Figure~\ref{fig:detection_prob} also show the corresponding two-dimensional detection probability maps based on the synthetic contrast curves. These detection probability maps, calculated using the Exoplanet Detection Map Calculator \citep[``Exo-DMC,''][]{b20}, effectively demonstrate the degree of ``completeness'' to planetary mass companions as a function of mass and semi-major axis for a given coronagraphic observation. 
For a given planet mass and semi-major axis, Exo-DMC simulates an ensemble of orbital geometries with varying planetary eccentricity and inclination to determine which fraction of those realizations, identified by the various completeness contours, would be detected based on the supplied single contrast curve, thus providing a measure of the completeness of the observations. 
This calculation uses the evolutionary models of \citet{ptb20} assuming an age of 14\,Myr to convert from a contrast to a physical mass, and so these plots are dependent on the underlying evolutionary models as well the formulation of atmospheric process that they employ.

\subsection{Spectroscopy of a Planetary Mass Companion}
VHS\,1256b \citep{gbp15, msb18} is a wide separation ($\sim$103\,AU), substellar companion (19$\pm$5\,M$_\mathrm{Jup}$) to a young M7.5 binary star \citep{ssk16, rcw16}. While the VHS\,1256 system is not a member of any known kinematic young moving groups, \citep{gbp15} derive an age of 150-300 Myr, consistent with its low surface gravity.  With a spectral type of L7, infrared parallax measurements \citep{dlm20} show that VHS\,1256b shares a region in a colour-magnitude diagram with other planetary mass companions that are near the L/T transition, and close to the deuterium burning limit such as HR 8799b, HD 203030B, and 2MASS\,J22362452+4751425\,b \citep{mmb08,mh06,blm17}.  

As shown in Table 2, we will observe VHS\,1256b using the NIRSpec IFS from 1.66 to 5.27\,$\mu$m with the G140H, G235H,  G395H gratings ($\lambda/\Delta\lambda\sim$1500-3500). Our selected exposure time provides a signal-to-noise ratio (SNR) of $>$20 across the majority of the wavelength range with many features detectable to a SNR of $>$100. However, these estimations are based on initial simulations, and the actual data may indeed have different SNR values. 
To probe silicate cloud features at 8 to 12\,$\mu$m for the first time ever for low surface gravity objects, we will also perform mid-infrared spectroscopy on VHS\,1256b from $\sim$5 to 28\,$\mu$m using the MIRI MRS in all four channels with all three dispersers. 
To improve the sensitivity of the observations and minimize detector effects, we will utilize four dither positions for both the NIRSpec and MIRI observations. 

Historically, the main obstacle to obtaining unbiased spectroscopy of planetary mass companions from ground-based spectrographs coupled to AO systems has been contamination from uncontrolled, residual scattered starlight within $\sim$1-2$^{\prime\prime}$ \citep{hos07,mld06,p16}.  The relatively large angular separation ($\sim$8$^{\prime\prime}$) of VHS\,1256b from its host binary pair means that our JWST observations should have minimal contamination of residual scattered starlight. 
Thus, JWST Spectroscopy of this object will allow high-fidelity detections of molecules from $\sim$2 to 28\,$\mu$m that will be vital for the understanding of cloud physics driven by dust or thermochemical instability under low gravity conditions, as well as searching for evidence of silicate clouds at 8 to 12\,$\mu$m. 
The VHS\,1256b spectroscopy should also serve as a valuable community dataset for testing spectral extraction algorithms for NIRSpec and MIRI.

Figure~\ref{fig:vhs1256} shows an example of the signal-to-noise and wavelength coverage for the spectroscopy of the VHS\,1256b companion that will be obtained with our program.  The figure shows a theoretical model atmosphere of a substellar object with the currently best-fit derived effective temperature and gravity  parameters for VHS 1256b \citep{zbm20}.  To calculate the effective uncertainty in the model as a function of wavelength, this model was then used as input into the \textit{JWST} ETC, which in turn provides a measure of the expected signal-to-noise as a function of wavelength.  Dividing the original model spectrum by this signal-to-noise spectrum effectively provides a measure of the noise as a function of wavelength, and this noise is represented by the purple shaded regions in the figure. The lower panel of Figure~\ref{fig:vhs1256} shows the SNR as a function of wavelength calculated using the ETC.   

\subsubsection{A Suite of Atmospheric retrievals for VHS\,1256b}\label{sec:retrievals}
We will employ an extensive retrieval analysis to the reduced spectroscopic data for VHS\,1256b, testing various approaches. 
Exploring a range of atmospheric chemistry parameters will probe whether the atmosphere is in chemical equilibrium versus disequilibrium \citep{msm20, msl20}.  
We expect that the quality of our spectra will easily be sensitive enough to fully sample the prominent CO features at $\sim$2.3 and $\sim$4.3\,$\mu$m, as well as numerous H$_2$O and CH$_4$ features across the majority of the spectral range. By unlocking the mid-IR we will also be able to robustly explore potential cloud structures \citep{mkv21} and cloud species present \citep{bfg21}. Due to the extensive variability present on VHS 1256b \citep{bzm20}, and as the data will be collected over a period of hours, we will also probe signatures of variability within our retrieval analysis. 

To simulate the outcome of using an atmospheric retrieval-based approach, Figure~\ref{fig:niall_retrieval} shows a demonstration of a simulated retrieval analysis of VHS\,1256b comparing \textit{HST WFC3} against simulated \textit{JWST} NIRSpec spectroscopy. For this, employing a forward model created via TauREx3.1 \citep{wtr15, wrt15, acw21, acv22} and FastChem \citep{ks18}, we simulated realistic NIRSpec spectroscopy with SNR ranging from $\sim$50 to $\sim$250 generated with the \textit{JWST} ETC.  We also simulated a WFC3 dataset with a uniform SNR of 50. We then ran retrievals using these simulated observations and wide uniform priors on all model parameters. Figure~\ref{fig:niall_retrieval} highlights the novel power JWST can offer for precisely and accurately constraining crucial model parameters within retrievals.  
However, the improved precision shown in Figure~\ref{fig:niall_retrieval} is probably slightly optimistic as:~1) the retrieval algorithm is retrieving its own model input; 2) these simulations use a simplistic forward model
; and 3) we assume perfect data reduction/calibration ignoring any photometric variability during the observations. 
However, we stress that this dramatic increase in precision is driven primarily by the order of magnitude increase in wavelength coverage and spectral points as well as improved SNR.  Nonetheless, Figure~\ref{fig:niall_retrieval} gives a qualitative sense of the level of improvement that can be gained with the advent of \textit{JWST} spectroscopy.

\begin{figure*}
  \centering
  \includegraphics[width=1.0\textwidth]{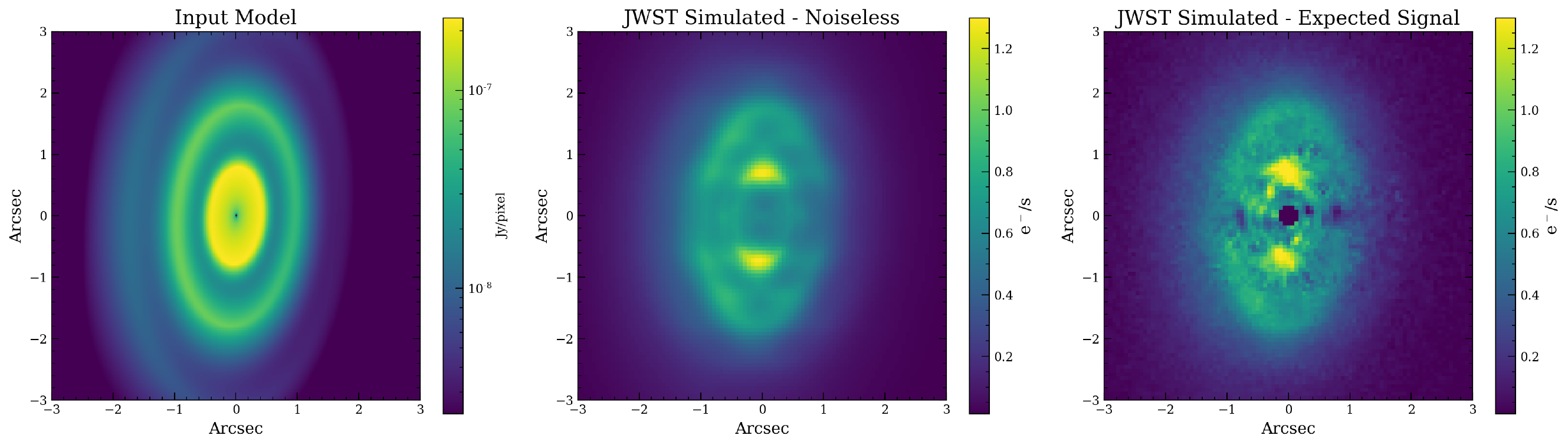}
  \caption{Simulated NIRCam data for our planned observations of HD\,141569A using the F300M filter. The left figure shows our input model of the three rings that comprise the HD\,141569A disk and described in Table 3, while the middle panel shows a representative NIRCam image free of noise.  The right panel shows a simulated image, now including a realistic estimate of the NIRCam noise that we can expect with our ERS observations. }
  \label{fig:hd141569}
\end{figure*}

\subsubsection{Variability Monitoring of VHS\,1256b}
 Quasiperiodic brightness variability is commonly found in field L and T type brown dwarfs, with periods from 1.5-30 hours, and amplitudes ranging from 0.01-27\% \citep{rlj14,mha15}.  Young directly imaged exoplanets are likely even more variable, as their free-floating counterparts appear  to  be  particularly  variable \citep{bvb15,aks16}. VHS\,1256b is the most variable of this group; with measured variability amplitudes of $>$25\% over an 8-hour long \textit{HST} observation \citep{bzm20,zbm20}, it may be the most variable substellar object ever observed. Indeed, VHS 1256b is so highly variable that it may vary by up to 10\% over our three-hour ERS observation. However, it should be noted that the variability of substellar objects is wavelength dependent \citep{mha15, vba20}, and this has been quantified for VHS\,1256b in \cite{zbm20}.  Over $\sim$1.2 hours on-sky, our program will obtain three consecutive NIRSpec IFS spectra covering 1-5 $\mu$m in the following order: 1.66-3.17 $\mu$m, 2.37-5.27 $\mu$m, and 0.97-1.89 $\mu$m. Each spectrum will be taken at four dither positions, providing two temporal samples for each spectral range, separated by $\sim$12 minutes from the central time point of each spectrum. It is worth noting that we will also record ``up-the-ramp'' flux measurements, and thus our observations may be sensitive to variability on even shorter timescales.  In the extreme case, if VHS\,1256b has a similar 1-5\,$\mu$m variability amplitude in this epoch as in \citet{bzm20}, this would translate to a $\sim$few percent change in overall flux across the NIRSpec observation, sufficient to cause non-negligible issues with spectral fitting when stitching together, e.g., the 1.66-3.17 $\mu$m spectrum with the 0.97-1.89 $\mu$m spectrum (the muted variability in the water bands, however, will facilitate stitching together the spectra). This effect is expected to be even more pronounced with the MIRI MRS observations, which occur over a longer timescale. To address this risk, we have initiated a coordinated ground-based observing campaign that will utilize a suite of longitudinally-distributed observing sites to observe prior to, and ideally in parallel with, the ERS observations.  These observations will allow us to correctly stitch together the ERS spectroscopy, and constrain the wavelength-by-wavelength amplitudes and phase shifts.  

\subsection{Coronagraphy of a Young Circumstellar Disk}
HD\,141569A is a young circumstellar disk \citep[5$\pm$3\,Myr,][]{wrb00} that has been a well-studied target of past space-based \citep[e.g.,][]{augereau99, wbs99, kgs16} as well as ground-based \citep{blr15,pbp16,cgc16, bmd20, sbb21} high contrast imaging efforts.  
HD\,141569A is also a unique system for our understanding of both protoplanetary disk dispersal processes as well as the early stages of debris disk systems.  In terms of disk luminosity HD\,141569A sits at the transition between protoplanetary and debris disks \citep[][]{wpk15}, and so is a rare object caught in transition between the two phases. Thus, from a purely scientific point of view, our ERS observations will help to illuminate whether e.g.~the dust within the disk is primordial or secondary, and how the observed morphological structures relate to planet formation.  Our observations of HD\,141569A will demonstrate the power \textit{JWST} holds for characterizing circumstellar debris disks, allowing us to determine the disk's dust size distribution and composition as a function of radius, and search for planet-induced structures in the mid-infrared for the first time.

Given that it is comprised of at least three concentric disk rings, HD\,141569A is also an ideal target to test the sensitivity of \textit{JWST} to extended structures as a function of stellocentric angle, as well as study the effects of various PSF subtraction post-processing algorithms on diffuse structures. We will obtain coronagraphic imaging of HD\,141569A from 3.0 to 15.5\,$\mu$m using both NIRCam and MIRI (Table 2).  By utilizing the F300M and F360M NIRCam filters, We will image the disk on and off the 3.0\,$\mu$m H$_2$O ice absorption band that is difficult to probe with ground-based telescopes.  In addition to using the 10.65$\mu$m and 11.4$\mu$m MIRI filters, we will image the disk at 15.5 $\mu$m, one of the primary wavelengths expected to be used for \textit{JWST} disk imaging going forward. As with the case of the HIP\,65426b observations, for a calibration strategy a brighter calibrator star (HD\,140986, Table 2) will be used to generate a significant volume of PSF calibration images without a prohibitively large amount of exposure time.  The enhanced brightness of HD\,140986 (K=3.64 versus K=6.82 for HD\,141569A) means that we will require 5.4 hours of calibration observations using a five-point dither strategy, compared to 8.1 hours for HD\,141569A.

To extract the signal of the disk in the NIRCam data, we will use the KLIP algorithm to subtract the residual scattered starlight using the dithered images of the calibrator star. We will also use recently-developed PSF subtraction algorithms that are known to better preserve the absolute flux of bright extended structures than PCA-based methods, such as the Non-negative Matrix Factorization \citep{Ren2018} and Data Imputation \citep{Ren2020}, and present a comparison of the results obtained with the different algorithms. The brightness ratio of the disk between the F300M and F360M filters will allow us to constrain the composition of the dust, in particular the ice fraction in the different belts of the system. This will reveal whether the dust is cometary or asteroid-like in nature, and will provide direct observations of how water is radially distributed in the outer parts ($\gtrsim$50\,AU) of this young system. For this analysis, we will use both a forward modeling approach as well as an empirical approach to measure the surface brightnesses and scattering phase functions in both filters. In both cases, we will compare the data to models created with the MCFOST radiative transfer code \citep{Pinte2006} which can simulate dust belts with various chemical compositions and grain size distributions.

In the mid-infrared, where the existing resolved images of the HD\,141569A disk are from ground-based instruments and only sensitive to the warm inner disk at $\sim$50\,AU \citep{mtp10, Thi2014, Mawet2017}, very little is known about the thermal properties of the outer disk components at $\sim$200 and $\sim$400\,AU. Similarly, the SED of the system is dominated by the emission of the inner disk between $\sim$8\,$\mu$m and $\sim$40\,$\mu$m \cite[see][]{Mawet2017}, which also prevents any constraints on the thermal properties of the outer belts. Our MIRI imaging will be sensitive for the first time to the mid-IR emission of the outer belts and spatially separate them from the inner belt emission. These independent measurements of each disk component will allow us to disentangle their individual thermal properties. Simulations of the data we will collect with MIRI based on the model from \citet{Thi2014}, indicate that the two outer belts should be directly visible in the three MIRI filters, with little contamination from the star, while the innermost belt will be impacted by both the residual host starlight and obscuration from the 4QPM coronagraph. Given this, HD\,141569A thus has the added bonus that this target will serve as a good feasibility test for the MIRI coronagraphs to check whether future observations of protoplanetary disks might be limited by the (nearly) unresolved warmer inner ring components.

\begin{table}
  \caption{Parameters for HD\,141569A model}\label{tabModel}
  \centering 
  \begin{tabular}{l|rrr}
  \hline\hline
System Parameters                    &      &           &       \\
\hline
Distance (pc) 						 &      &   110.6   &       \\
Inclination (\degr)					 &      &   56.9    &       \\
Position angle (\degr) 				 &      &   356.1   &       \\
Vertical aspect ratio $h$ 			 &      &   0.04    &       \\
F300M Stellar flux (Jy) 		     &      &   0.71    &       \\
F360M Stellar flux (Jy)      	     &      &   0.50    &       \\
                         		     &      &           &       \\
\underline{Ring Parameters}                      & \underline{Ring 1} & \underline{Ring 2} & \underline{Ring 3} \\
Semi-major axis (\arcsec) 			 & 0.8  & 1.8       & 3.2    \\
Critical radius (au) 				 & 88   & 199       & 354    \\
Inward  density index $\alpha_{in}$  & 1.5  & 5.5       & 5.5    \\
Outward density index $\alpha_{out}$ & -7.5 & -7.5      & -3.5   \\
HG parameter $g$ 	                 & 0.0  & 0.1       & 0.3    \\
Peak reflectance ($\times10^{-4}$~\arcsec$^{-2}$) 		& 8.0 & 2.8 & 0.4 \\
  \hline
  \end{tabular}
\label{disktable}
\end{table}

After removing the residual starlight using the reference star observations, we will use the MIRI imaging for three main applications. First, the 10.65\,$\mu$m image will let us investigate the transition between the scattered-light and thermal regimes in each belt. While the inner disk 10.65\,$\mu$m image should be largely dominated by thermal emission, scattered-light may be significant in the images of the outer disks. We will use both the MIRI F1065C image and the F300M/F360M NIRCam images to disentangle the scattered-light and thermal contributions of each belt. Second, we will use the 10.65\,$\mu$m and 11.4\,$\mu$m images to measure the radial distribution of Polycyclic Aromatic Hydrocarbons (PAH). Unresolved Spitzer IRS observations have previously detected PAH in this system \citep{skf05, bjf09}, but it is currently unclear if they are distributed uniformly throughout the disk, in only parts of the disk, or in an outflowing halo of small grains. 
The F1065C and F1140C datasets will be sensitive to the broad PAH emission feature at 11.3\,$\mu$m, allowing us to constrain the PAH abundances in each ring. Third, we will use the 10.65\,$\mu$m and 15.5\,$\mu$m images to measure the continuum thermal emission of each ring individually and construct a temperature map of the whole system. This will allow us to derive the emissivity of the grains and thus the minimum grain size of each ring through radiative transfer modeling. The distribution of small micron-sized dust inferred from these observations will be particularly powerful when coupled with information about the distribution of larger millimeter-sized dust bodies and CO inferred from ALMA observations \citep[e.g.,][]{dpd20}. This will give the grain size distribution over a wider range of particle sizes, and allow us to test for models of grain radial transport through interaction with a remnant gas disk \citep[e.g.,][]{ta01}. 

Finally, in all NIRCam and MIRI images of HD 141569A, we will scrutinize the disk morphology to look for direct and indirect evidence of planets. The system shows large cavities that suggest dynamical clearing by planets \citep[e.g.,][]{w05, kgs16, pbp16}, and asymmetries in the dust density distribution possibly caused by a massive collision between planetesimals \citep{sbb21}.  The longer wavelengths of JWST offer a more favorable contrast regime to probe for the presence of giant planets than existing shorter wavelength observations, as discussed in \S1.2. The NIRCam F360M  and the MIRI F1065C images will be the best suited to look for planets within the gaps of the system, as they offer the best compromise between contrast limits and spatial resolution.

To simulate the HD\,141569A NIRCam data, we use the basic physical parameters listed in Table 3. 
We first generate a model composed of three rings, of respective semi-major axes 0\farcs8, 1\farcs8, and 3\farcs2. Assuming a distance of 110.6~pc \citep{gbv18}, these values correspond to radii of 88~AU, 199~AU, and 354~AU, based on the findings of \citet{kgs16}, and \citet{mla01}. We simulate a common 56.9\degr{} inclination from a face-on configuration and a common position angle 356.1\degr, east of north for the three rings \citep{pbp16}. We use the ring parametric model described in \citet{alm99} to model the dust density distribution. The radial profile is modeled with an increasing power law (index $\alpha_{in}$) inward of the critical radius quoted above and a decreasing power law (index $\alpha_{out}$) outward from this radius. The power law indices used to model each of the three rings are displayed in Table~\ref{disktable}. The vertical profile of the disks follows a Gaussian profile with a standard deviation increasing radially with an aspect ratio $h=0.04$ \citep{t09}. 


We simulate the scattering phase function of the rings using a Henyey-Greenstein (HG) in parametric model with an asymmetry parameter $g$ of 0.0 for the inner ring \citep{kgs16}, 0.1 for the middle ring \citep{wbs99}, and 0.3 for the outer ring \cite{cka03}. We use optically thin scattered light propagation from the central star, computed with the same code as in \citet{mgp15} to simulate synthetic images of the three rings. We then scale the peak surface brightness of each disk to the product between the stellar flux at the considered wavelength and the disk peak reflectance. The left panel of Figure~\ref{fig:hd141569} shows this model, free of any detector noise.  We obtain the peak reflectance of each ring from previous \textit{HST/NICMOS} data obtained in the F160W filter (${\sim}1.6\,\mu$m) from the ALICE archival re-analysis \citep{hcs18}, assuming that the dust reflectance is identical at the longer NIRCam wavelengths (grey scattering). We compute the stellar flux in the NIRCam F300M and F360M filters from the \textit{JWST} online ETC using a Phoenix A1V spectrum scaled to the $K=6.82$ HD\,141569A magnitude \citep{csv03}. The stellar fluxes in the two filters and the peak reflectance of the three rings are reported in Table~\ref{disktable}. Figure~\ref{fig:hd141569} also shows a simulated image of the data we can expect to receive from our NIRCam imaging of this system, incorporating realistic noise sources.

\subsection{Aperture Masking Interferometry}
\begin{figure}
  \centering
  \includegraphics[width=0.5\textwidth]{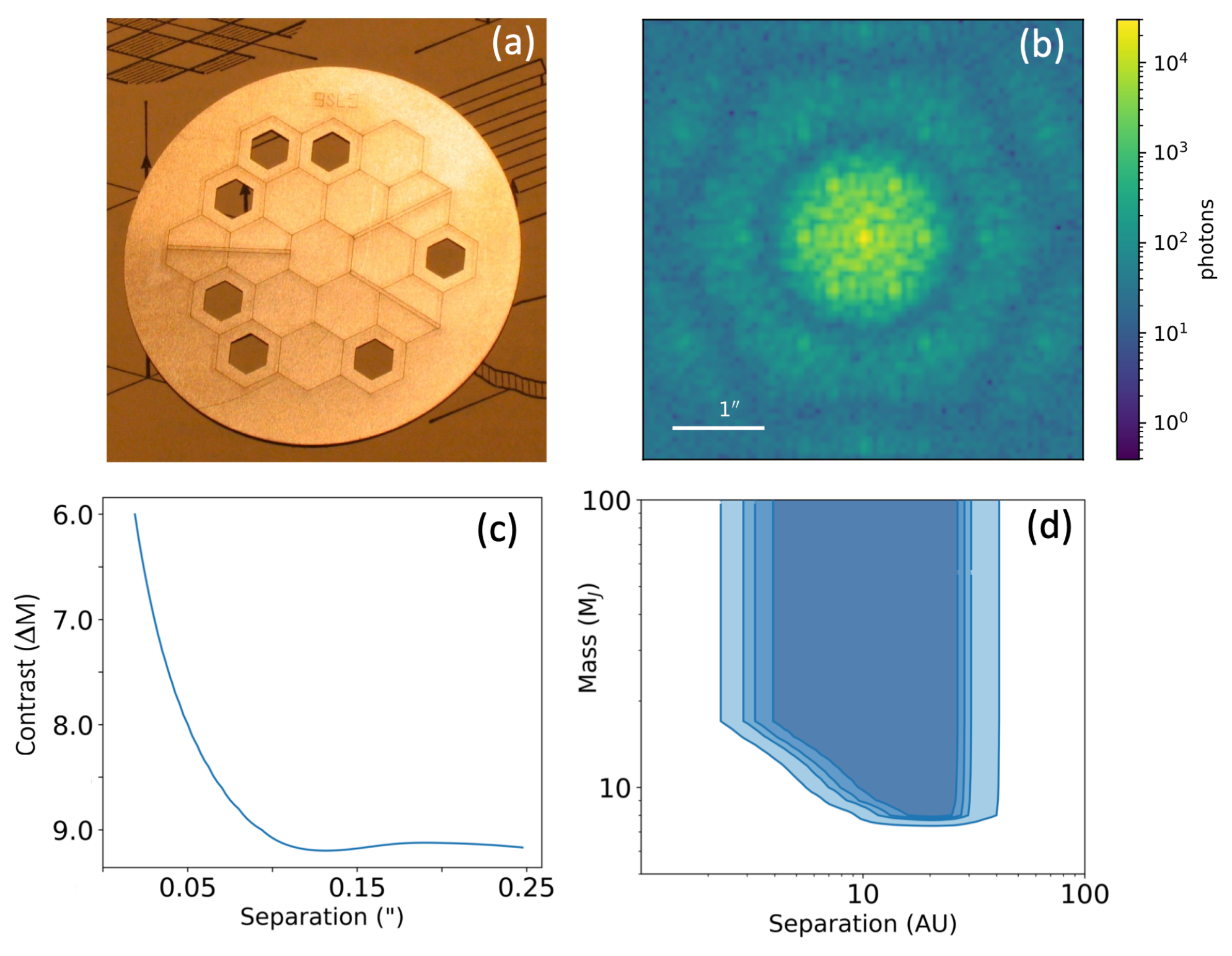}
  \caption{\textit{(a):} An image of the sparse aperture mask installed within the NIRISS instrument \citep{slt10} for the purpose of performing Aperture Masking Interferometry \citep{asg14, gps15, ss19}.  \textit{(b):} A simulated NIRISS interferogram in the F380M filter for an integration time of $\sim$1.28\,s.  \textit{(c):} a simulated contrast curve calculated using the methods described in \citep{ss19}, and \textit{(d)} the corresponding detection probability map showing the 30, 60, 70 and 80\% detection probability contours.}
  \label{fig:niriss_ami}
\end{figure}

Given its 6.5m effective aperture, for the purpose of gaining direct images of self-luminous exoplanets, \textit{JWST} will be ultimately limited by its inner working angle defined by its diffraction limit of $\lambda/D{\sim}$120-140 milliarcseconds at 3-4\,$\mu$m. For the nearest star forming regions where the planet formation process is still likely ongoing \citep[e.g., Taurus, $\sim$2\,Myr,][]{kim11} or has recently ceased, \citep[e.g., the Scorpius-Centaurus Association, $\sim$11-16\,Myr,][]{dhd99,pm08} with characteristic distances of $\sim$\,140\,pc, this angular resolution corresponds to physical separations of 17-20\,AU, well outside the typical water ice line separations of $\sim$2-3\,AU where planet formation is thought to be most efficient \citep{fjm19,fmp19, frh21}. By dividing the full \textit{JWST} pupil into a set of smaller sub apertures, the technique of Aperture Masking Interferometry \citep[``AMI,''][]{tmd00,i13,ss19} can provide modest sensitivity inside the diffraction limit of \textit{JWST}. This technique has been used extensively from the ground \citep[e.g.][]{hci11,ki12,hki15,sfe15}, but never in space, and so obtaining a representative dataset as early as possible is critical for future planning. The NIRISS instrument is equipped with a sparse aperture mask \citep{slt10} containing seven holes, providing 21 distinct ``non-redundant'' baselines that can be used with the medium-band NIRISS filters from 3.8-4.8$\mu$m (F380M, F430M, and F480M). Figure~\ref{fig:niriss_ami} shows an image of the seven-hole mask, as well as a simulation of the resulting interferogram on the detector using the NIRISS F380M filter generated using the methodology described in \cite{ss19}, assuming a point source for a  total integration time of 1.28\,s.  With the goal of identifying the limiting systematics inherent in AMI observations we will obtain 3.8\,$\mu$m sparse aperture masking observations of the HIP\,65426 system. Past studies have demonstrated that systems with already-identified wide separation planets may also have additional planets at smaller angular separations \citep{mzk10, wak19,lrn20,nll20}, and this effort will place constraints on any additional companions in this system. Our proposed observations of the HIP\,65426 system are predicted to reach a contrast ranging from 6 to $\sim$9 magnitudes within the 120\,mas \textit{JWST} diffraction limit at 3.8\,$\mu$m, providing sensitivity to substellar companions within $\sim$14\,AU for this system (Figure~\ref{fig:niriss_ami}, lower panels).   

\section{Analysis Plan \& Science Enabling Products for the Broader Community}
Our consortium is comprised of a set of four working subgroups, corresponding to: 1) exoplanet coronagraphic imaging with NIRCam and MIRI; 2) debris disk imaging with the same two instruments; 3) spectroscopy of substellar companions with NIRSpec and MIRI; and 4) NIRISS aperture masking interferometry. 

Our team will be responsible for delivering a set of SEPs described in more detail below that will be widely available, and will assist the community in preparing future \textit{JWST} proposals.  A final working group will have the task of ensuring the quality of the SEPs. This working group will validate the utility of our internal SEPs for preparing future proposals and provide immediate feedback to the individual working groups regarding the quality of the SEPs. 


\subsection{Science-Enabling Products}
In this section we describe in detail each of the SEPs our consortium will deliver: 

\textbf{Coronagraphic imaging simulator:} We will widely distribute an update \citep{csd21} to the \textit{JWST} simulation tool PanCAKE \citep{ppv18, gbb18} that is an adaptation of the \textit{JWST} ETC \textit{Pandeia}.  This tool will model the impact on coronagraphic imaging due to details of the observatory such as target acquisition offsets, time evolution of the incoming stellar wave front, spacecraft rolls, and small grid dithers. This tool will also preserve all the capabilities of the \textit{Pandeia} simulator for generating coronagraphic scenes with MIRI and NIRCam, but with added accessibility and functionality.   

\textbf{High-Contrast Imaging Analysis Pipeline:} We will release a publicly available Python-based high-contrast imaging pipeline to process \textit{JWST} NIRCam and MIRI coronagraphic imaging data and generate contrast curves based on the existing PyKLIP pipeline \citep{wrd15}, a Python adaption of the commonly used KLIP algorithm that implements several sophisticated state-of-the-art methods for PSF subtraction and forward modelling of systematics. 
We will test our pipeline against other leading high-contrast pipelines to ensure consistent results. An online manual will describe the algorithm and provide a tutorial to run the pipeline using the ERS HIP\,65426 data. This pipeline will be optimized for detecting point-sources, but will also produce scientifically usable images of resolved structures, which we will test and document using the HD\,141569A images. By the Cycle 2 proposal deadline, users will be able to simulate realistic scenes with the modified version of PanCAKE that will be provided, which can then be interfaced with PyKLIP to produce realistic simulated reductions. The pipeline will also enable immediate analysis of data from future \textit{JWST} cycles. 

\textbf{Contrast Curves:} We will use our coronagraphic NIRCam/MIRI imaging and NIRISS aperture masking data of HIP 65426 to provide contrast curves showing the point source sensitivity of \textit{JWST} as a function of wavelength and stellocentric angle. To reflect various post-processing strategies, we will provide multiple curves per bandpass of contrast (in units of delta-magnitude) vs. angular separation. 

\textbf{Aperture Masking Analysis Pipeline:} We will release a publicly available Python-based pipeline to process simulated NIRISS AMI data generating a contrast curve and detection tests for single point-source companions. 
This will be in place prior to the Cycle 2 proposal deadline along with an online tutorial to guide users to reproduce our team’s results. The pipeline will enable rapid sensitivity estimates for proposers in future cycles, as well as fast analysis of future datasets.

\textbf{Communication of Best Practices:} The wide variety of \textit{JWST's} high-contrast modes will spawn numerous technical questions about best practices for observations and data post processing. For example, the best strategy for utilizing \textit{JWST} in its coronagraphic mode for various point source science cases is \textit{a priori} not obvious. This same question has been an active area of research in the high-contrast imaging community for the past decade, even for ground-based efforts \citep[e.g.,][]{xmn18, wmr21}. 
For \textit{JWST}, there is an expectation that RDI will be more effective than ADI due to its limited ability to achieve large roll angles, but it is a clear goal of this program to establish the point in an observation sequence where one method becomes beneficial relative to another.   While our ERS program is not intended to be comprehensive, we will make clear recommendations to future \textit{JWST} proposers about the following:

\begin{enumerate}
\item For what observing scenarios does RDI become more effective than ADI? We will also establish whether this crossover point depends on the  variety of PSF subtraction, e.g.~the Locally Optimized Combination of Images versus KLIP \citep{lmd07, spl12}. 
\item At what angular separations does AMI achieve contrast that is superior to coronagraphic imaging?
\item To what extent do small grid dithers for PSF reference stars improve the final contrasts?
\end{enumerate}


\vspace{0.25in}
\section{Summary \& Conclusions}
The ERS program described in this work will utilize all four \textit{JWST} instruments to carry out some of the first \textit{direct} characterisation of planetary mass companions at wavelengths beyond $\sim$5$\mu$m, as well as image a circumstellar disk in the mid-infrared with unprecedented sensitivity. Specifically, our program will carry out: 

\begin{itemize}
    \item\textbf{Coronagraphy of an Extrasolar Planet:} Our program will gather photometric measurements of an extrasolar planet with unprecedented sensitivity from 3-5\,$\mu$m, as well as gather the first-ever coronagraphic observations beyond 5\,$\mu$m.  Our team will produce tools to aid the community in preparing future \textit{JWST} proposals such as an update to the \textit{JWST} coronagraphic imaging simulator PanCAKE, a standalone high-contrast imaging analysis pipeline, as well as contrast curves that define the coronagraphic performance of \textit{JWST}.  
    \item\textbf{Spectroscopy of a planetary mass companion:} We will obtain spectroscopy beyond 20\,$\mu$m of the planetary mass companion VHS\,1256b. This dataset will represent an accurate performance of the spectroscopic capabilities of \textit{JWST} using NIRSpec and MIRI, and will be a valuable resource for the community to prepare proposals related to spectroscopy going forward. 
    \item\textbf{Coronagraphy of a circumstellar disk:} We will obtain corongraphy out to 15.5\,$\mu$m of HD\,141569A, a young circumstellar disk, providing an understanding of \textit{JWST's} sensitivity to extended structures as a function of stellocentric angle, and sampling the disk brightness on and off the 3.0\,$\mu$m H$_2$O ice feature.  
    \item\textbf{Aperture Masking Interferometry:} To evaluate the sensitivity of \textit{JWST} near its fundamental diffraction limit of $\lambda/D{\sim}$120-140 milliarcseconds at wavelengths of 3-4\,$\mu$m, we will carry out observations of the HIP\,65426 system. These observations will have sensitivity to substellar companions in this system within $\sim$14\,AU, as well as identify the limiting systematics inherent in AMI observations.  
\end{itemize}

We have assembled a broad and representative consortium that comprises a large fraction of the entire high contrast imaging community. Our team members range from senior experts with decades of experience in exoplanet and disk imaging with \textit{HST}, infrared astronomy (e.g., \textit{Spitzer}) and adaptive optics, to early career researchers advancing the state-of-the-art in high contrast image processing and exoplanet atmosphere modelling. 
Through a series of calls and open consortium in-person meetings, we have drawn inclusively on a wide range of perspectives in identifying the highest priority investigations needed prior to Cycle 2. 
In addition to the scientific outcomes highlighted in this work, our program will also (critically) assess the performance of the observatory in the key modes expected to be commonly used for exoplanet direct imaging and spectroscopy, optimize data calibration and processing, and generate representative datasets that will enable a broad user base to effectively plan for general observing programs in future cycles. While this ERS program directly addresses the direct imaging of young, Jovian-mass planets with \textit{JWST}, this effort marks an important milestone in the long term effort to image much lower mass, terrestrial planets with future large-scale space missions. 

\acknowledgments
We thank the anonymous referee for numerous helpful suggestions, which significantly improved the manuscript.  We acknowledge the significant harm caused to members of the LGBTQIA+ community in the Department of State and NASA, while under the leadership of James Webb as Under Secretary of State and NASA Administrator, respectively.
This project was supported by a grant from STScI (JWST-ERS-01386) under NASA contract NAS5-03127. 
M.T.is supported by JSPS KAKENHI grant No.18H05442.
MB acknowledges support from the Deutsche Forschungsgemeinschaft (DFG) through grant Kr 2164/13-2.
K.W. acknowledges support from NASA through the NASA Hubble Fellowship grant HST-HF2-51472.001-A awarded by the Space Telescope Science Institute, which is operated by the Association of Universities for Research in Astronomy, Incorporated, under NASA contract NAS5-26555.

\bibliography{MasterBiblio_Sasha}

\begin{thebibliography}{}
\expandafter\ifx\csname natexlab\endcsname\relax\def\natexlab#1{#1}\fi
\providecommand{\url}[1]{\href{#1}{#1}}
\providecommand{\dodoi}[1]{}
\providecommand{\doarXiv}[1]{\href{https://arxiv.org/abs/#1}{\nolinkurl{https://arxiv.org/abs/#1}}}

\bibitem[{{Al-Refaie} {et~al.}(2022){Al-Refaie}, {Changeat}, {Venot},
  {Waldmann}, \& {Tinetti}}]{acv22}
{Al-Refaie}, A.~F., {Changeat}, Q., {Venot}, O., {et~al.} 2022,
  \href{http://dx.doi.org/10.3847/1538-4357/ac6dcd}{\color{magenta}\apj},
  \href{https://ui.adsabs.harvard.edu/abs/2022ApJ...932..123A}{\color{blue}932},
  \href{https://ui.adsabs.harvard.edu/abs/2022ApJ...932..123A}{\color{blue}123}

\bibitem[{{Al-Refaie} {et~al.}(2021){Al-Refaie}, {Changeat}, {Waldmann}, \&
  {Tinetti}}]{acw21}
{Al-Refaie}, A.~F., {Changeat}, Q., {Waldmann}, I.~P., \& {Tinetti}, G. 2021,
  \href{http://dx.doi.org/10.3847/1538-4357/ac0252}{\color{magenta}\apj},
  \href{https://ui.adsabs.harvard.edu/abs/2021ApJ...917...37A}{\color{blue}917},
  \href{https://ui.adsabs.harvard.edu/abs/2021ApJ...917...37A}{\color{blue}37}

\bibitem[{{Amara} \& {Quanz}(2012)}]{aq12}
{Amara}, A., \& {Quanz}, S.~P. 2012,
  \href{http://dx.doi.org/10.1111/j.1365-2966.2012.21918.x}{\color{magenta}\mnras},
  \href{https://ui.adsabs.harvard.edu/abs/2012MNRAS.427..948A}{\color{blue}427},
  \href{https://ui.adsabs.harvard.edu/abs/2012MNRAS.427..948A}{\color{blue}948}

\bibitem[{{Apai} {et~al.}(2016){Apai}, {Kasper}, {Skemer}, {Hanson},
  {Lagrange}, {Biller}, {Bonnefoy}, {Buenzli}, \& {Vigan}}]{aks16}
{Apai}, D., {Kasper}, M., {Skemer}, A., {et~al.} 2016,
  \href{http://dx.doi.org/10.3847/0004-637X/820/1/40}{\color{magenta}\apj},
  \href{https://ui.adsabs.harvard.edu/abs/2016ApJ...820...40A}{\color{blue}820},
  \href{https://ui.adsabs.harvard.edu/abs/2016ApJ...820...40A}{\color{blue}40}

\bibitem[{{Artigau} {et~al.}(2014){Artigau}, {Sivaramakrishnan}, {Greenbaum},
  {Doyon}, {Goudfrooij}, {Fullerton}, {Lafreni{\`e}re}, {Volk}, {Albert},
  {Martel}, {Ford}, \& {McKernan}}]{asg14}
{Artigau}, {\'E}., {Sivaramakrishnan}, A., {Greenbaum}, A.~Z., {et~al.} 2014,
  \href{http://dx.doi.org/10.1117/12.2055191}{\color{magenta}Proc.~SPIE},
  \href{https://ui.adsabs.harvard.edu/abs/2014SPIE.9143E..40A}{\color{blue}9143},
  \href{https://ui.adsabs.harvard.edu/abs/2014SPIE.9143E..40A}{\color{blue}914340}

\bibitem[{{Augereau} {et~al.}(1999{\natexlab{a}}){Augereau}, {Lagrange},
  {Mouillet}, \& {M{\'e}nard}}]{augereau99}
{Augereau}, J.~C., {Lagrange}, A.~M., {Mouillet}, D., \& {M{\'e}nard}, F.
  1999{\natexlab{a}}, \aap,
  \href{https://ui.adsabs.harvard.edu/abs/1999A&A...350L..51A}{\color{blue}350},
  \href{https://ui.adsabs.harvard.edu/abs/1999A&A...350L..51A}{\color{blue}L51}

\bibitem[{{Augereau} {et~al.}(1999{\natexlab{b}}){Augereau}, {Lagrange},
  {Mouillet}, {Papaloizou}, \& {Grorod}}]{alm99}
{Augereau}, J.~C., {Lagrange}, A.~M., {Mouillet}, D., {et~al.}
  1999{\natexlab{b}}, \aap,
  \href{http://adsabs.harvard.edu/abs/1999A%26A...348..557A}{\color{blue}348},
  \href{http://adsabs.harvard.edu/abs/1999A%26A...348..557A}{\color{blue}557}

\bibitem[{{Avenhaus} {et~al.}(2018){Avenhaus}, {Quanz}, {Garufi}, {Perez},
  {Casassus}, {Pinte}, {Bertrang}, {Caceres}, {Benisty}, \& {Dominik}}]{aqg18}
{Avenhaus}, H., {Quanz}, S.~P., {Garufi}, A., {et~al.} 2018,
  \href{http://dx.doi.org/10.3847/1538-4357/aab846}{\color{magenta}\apj},
  \href{https://ui.adsabs.harvard.edu/abs/2018ApJ...863...44A}{\color{blue}863},
  \href{https://ui.adsabs.harvard.edu/abs/2018ApJ...863...44A}{\color{blue}44}

\bibitem[{{Bagnasco} {et~al.}(2007){Bagnasco}, {Kolm}, {Ferruit}, {Honnen},
  {Koehler}, {Lemke}, {Maschmann}, {Melf}, {Noyer}, {Rumler}, {Salvignol},
  {Strada}, \& {Te Plate}}]{bkf07}
{Bagnasco}, G., {Kolm}, M., {Ferruit}, P., {et~al.} 2007,
  \href{http://dx.doi.org/10.1117/12.735602}{\color{magenta}Proc.~SPIE},
  \href{https://ui.adsabs.harvard.edu/abs/2007SPIE.6692E..0MB}{\color{blue}6692},
  \href{https://ui.adsabs.harvard.edu/abs/2007SPIE.6692E..0MB}{\color{blue}66920M}

\bibitem[{{Barman} {et~al.}(2015){Barman}, {Konopacky}, {Macintosh}, \&
  {Marois}}]{bkm15}
{Barman}, T.~S., {Konopacky}, Q.~M., {Macintosh}, B., \& {Marois}, C. 2015,
  \href{http://dx.doi.org/10.1088/0004-637X/804/1/61}{\color{magenta}ApJ},
  \href{https://ui.adsabs.harvard.edu/abs/2015ApJ...804...61B}{\color{blue}804},
  \href{https://ui.adsabs.harvard.edu/abs/2015ApJ...804...61B}{\color{blue}61}

\bibitem[{{Barman} {et~al.}(2011){Barman}, {Macintosh}, {Konopacky}, \&
  {Marois}}]{bmk11}
{Barman}, T.~S., {Macintosh}, B., {Konopacky}, Q.~M., \& {Marois}, C. 2011,
  \href{http://dx.doi.org/10.1088/0004-637X/733/1/65}{\color{magenta}\apj},
  \href{http://adsabs.harvard.edu/abs/2011ApJ...733...65B}{\color{blue}733},
  \href{http://adsabs.harvard.edu/abs/2011ApJ...733...65B}{\color{blue}65}

\bibitem[{{Baudoz} {et~al.}(2006){Baudoz}, {Boccaletti}, {Riaud}, {Cavarroc},
  {Baudrand}, {Reess}, \& {Rouan}}]{bbr06}
{Baudoz}, P., {Boccaletti}, A., {Riaud}, P., {et~al.} 2006,
  \href{http://dx.doi.org/10.1086/503124}{\color{magenta}\pasp},
  \href{https://ui.adsabs.harvard.edu/abs/2006PASP..118..765B}{\color{blue}118},
  \href{https://ui.adsabs.harvard.edu/abs/2006PASP..118..765B}{\color{blue}765}

\bibitem[{{Beichman} {et~al.}(2012){Beichman}, {Rieke}, {Eisenstein}, {Greene},
  {Krist}, {McCarthy}, {Meyer}, \& {Stansberry}}]{bre12}
{Beichman}, C.~A., {Rieke}, M., {Eisenstein}, D., {et~al.} 2012,
  \href{http://dx.doi.org/10.1117/12.925447}{\color{magenta}Proc.~SPIE},
  \href{https://ui.adsabs.harvard.edu/abs/2012SPIE.8442E..2NB}{\color{blue}8442},
  \href{https://ui.adsabs.harvard.edu/abs/2012SPIE.8442E..2NB}{\color{blue}84422N}

\bibitem[{{Beichman} {et~al.}(2010){Beichman}, {Krist}, {Trauger}, {Greene},
  {Oppenheimer}, {Sivaramakrishnan}, {Doyon}, {Boccaletti}, {Barman}, \&
  {Rieke}}]{bkt10}
{Beichman}, C.~A., {Krist}, J., {Trauger}, J.~T., {et~al.} 2010,
  \href{http://dx.doi.org/10.1086/651057}{\color{magenta}\pasp},
  \href{http://adsabs.harvard.edu/abs/2010PASP..122..162B}{\color{blue}122},
  \href{http://adsabs.harvard.edu/abs/2010PASP..122..162B}{\color{blue}162}

\bibitem[{{Bern{\'e}} {et~al.}(2009){Bern{\'e}}, {Joblin}, {Fuente}, \&
  {M{\'e}nard}}]{bjf09}
{Bern{\'e}}, O., {Joblin}, C., {Fuente}, A., \& {M{\'e}nard}, F. 2009,
  \href{http://dx.doi.org/10.1051/0004-6361:200810559}{\color{magenta}\aap},
  \href{https://ui.adsabs.harvard.edu/abs/2009A&A...495..827B}{\color{blue}495},
  \href{https://ui.adsabs.harvard.edu/abs/2009A&A...495..827B}{\color{blue}827}

\bibitem[{{Beuzit} {et~al.}(2019){Beuzit}, {Vigan}, {Mouillet}, {Dohlen},
  {Gratton}, {Boccaletti}, {Sauvage}, {Schmid}, {Langlois}, {Petit},
  {Baruffolo}, {Feldt}, {Milli}, {Wahhaj}, {Abe}, {Anselmi}, {Antichi},
  {Barette}, {Baudrand}, {Baudoz}, {Bazzon}, {Bernardi}, {Blanchard}, {Brast},
  {Bruno}, {Buey}, {Carbillet}, {Carle}, {Cascone}, {Chapron}, {Charton},
  {Chauvin}, {Claudi}, {Costille}, {De Caprio}, {de Boer}, {Delboulb{\'e}},
  {Desidera}, {Dominik}, {Downing}, {Dupuis}, {Fabron}, {Fantinel}, {Farisato},
  {Feautrier}, {Fedrigo}, {Fusco}, {Gigan}, {Ginski}, {Girard}, {Giro},
  {Gisler}, {Gluck}, {Gry}, {Henning}, {Hubin}, {Hugot}, {Incorvaia}, {Jaquet},
  {Kasper}, {Lagadec}, {Lagrange}, {Le Coroller}, {Le Mignant}, {Le Ruyet},
  {Lessio}, {Lizon}, {Llored}, {Lundin}, {Madec}, {Magnard}, {Marteaud},
  {Martinez}, {Maurel}, {M{\'e}nard}, {Mesa}, {M{\"o}ller-Nilsson}, {Moulin},
  {Moutou}, {Orign{\'e}}, {Parisot}, {Pavlov}, {Perret}, {Pragt}, {Puget},
  {Rabou}, {Ramos}, {Reess}, {Rigal}, {Rochat}, {Roelfsema}, {Rousset}, {Roux},
  {Saisse}, {Salasnich}, {Santambrogio}, {Scuderi}, {Segransan}, {Sevin},
  {Siebenmorgen}, {Soenke}, {Stadler}, {Suarez}, {Tiph{\`e}ne}, {Turatto},
  {Udry}, {Vakili}, {Waters}, {Weber}, {Wildi}, {Zins}, \& {Zurlo}}]{bvm19}
{Beuzit}, J.~L., {Vigan}, A., {Mouillet}, D., {et~al.} 2019,
  \href{http://dx.doi.org/10.1051/0004-6361/201935251}{\color{magenta}\aap},
  \href{https://ui.adsabs.harvard.edu/abs/2019A&A...631A.155B}{\color{blue}631},
  \href{https://ui.adsabs.harvard.edu/abs/2019A&A...631A.155B}{\color{blue}A155}

\bibitem[{{Biller} {et~al.}(2015{\natexlab{a}}){Biller}, {Vos}, {Bonavita},
  {Buenzli}, {Baxter}, {Crossfield}, {Allers}, {Liu}, {Bonnefoy}, {Deacon},
  {Brandner}, {Schlieder}, {Dupuy}, {Kopytova}, {Manjavacas}, {Allard},
  {Homeier}, \& {Henning}}]{bvb15}
{Biller}, B.~A., {Vos}, J., {Bonavita}, M., {et~al.} 2015{\natexlab{a}},
  \href{http://dx.doi.org/10.1088/2041-8205/813/2/L23}{\color{magenta}\apjl},
  \href{https://ui.adsabs.harvard.edu/abs/2015ApJ...813L..23B}{\color{blue}813},
  \href{https://ui.adsabs.harvard.edu/abs/2015ApJ...813L..23B}{\color{blue}L23}

\bibitem[{{Biller} {et~al.}(2015{\natexlab{b}}){Biller}, {Liu}, {Rice},
  {Wahhaj}, {Nielsen}, {Hayward}, {Kuchner}, {Close}, {Chun}, {Ftaclas}, \&
  {Toomey}}]{blr15}
{Biller}, B.~A., {Liu}, M.~C., {Rice}, K., {et~al.} 2015{\natexlab{b}},
  \href{http://dx.doi.org/10.1093/mnras/stv870}{\color{magenta}\mnras},
  \href{https://ui.adsabs.harvard.edu/abs/2015MNRAS.450.4446B}{\color{blue}450},
  \href{https://ui.adsabs.harvard.edu/abs/2015MNRAS.450.4446B}{\color{blue}4446}

\bibitem[{{Birkmann} {et~al.}(2022){Birkmann}, {Ferruit}, {Giardino},
  {Nielsen}, {Garc{\'\i}a Mu{\~n}oz}, {Kendrew}, {Rauscher}, {Beck}, {Keyes},
  {Valenti}, {Jakobsen}, {Dorner}, {Alves de Oliveira}, {Arribas}, {B{\"o}ker},
  {Bunker}, {Charlot}, {de Marchi}, {Kumari}, {L{\'o}pez-Caniego},
  {L{\"u}tzgendorf}, {Maiolino}, {Manjavacas}, {Marston}, {Moseley}, {Prizkal},
  {Proffitt}, {Rawle}, {Rix}, {te Plate}, {Sabbi}, {Sirianni}, {Willott}, \&
  {Zeidler}}]{bfg22}
{Birkmann}, S.~M., {Ferruit}, P., {Giardino}, G., {et~al.} 2022,
  \href{http://dx.doi.org/10.1051/0004-6361/202142592}{\color{magenta}\aap},
  \href{https://ui.adsabs.harvard.edu/abs/2022A&A...661A..83B}{\color{blue}661},
  \href{https://ui.adsabs.harvard.edu/abs/2022A&A...661A..83B}{\color{blue}A83}

\bibitem[{{Boccaletti} {et~al.}(2005){Boccaletti}, {Baudoz}, {Baudrand},
  {Reess}, \& {Rouan}}]{bbb05}
{Boccaletti}, A., {Baudoz}, P., {Baudrand}, J., {et~al.} 2005,
  \href{http://dx.doi.org/10.1016/j.asr.2005.04.107}{\color{magenta}Advances in
  Space Research},
  \href{https://ui.adsabs.harvard.edu/abs/2005AdSpR..36.1099B}{\color{blue}36},
  \href{https://ui.adsabs.harvard.edu/abs/2005AdSpR..36.1099B}{\color{blue}1099}

\bibitem[{{Boccaletti} {et~al.}(2015){Boccaletti}, {Lagage}, {Baudoz},
  {Beichman}, {Bouchet}, {Cavarroc}, {Dubreuil}, {Glasse}, {Glauser}, {Hines},
  {Lajoie}, {Lebreton}, {Perrin}, {Pueyo}, {Reess}, {Rieke}, {Ronayette},
  {Rouan}, {Soummer}, \& {Wright}}]{blb15}
{Boccaletti}, A., {Lagage}, P.~O., {Baudoz}, P., {et~al.} 2015,
  \href{http://dx.doi.org/10.1086/682256}{\color{magenta}\pasp},
  \href{https://ui.adsabs.harvard.edu/abs/2015PASP..127..633B}{\color{blue}127},
  \href{https://ui.adsabs.harvard.edu/abs/2015PASP..127..633B}{\color{blue}633}

\bibitem[{{Bohn} {et~al.}(2020){Bohn}, {Kenworthy}, {Ginski}, {Rieder},
  {Mamajek}, {Meshkat}, {Pecaut}, {Reggiani}, {de Boer}, {Keller}, {Snik}, \&
  {Southworth}}]{bkg20}
{Bohn}, A.~J., {Kenworthy}, M.~A., {Ginski}, C., {et~al.} 2020,
  \href{http://dx.doi.org/10.3847/2041-8213/aba27e}{\color{magenta}\apjl},
  \href{https://ui.adsabs.harvard.edu/abs/2020ApJ...898L..16B}{\color{blue}898},
  \href{https://ui.adsabs.harvard.edu/abs/2020ApJ...898L..16B}{\color{blue}L16}

\bibitem[{{Bonavita}(2020)}]{b20}
{Bonavita}, M. 2020, {Exo-DMC: Exoplanet Detection Map Calculator}

\bibitem[{{Bowler}(2016)}]{b16}
{Bowler}, B.~P. 2016,
  \href{http://dx.doi.org/10.1088/1538-3873/128/968/102001}{\color{magenta}\pasp},
  \href{http://adsabs.harvard.edu/abs/2016PASP..128j2001B}{\color{blue}128},
  \href{http://adsabs.harvard.edu/abs/2016PASP..128j2001B}{\color{blue}102001}

\bibitem[{{Bowler} {et~al.}(2010){Bowler}, {Liu}, {Dupuy}, \&
  {Cushing}}]{bld10}
{Bowler}, B.~P., {Liu}, M.~C., {Dupuy}, T.~J., \& {Cushing}, M.~C. 2010,
  \href{http://dx.doi.org/10.1088/0004-637X/723/1/850}{\color{magenta}\apj},
  \href{http://adsabs.harvard.edu/abs/2010ApJ...723..850B}{\color{blue}723},
  \href{http://adsabs.harvard.edu/abs/2010ApJ...723..850B}{\color{blue}850}

\bibitem[{{Bowler} {et~al.}(2020){Bowler}, {Zhou}, {Morley}, {Kataria},
  {Bryan}, {Benneke}, \& {Batygin}}]{bzm20}
{Bowler}, B.~P., {Zhou}, Y., {Morley}, C.~V., {et~al.} 2020,
  \href{http://dx.doi.org/10.3847/2041-8213/ab8197}{\color{magenta}\apjl},
  \href{https://ui.adsabs.harvard.edu/abs/2020ApJ...893L..30B}{\color{blue}893},
  \href{https://ui.adsabs.harvard.edu/abs/2020ApJ...893L..30B}{\color{blue}L30}

\bibitem[{{Bowler} {et~al.}(2017){Bowler}, {Liu}, {Mawet}, {Ngo}, {Malo},
  {Mace}, {McLane}, {Lu}, {Tristan}, {Hinkley}, {Hillenbrand}, {Shkolnik},
  {Benneke}, \& {Best}}]{blm17}
{Bowler}, B.~P., {Liu}, M.~C., {Mawet}, D., {et~al.} 2017,
  \href{http://dx.doi.org/10.3847/1538-3881/153/1/18}{\color{magenta}\aj},
  \href{https://ui.adsabs.harvard.edu/abs/2017AJ....153...18B}{\color{blue}153},
  \href{https://ui.adsabs.harvard.edu/abs/2017AJ....153...18B}{\color{blue}18}

\bibitem[{{Brande} {et~al.}(2020){Brande}, {Barclay}, {Schlieder}, {Lopez}, \&
  {Quintana}}]{bbs20}
{Brande}, J., {Barclay}, T., {Schlieder}, J.~E., {et~al.} 2020,
  \href{http://dx.doi.org/10.3847/1538-3881/ab5444}{\color{magenta}\aj},
  \href{https://ui.adsabs.harvard.edu/abs/2020AJ....159...18B}{\color{blue}159},
  \href{https://ui.adsabs.harvard.edu/abs/2020AJ....159...18B}{\color{blue}18}

\bibitem[{{Bruzzone} {et~al.}(2020){Bruzzone}, {Metchev}, {Duch{\^e}ne},
  {Millar-Blanchaer}, {Dong}, {Esposito}, {Wang}, {Graham}, {Mazoyer}, {Wolff},
  {Ammons}, {Schneider}, {Greenbaum}, {Matthews}, {Arriaga}, {Bailey},
  {Barman}, {Bulger}, {Chilcote}, {Cotten}, {De Rosa}, {Doyon}, {Fitzgerald},
  {Follette}, {Gerard}, {Goodsell}, {Hibon}, {Hom}, {Hung}, {Ingraham},
  {Kalas}, {Konopacky}, {Larkin}, {Macintosh}, {Maire}, {Marchis}, {Marois},
  {Morzinski}, {Nielsen}, {Oppenheimer}, {Palmer}, {Patel}, {Patience},
  {Perrin}, {Poyneer}, {Pueyo}, {Rajan}, {Rameau}, {Rantakyr{\"o}},
  {Savransky}, {Sivaramakrishnan}, {Song}, {Soummer}, {Thomas}, {Wallace},
  {Ward-Duong}, \& {Wiktorowicz}}]{bmd20}
{Bruzzone}, J.~S., {Metchev}, S., {Duch{\^e}ne}, G., {et~al.} 2020,
  \href{http://dx.doi.org/10.3847/1538-3881/ab5d2e}{\color{magenta}\aj},
  \href{https://ui.adsabs.harvard.edu/abs/2020AJ....159...53B}{\color{blue}159},
  \href{https://ui.adsabs.harvard.edu/abs/2020AJ....159...53B}{\color{blue}53}

\bibitem[{{Burningham} {et~al.}(2021){Burningham}, {Faherty}, {Gonzales},
  {Marley}, {Visscher}, {Lupu}, {Gaarn}, {Fabienne Bieger}, {Freedman}, \&
  {Saumon}}]{bfg21}
{Burningham}, B., {Faherty}, J.~K., {Gonzales}, E.~C., {et~al.} 2021,
  \href{http://dx.doi.org/10.1093/mnras/stab1361}{\color{magenta}\mnras},
  \href{https://ui.adsabs.harvard.edu/abs/2021MNRAS.506.1944B}{\color{blue}506},
  \href{https://ui.adsabs.harvard.edu/abs/2021MNRAS.506.1944B}{\color{blue}1944}

\bibitem[{{Carter} {et~al.}(2021{\natexlab{a}}){Carter}, {Skemer}, {Danielski},
  {Leisenring}, {Wang}, {Van Gorkom}, {York}, {Adams}, {Biller}, {Girard},
  {Hinkley}, {Nickson}, {Perrin}, \& {Pueyo}}]{csd21}
{Carter}, A.~L., {Skemer}, A. J.~I., {Danielski}, C., {et~al.}
  2021{\natexlab{a}},
  \href{http://dx.doi.org/10.1117/12.2594501}{\color{magenta}Proc.~SPIE},
  \href{https://ui.adsabs.harvard.edu/abs/2021SPIE11823E..0HC}{\color{blue}11823},
  \href{https://ui.adsabs.harvard.edu/abs/2021SPIE11823E..0HC}{\color{blue}118230H}

\bibitem[{{Carter} {et~al.}(2021{\natexlab{b}}){Carter}, {Hinkley}, {Bonavita},
  {Phillips}, {Girard}, {Perrin}, {Pueyo}, {Vigan}, {Gagn{\'e}}, \&
  {Skemer}}]{chb21}
{Carter}, A.~L., {Hinkley}, S., {Bonavita}, M., {et~al.} 2021{\natexlab{b}},
  \href{http://dx.doi.org/10.1093/mnras/staa3579}{\color{magenta}\mnras},
  \href{https://ui.adsabs.harvard.edu/abs/2021MNRAS.501.1999C}{\color{blue}501},
  \href{https://ui.adsabs.harvard.edu/abs/2021MNRAS.501.1999C}{\color{blue}1999}

\bibitem[{{Cassan} {et~al.}(2012){Cassan}, {Kubas}, {Beaulieu}, {Dominik},
  {Horne}, {Greenhill}, {Wambsganss}, {Menzies}, {Williams}, {J{\o}rgensen},
  {Udalski}, {Bennett}, {Albrow}, {Batista}, {Brillant}, {Caldwell}, {Cole},
  {Coutures}, {Cook}, {Dieters}, {Dominis Prester}, {Donatowicz}, {Fouqu{\'e}},
  {Hill}, {Kains}, {Kane}, {Marquette}, {Martin}, {Pollard}, {Sahu}, {Vinter},
  {Warren}, {Watson}, {Zub}, {Sumi}, {Szyma{\'n}ski}, {Kubiak}, {Poleski},
  {Soszynski}, {Ulaczyk}, {Pietrzy{\'n}ski}, \& {Wyrzykowski}}]{ckb12}
{Cassan}, A., {Kubas}, D., {Beaulieu}, J.~P., {et~al.} 2012,
  \href{http://dx.doi.org/10.1038/nature10684}{\color{magenta}\nat},
  \href{https://ui.adsabs.harvard.edu/abs/2012Natur.481..167C}{\color{blue}481},
  \href{https://ui.adsabs.harvard.edu/abs/2012Natur.481..167C}{\color{blue}167}

\bibitem[{{Cavarroc} {et~al.}(2008){Cavarroc}, {Amiaux}, {Baudoz},
  {Boccaletti}, {Bouchet}, {Dubreuil}, {Lagage}, {Moreau}, {Pantin}, {Reess},
  {Ronayette}, \& {Wright}}]{cab08}
{Cavarroc}, C., {Amiaux}, J., {Baudoz}, P., {et~al.} 2008,
  \href{http://dx.doi.org/10.1117/12.789089}{\color{magenta}Proc.~SPIE},
  \href{https://ui.adsabs.harvard.edu/abs/2008SPIE.7010E..0WC}{\color{blue}7010},
  \href{https://ui.adsabs.harvard.edu/abs/2008SPIE.7010E..0WC}{\color{blue}70100W}

\bibitem[{{Chabrier} {et~al.}(2007){Chabrier}, {Baraffe}, {Selsis}, {Barman},
  {Hennebelle}, \& {Alibert}}]{cbs07}
{Chabrier}, G., {Baraffe}, I., {Selsis}, F., {et~al.} 2007, Protostars and
  Planets V,
  \href{http://adsabs.harvard.edu/abs/2007prpl.conf..623C}{\color{blue}623}

\bibitem[{{Chauvin} {et~al.}(2004){Chauvin}, {Lagrange}, {Dumas}, {Zuckerman},
  {Mouillet}, {Song}, {Beuzit}, \& {Lowrance}}]{cld04}
{Chauvin}, G., {Lagrange}, A., {Dumas}, C., {et~al.} 2004,
  \href{http://dx.doi.org/10.1051/0004-6361:200400056}{\color{magenta}\aap},
  \href{http://adsabs.harvard.edu/abs/2004A%26A...425L..29C}{\color{blue}425},
  \href{http://adsabs.harvard.edu/abs/2004A%26A...425L..29C}{\color{blue}L29}

\bibitem[{{Chauvin} {et~al.}(2017){Chauvin}, {Desidera}, {Lagrange}, {Vigan},
  {Gratton}, {Langlois}, {Bonnefoy}, {Beuzit}, {Feldt}, {Mouillet}, {Meyer},
  {Cheetham}, {Biller}, {Boccaletti}, {D'Orazi}, {Galicher}, {Hagelberg},
  {Maire}, {Mesa}, {Olofsson}, {Samland}, {Schmidt}, {Sissa}, {Bonavita},
  {Charnay}, {Cudel}, {Daemgen}, {Delorme}, {Janin-Potiron}, {Janson},
  {Keppler}, {Le Coroller}, {Ligi}, {Marleau}, {Messina}, {Molli{\`e}re},
  {Mordasini}, {M{\"u}ller}, {Peretti}, {Perrot}, {Rodet}, {Rouan}, {Zurlo},
  {Dominik}, {Henning}, {Menard}, {Schmid}, {Turatto}, {Udry}, {Vakili}, {Abe},
  {Antichi}, {Baruffolo}, {Baudoz}, {Baudrand}, {Blanchard}, {Bazzon}, {Buey},
  {Carbillet}, {Carle}, {Charton}, {Cascone}, {Claudi}, {Costille}, {Deboulbe},
  {De Caprio}, {Dohlen}, {Fantinel}, {Feautrier}, {Fusco}, {Gigan}, {Giro},
  {Gisler}, {Gluck}, {Hubin}, {Hugot}, {Jaquet}, {Kasper}, {Madec}, {Magnard},
  {Martinez}, {Maurel}, {Le Mignant}, {M{\"o}ller-Nilsson}, {Llored}, {Moulin},
  {Orign{\'e}}, {Pavlov}, {Perret}, {Petit}, {Pragt}, {Puget}, {Rabou},
  {Ramos}, {Rigal}, {Rochat}, {Roelfsema}, {Rousset}, {Roux}, {Salasnich},
  {Sauvage}, {Sevin}, {Soenke}, {Stadler}, {Suarez}, {Weber}, {Wildi},
  {Antoniucci}, {Augereau}, {Baudino}, {Brandner}, {Engler}, {Girard}, {Gry},
  {Kral}, {Kopytova}, {Lagadec}, {Milli}, {Moutou}, {Schlieder},
  {Szul{\'a}gyi}, {Thalmann}, \& {Wahhaj}}]{cdl17}
{Chauvin}, G., {Desidera}, S., {Lagrange}, A.-M., {et~al.} 2017,
  \href{http://dx.doi.org/10.1051/0004-6361/201731152}{\color{magenta}\aap},
  \href{http://adsabs.harvard.edu/abs/2017A%26A...605L...9C}{\color{blue}605},
  \href{http://adsabs.harvard.edu/abs/2017A%26A...605L...9C}{\color{blue}L9}

\bibitem[{{Cheetham} {et~al.}(2019){Cheetham}, {Samland}, {Brems}, {Launhardt},
  {Chauvin}, {S{\'e}gransan}, {Henning}, {Quirrenbach}, {Avenhaus}, {Cugno},
  {Girard}, {Godoy}, {Kennedy}, {Maire}, {Metchev}, {M{\"u}ller}, {Musso
  Barcucci}, {Olofsson}, {Pepe}, {Quanz}, {Queloz}, {Reffert}, {Rickman}, {van
  Boekel}, {Boccaletti}, {Bonnefoy}, {Cantalloube}, {Charnay}, {Delorme},
  {Janson}, {Keppler}, {Lagrange}, {Langlois}, {Lazzoni}, {Menard}, {Mesa},
  {Meyer}, {Schmidt}, {Sissa}, {Udry}, \& {Zurlo}}]{csb19}
{Cheetham}, A.~C., {Samland}, M., {Brems}, S.~S., {et~al.} 2019,
  \href{http://dx.doi.org/10.1051/0004-6361/201834112}{\color{magenta}\aap},
  \href{https://ui.adsabs.harvard.edu/abs/2019A&A...622A..80C}{\color{blue}622},
  \href{https://ui.adsabs.harvard.edu/abs/2019A&A...622A..80C}{\color{blue}A80}

\bibitem[{{Chiang} {et~al.}(2009){Chiang}, {Kite}, {Kalas}, {Graham}, \&
  {Clampin}}]{ckk09}
{Chiang}, E., {Kite}, E., {Kalas}, P., {et~al.} 2009,
  \href{http://dx.doi.org/10.1088/0004-637X/693/1/734}{\color{magenta}\apj},
  \href{http://adsabs.harvard.edu/abs/2009ApJ...693..734C}{\color{blue}693},
  \href{http://adsabs.harvard.edu/abs/2009ApJ...693..734C}{\color{blue}734}

\bibitem[{{Chilcote} {et~al.}(2020){Chilcote}, {Konopacky}, {De Rosa},
  {Hamper}, {Macintosh}, {Marois}, {Perrin}, {Savransky}, {Soummer},
  {V{\'e}ran}, {Agapito}, {Aleman}, {Ammons}, {Bonaglia}, {Boucher}, {Curliss},
  {Dunn}, {Esposito}, {Filion}, {Fitzsimmons}, {Kain}, {Kerley}, {Landry},
  {Lardiere}, {Lemoine-Busserolle}, {Li}, {Limbach}, {Madurowicz}, {Maire},
  {N'Diaye}, {Nielsen}, {Poyneer}, {Pueyo}, {Summey}, \& {Thomas}}]{ckd20}
{Chilcote}, J., {Konopacky}, Q., {De Rosa}, R.~J., {et~al.} 2020,
  \href{http://dx.doi.org/10.1117/12.2562578}{\color{magenta}Proc.~SPIE},
  \href{https://ui.adsabs.harvard.edu/abs/2020SPIE11447E..1SC}{\color{blue}11447},
  \href{https://ui.adsabs.harvard.edu/abs/2020SPIE11447E..1SC}{\color{blue}114471S}

\bibitem[{{Choquet} {et~al.}(2016){Choquet}, {Perrin}, {Chen}, {Soummer},
  {Pueyo}, {Hagan}, {Gofas-Salas}, {Rajan}, {Golimowski}, {Hines}, {Schneider},
  {Mazoyer}, {Augereau}, {Debes}, {Stark}, {Wolff}, {N'Diaye}, \&
  {Hsiao}}]{cpc16}
{Choquet}, {\'E}., {Perrin}, M.~D., {Chen}, C.~H., {et~al.} 2016,
  \href{http://dx.doi.org/10.3847/2041-8205/817/1/L2}{\color{magenta}\apjl},
  \href{https://ui.adsabs.harvard.edu/abs/2016ApJ...817L...2C}{\color{blue}817},
  \href{https://ui.adsabs.harvard.edu/abs/2016ApJ...817L...2C}{\color{blue}L2}

\bibitem[{{Clampin} {et~al.}(2003){Clampin}, {Krist}, {Ardila}, {Golimowski},
  {Hartig}, {Ford}, {Illingworth}, {Bartko}, {Ben{\'{\i}}tez}, {Blakeslee},
  {Bouwens}, {Broadhurst}, {Brown}, {Burrows}, {Cheng}, {Cross}, {Feldman},
  {Franx}, {Gronwall}, {Infante}, {Kimble}, {Lesser}, {Martel}, {Menanteau},
  {Meurer}, {Miley}, {Postman}, {Rosati}, {Sirianni}, {Sparks}, {Tran},
  {Tsvetanov}, {White}, \& {Zheng}}]{cka03}
{Clampin}, M., {Krist}, J.~E., {Ardila}, D.~R., {et~al.} 2003, \aj,
  \href{http://adsabs.harvard.edu/abs/2003AJ....126..385C}{\color{blue}126},
  \href{http://adsabs.harvard.edu/abs/2003AJ....126..385C}{\color{blue}385}

\bibitem[{{Cridland} {et~al.}(2019){Cridland}, {Eistrup}, \& {van
  Dishoeck}}]{cev19}
{Cridland}, A.~J., {Eistrup}, C., \& {van Dishoeck}, E.~F. 2019,
  \href{http://dx.doi.org/10.1051/0004-6361/201834378}{\color{magenta}\aap},
  \href{https://ui.adsabs.harvard.edu/abs/2019A&A...627A.127C}{\color{blue}627},
  \href{https://ui.adsabs.harvard.edu/abs/2019A&A...627A.127C}{\color{blue}A127}

\bibitem[{{Cridland} {et~al.}(2016){Cridland}, {Pudritz}, \& {Alessi}}]{cpa16}
{Cridland}, A.~J., {Pudritz}, R.~E., \& {Alessi}, M. 2016,
  \href{http://dx.doi.org/10.1093/mnras/stw1511}{\color{magenta}\mnras},
  \href{https://ui.adsabs.harvard.edu/abs/2016MNRAS.461.3274C}{\color{blue}461},
  \href{https://ui.adsabs.harvard.edu/abs/2016MNRAS.461.3274C}{\color{blue}3274}

\bibitem[{{Currie} {et~al.}(2011){Currie}, {Burrows}, {Itoh}, {Matsumura},
  {Fukagawa}, {Apai}, {Madhusudhan}, {Hinz}, {Rodigas}, {Kasper}, {Pyo}, \&
  {Ogino}}]{cbi11}
{Currie}, T., {Burrows}, A., {Itoh}, Y., {et~al.} 2011,
  \href{http://dx.doi.org/10.1088/0004-637X/729/2/128}{\color{magenta}\apj},
  \href{http://adsabs.harvard.edu/abs/2011ApJ...729..128C}{\color{blue}729},
  \href{http://adsabs.harvard.edu/abs/2011ApJ...729..128C}{\color{blue}128}

\bibitem[{{Currie} {et~al.}(2016){Currie}, {Grady}, {Cloutier}, {Konishi},
  {Stassun}, {Debes}, {van der Marel}, {Muto}, {Jayawardhana}, \&
  {Ratzka}}]{cgc16}
{Currie}, T., {Grady}, C.~A., {Cloutier}, R., {et~al.} 2016,
  \href{http://dx.doi.org/10.3847/2041-8205/819/2/L26}{\color{magenta}\apjl},
  \href{https://ui.adsabs.harvard.edu/abs/2016ApJ...819L..26C}{\color{blue}819},
  \href{https://ui.adsabs.harvard.edu/abs/2016ApJ...819L..26C}{\color{blue}L26}

\bibitem[{{Currie} {et~al.}(2018){Currie}, {Brandt}, {Uyama}, {Nielsen},
  {Blunt}, {Guyon}, {Tamura}, {Marois}, {Mede}, {Kuzuhara}, {Groff},
  {Jovanovic}, {Kasdin}, {Lozi}, {Hodapp}, {Chilcote}, {Carson}, {Martinache},
  {Goebel}, {Grady}, {McElwain}, {Akiyama}, {Asensio-Torres}, {Hayashi},
  {Janson}, {Knapp}, {Kwon}, {Nishikawa}, {Oh}, {Schlieder}, {Serabyn},
  {Sitko}, \& {Skaf}}]{cbu18}
{Currie}, T., {Brandt}, T.~D., {Uyama}, T., {et~al.} 2018,
  \href{http://dx.doi.org/10.3847/1538-3881/aae9ea}{\color{magenta}\aj},
  \href{https://ui.adsabs.harvard.edu/abs/2018AJ....156..291C}{\color{blue}156},
  \href{https://ui.adsabs.harvard.edu/abs/2018AJ....156..291C}{\color{blue}291}

\bibitem[{{Cushing} {et~al.}(2006){Cushing}, {Roellig}, {Marley}, {Saumon},
  {Leggett}, {Kirkpatrick}, {Wilson}, {Sloan}, {Mainzer}, \& {Van
  Cleve}}]{crm06}
{Cushing}, M.~C., {Roellig}, T.~L., {Marley}, M.~S., {et~al.} 2006,
  \href{http://dx.doi.org/10.1086/505637}{\color{magenta}\apj},
  \href{https://ui.adsabs.harvard.edu/abs/2006ApJ...648..614C}{\color{blue}648},
  \href{https://ui.adsabs.harvard.edu/abs/2006ApJ...648..614C}{\color{blue}614}

\bibitem[{{Cutri} {et~al.}(2003){Cutri}, {Skrutskie}, {van Dyk}, {Beichman},
  {Carpenter}, {Chester}, {Cambresy}, {Evans}, {Fowler}, {Gizis}, {Howard},
  {Huchra}, {Jarrett}, {Kopan}, {Kirkpatrick}, {Light}, {Marsh}, {McCallon},
  {Schneider}, {Stiening}, {Sykes}, {Weinberg}, {Wheaton}, {Wheelock}, \&
  {Zacarias}}]{csv03}
{Cutri}, R.~M., {Skrutskie}, M.~F., {van Dyk}, S., {et~al.} 2003, {2MASS All
  Sky Catalog of point sources.}

\bibitem[{{De Rosa} {et~al.}(2016){De Rosa}, {Rameau}, {Patience}, {Graham},
  {Doyon}, {Lafreni{\`e}re}, {Macintosh}, {Pueyo}, {Rajan}, {Wang},
  {Ward-Duong}, {Hung}, {Maire}, {Nielsen}, {Ammons}, {Bulger}, {Cardwell},
  {Chilcote}, {Galvez}, {Gerard}, {Goodsell}, {Hartung}, {Hibon}, {Ingraham},
  {Johnson-Groh}, {Kalas}, {Konopacky}, {Marchis}, {Marois}, {Metchev},
  {Morzinski}, {Oppenheimer}, {Perrin}, {Rantakyr{\"o}}, {Savransky}, \&
  {Thomas}}]{drp16}
{De Rosa}, R.~J., {Rameau}, J., {Patience}, J., {et~al.} 2016,
  \href{http://dx.doi.org/10.3847/0004-637X/824/2/121}{\color{magenta}ApJ},
  \href{https://ui.adsabs.harvard.edu/abs/2016ApJ...824..121D}{\color{blue}824},
  \href{https://ui.adsabs.harvard.edu/abs/2016ApJ...824..121D}{\color{blue}121}

\bibitem[{{de Zeeuw} {et~al.}(1999){de Zeeuw}, {Hoogerwerf}, {de Bruijne},
  {Brown}, \& {Blaauw}}]{dhd99}
{de Zeeuw}, P.~T., {Hoogerwerf}, R., {de Bruijne}, J.~H.~J., {et~al.} 1999,
  \href{http://dx.doi.org/10.1086/300682}{\color{magenta}\aj},
  \href{http://adsabs.harvard.edu/abs/1999AJ....117..354D}{\color{blue}117},
  \href{http://adsabs.harvard.edu/abs/1999AJ....117..354D}{\color{blue}354}

\bibitem[{{Desidera} {et~al.}(2021){Desidera}, {Chauvin}, {Bonavita},
  {Messina}, {LeCoroller}, {Schmidt}, {Gratton}, {Lazzoni}, {Meyer},
  {Schlieder}, {Cheetham}, {Hagelberg}, {Bonnefoy}, {Feldt}, {Lagrange},
  {Langlois}, {Vigan}, {Tan}, {Hambsch}, {Millward}, {Alcal{\'a}}, {Benatti},
  {Brandner}, {Carson}, {Covino}, {Delorme}, {D'Orazi}, {Janson}, {Rigliaco},
  {Beuzit}, {Biller}, {Boccaletti}, {Dominik}, {Cantalloube}, {Fontanive},
  {Galicher}, {Henning}, {Lagadec}, {Ligi}, {Maire}, {Menard}, {Mesa},
  {M{\"u}ller}, {Samland}, {Schmid}, {Sissa}, {Turatto}, {Udry}, {Zurlo},
  {Asensio-Torres}, {Kopytova}, {Rickman}, {Abe}, {Antichi}, {Baruffolo},
  {Baudoz}, {Baudrand}, {Blanchard}, {Bazzon}, {Buey}, {Carbillet}, {Carle},
  {Charton}, {Cascone}, {Claudi}, {Costille}, {Deboulb{\'e}}, {De Caprio},
  {Dohlen}, {Fantinel}, {Feautrier}, {Fusco}, {Gigan}, {Giro}, {Gisler},
  {Gluck}, {Hubin}, {Hugot}, {Jaquet}, {Kasper}, {Madec}, {Magnard},
  {Martinez}, {Maurel}, {Le Mignant}, {M{\"o}ller-Nilsson}, {Llored}, {Moulin},
  {Orign{\'e}}, {Pavlov}, {Perret}, {Petit}, {Pragt}, {Puget}, {Rabou},
  {Ramos}, {Rigal}, {Rochat}, {Roelfsema}, {Rousset}, {Roux}, {Salasnich},
  {Sauvage}, {Sevin}, {Soenke}, {Stadler}, {Suarez}, {Weber}, \&
  {Wildi}}]{dcb21}
{Desidera}, S., {Chauvin}, G., {Bonavita}, M., {et~al.} 2021,
  \href{http://dx.doi.org/10.1051/0004-6361/202038806}{\color{magenta}\aap},
  \href{https://ui.adsabs.harvard.edu/abs/2021A&A...651A..70D}{\color{blue}651},
  \href{https://ui.adsabs.harvard.edu/abs/2021A&A...651A..70D}{\color{blue}A70}

\bibitem[{{Di Folco} {et~al.}(2020){Di Folco}, {P{\'e}ricaud}, {Dutrey},
  {Augereau}, {Chapillon}, {Guilloteau}, {Pi{\'e}tu}, \& {Boccaletti}}]{dpd20}
{Di Folco}, E., {P{\'e}ricaud}, J., {Dutrey}, A., {et~al.} 2020,
  \href{http://dx.doi.org/10.1051/0004-6361/201732243}{\color{magenta}\aap},
  \href{https://ui.adsabs.harvard.edu/abs/2020A&A...635A..94D}{\color{blue}635},
  \href{https://ui.adsabs.harvard.edu/abs/2020A&A...635A..94D}{\color{blue}A94}

\bibitem[{{Doyon} {et~al.}(2012){Doyon}, {Hutchings}, {Beaulieu}, {Albert},
  {Lafreni{\`e}re}, {Willott}, {Touahri}, {Rowlands}, {Maszkiewicz},
  {Fullerton}, {Volk}, {Martel}, {Chayer}, {Sivaramakrishnan}, {Abraham},
  {Ferrarese}, {Jayawardhana}, {Johnstone}, {Meyer}, {Pipher}, \&
  {Sawicki}}]{dhb12}
{Doyon}, R., {Hutchings}, J.~B., {Beaulieu}, M., {et~al.} 2012,
  \href{http://dx.doi.org/10.1117/12.926578}{\color{magenta}Proc.~SPIE},
  \href{https://ui.adsabs.harvard.edu/abs/2012SPIE.8442E..2RD}{\color{blue}8442},
  \href{https://ui.adsabs.harvard.edu/abs/2012SPIE.8442E..2RD}{\color{blue}84422R}

\bibitem[{{Dressing} \& {Charbonneau}(2013)}]{dc13}
{Dressing}, C.~D., \& {Charbonneau}, D. 2013,
  \href{http://dx.doi.org/10.1088/0004-637X/767/1/95}{\color{magenta}ApJ},
  \href{https://ui.adsabs.harvard.edu/abs/2013ApJ...767...95D}{\color{blue}767},
  \href{https://ui.adsabs.harvard.edu/abs/2013ApJ...767...95D}{\color{blue}95}

\bibitem[{{Dupuy} {et~al.}(2020){Dupuy}, {Liu}, {Magnier}, {Best}, {Baraffe},
  {Chabrier}, {Forveille}, {Metchev}, \& {Tremblin}}]{dlm20}
{Dupuy}, T.~J., {Liu}, M.~C., {Magnier}, E.~A., {et~al.} 2020,
  \href{http://dx.doi.org/10.3847/2515-5172/ab8942}{\color{magenta}Research
  Notes of the American Astronomical Society},
  \href{https://ui.adsabs.harvard.edu/abs/2020RNAAS...4...54D}{\color{blue}4},
  \href{https://ui.adsabs.harvard.edu/abs/2020RNAAS...4...54D}{\color{blue}54}

\bibitem[{{Esposito} {et~al.}(2020){Esposito}, {Kalas}, {Fitzgerald},
  {Millar-Blanchaer}, {Duch{\^e}ne}, {Patience}, {Hom}, {Perrin}, {De Rosa},
  {Chiang}, {Czekala}, {Macintosh}, {Graham}, {Ansdell}, {Arriaga}, {Bruzzone},
  {Bulger}, {Chen}, {Cotten}, {Dong}, {Draper}, {Follette}, {Hung}, {Lopez},
  {Matthews}, {Mazoyer}, {Metchev}, {Rameau}, {Ren}, {Rice}, {Song}, {Stahl},
  {Wang}, {Wolff}, {Zuckerman}, {Ammons}, {Bailey}, {Barman}, {Chilcote},
  {Doyon}, {Gerard}, {Goodsell}, {Greenbaum}, {Hibon}, {Hinkley}, {Ingraham},
  {Konopacky}, {Maire}, {Marchis}, {Marley}, {Marois}, {Nielsen},
  {Oppenheimer}, {Palmer}, {Poyneer}, {Pueyo}, {Rajan}, {Rantakyr{\"o}},
  {Ruffio}, {Savransky}, {Schneider}, {Sivaramakrishnan}, {Soummer}, {Thomas},
  \& {Ward-Duong}}]{ekf20}
{Esposito}, T.~M., {Kalas}, P., {Fitzgerald}, M.~P., {et~al.} 2020,
  \href{http://dx.doi.org/10.3847/1538-3881/ab9199}{\color{magenta}\aj},
  \href{https://ui.adsabs.harvard.edu/abs/2020AJ....160...24E}{\color{blue}160},
  \href{https://ui.adsabs.harvard.edu/abs/2020AJ....160...24E}{\color{blue}24}

\bibitem[{{Faherty} {et~al.}(2016){Faherty}, {Riedel}, {Cruz}, {Gagne},
  {Filippazzo}, {Lambrides}, {Fica}, {Weinberger}, {Thorstensen}, {Tinney},
  {Baldassare}, {Lemonier}, \& {Rice}}]{frc16}
{Faherty}, J.~K., {Riedel}, A.~R., {Cruz}, K.~L., {et~al.} 2016,
  \href{http://dx.doi.org/10.3847/0067-0049/225/1/10}{\color{magenta}\apjs},
  \href{https://ui.adsabs.harvard.edu/abs/2016ApJS..225...10F}{\color{blue}225},
  \href{https://ui.adsabs.harvard.edu/abs/2016ApJS..225...10F}{\color{blue}10}

\bibitem[{{Fernandes} {et~al.}(2019){Fernandes}, {Mulders}, {Pascucci},
  {Mordasini}, \& {Emsenhuber}}]{fmp19}
{Fernandes}, R.~B., {Mulders}, G.~D., {Pascucci}, I., {et~al.} 2019,
  \href{http://dx.doi.org/10.3847/1538-4357/ab0300}{\color{magenta}\apj},
  \href{https://ui.adsabs.harvard.edu/abs/2019ApJ...874...81F}{\color{blue}874},
  \href{https://ui.adsabs.harvard.edu/abs/2019ApJ...874...81F}{\color{blue}81}

\bibitem[{{Filacchione} {et~al.}(2014){Filacchione}, {Ciarniello},
  {Capaccioni}, {Clark}, {Nicholson}, {Hedman}, {Cuzzi}, {Cruikshank}, {Dalle
  Ore}, \& {Brown}}]{fcc14}
{Filacchione}, G., {Ciarniello}, M., {Capaccioni}, F., {et~al.} 2014,
  \href{http://dx.doi.org/10.1016/j.icarus.2014.06.001}{\color{magenta}icarus},
  \href{https://ui.adsabs.harvard.edu/abs/2014Icar..241...45F}{\color{blue}241},
  \href{https://ui.adsabs.harvard.edu/abs/2014Icar..241...45F}{\color{blue}45}

\bibitem[{{Forgan} \& {Rice}(2013)}]{fr13}
{Forgan}, D., \& {Rice}, K. 2013,
  \href{http://dx.doi.org/10.1093/mnras/stt672}{\color{magenta}\mnras},
  \href{https://ui.adsabs.harvard.edu/abs/2013MNRAS.432.3168F}{\color{blue}432},
  \href{https://ui.adsabs.harvard.edu/abs/2013MNRAS.432.3168F}{\color{blue}3168}

\bibitem[{{Fortney} {et~al.}(2013){Fortney}, {Mordasini}, {Nettelmann},
  {Kempton}, {Greene}, \& {Zahnle}}]{fmn13}
{Fortney}, J.~J., {Mordasini}, C., {Nettelmann}, N., {et~al.} 2013,
  \href{http://dx.doi.org/10.1088/0004-637X/775/1/80}{\color{magenta}\apj},
  \href{https://ui.adsabs.harvard.edu/abs/2013ApJ...775...80F}{\color{blue}775},
  \href{https://ui.adsabs.harvard.edu/abs/2013ApJ...775...80F}{\color{blue}80}

\bibitem[{{Frelikh} {et~al.}(2019){Frelikh}, {Jang}, {Murray-Clay}, \&
  {Petrovich}}]{fjm19}
{Frelikh}, R., {Jang}, H., {Murray-Clay}, R.~A., \& {Petrovich}, C. 2019,
  \href{http://dx.doi.org/10.3847/2041-8213/ab4a7b}{\color{magenta}\apjl},
  \href{https://ui.adsabs.harvard.edu/abs/2019ApJ...884L..47F}{\color{blue}884},
  \href{https://ui.adsabs.harvard.edu/abs/2019ApJ...884L..47F}{\color{blue}L47}

\bibitem[{{Freytag} {et~al.}(1996){Freytag}, {Ludwig}, \& {Steffen}}]{fls96}
{Freytag}, B., {Ludwig}, H.~G., \& {Steffen}, M. 1996, \aap,
  \href{https://ui.adsabs.harvard.edu/abs/1996A&A...313..497F}{\color{blue}313},
  \href{https://ui.adsabs.harvard.edu/abs/1996A&A...313..497F}{\color{blue}497}

\bibitem[{{Fulton} {et~al.}(2021){Fulton}, {Rosenthal}, {Hirsch}, {Isaacson},
  {Howard}, {Dedrick}, {Sherstyuk}, {Blunt}, {Petigura}, {Knutson}, {Behmard},
  {Chontos}, {Crepp}, {Crossfield}, {Dalba}, {Fischer}, {Henry}, {Kane},
  {Kosiarek}, {Marcy}, {Rubenzahl}, {Weiss}, \& {Wright}}]{frh21}
{Fulton}, B.~J., {Rosenthal}, L.~J., {Hirsch}, L.~A., {et~al.} 2021,
  \href{http://dx.doi.org/10.3847/1538-4365/abfcc1}{\color{magenta}\apjs},
  \href{https://ui.adsabs.harvard.edu/abs/2021ApJS..255...14F}{\color{blue}255},
  \href{https://ui.adsabs.harvard.edu/abs/2021ApJS..255...14F}{\color{blue}14}

\bibitem[{{Gaia Collaboration} {et~al.}(2018){Gaia Collaboration}, {Brown},
  {Vallenari}, {Prusti}, {de Bruijne}, {Babusiaux}, {Bailer-Jones}, {Biermann},
  {Evans}, {Eyer}, {Jansen}, {Jordi}, {Klioner}, {Lammers}, {Lindegren},
  {Luri}, {Mignard}, {Panem}, {Pourbaix}, {Randich}, {Sartoretti}, {Siddiqui},
  {Soubiran}, {van Leeuwen}, {Walton}, {Arenou}, {Bastian}, {Cropper},
  {Drimmel}, {Katz}, {Lattanzi}, {Bakker}, {Cacciari}, {Casta{\~n}eda},
  {Chaoul}, {Cheek}, {De Angeli}, {Fabricius}, {Guerra}, {Holl}, {Masana},
  {Messineo}, {Mowlavi}, {Nienartowicz}, {Panuzzo}, {Portell}, {Riello},
  {Seabroke}, {Tanga}, {Th{\'e}venin}, {Gracia-Abril}, {Comoretto},
  {Garcia-Reinaldos}, {Teyssier}, {Altmann}, {Andrae}, {Audard},
  {Bellas-Velidis}, {Benson}, {Berthier}, {Blomme}, {Burgess}, {Busso},
  {Carry}, {Cellino}, {Clementini}, {Clotet}, {Creevey}, {Davidson}, {De
  Ridder}, {Delchambre}, {Dell'Oro}, {Ducourant},
  {Fern{\'a}ndez-Hern{\'a}ndez}, {Fouesneau}, {Fr{\'e}mat}, {Galluccio},
  {Garc{\'\i}a-Torres}, {Gonz{\'a}lez-N{\'u}{\~n}ez}, {Gonz{\'a}lez-Vidal},
  {Gosset}, {Guy}, {Halbwachs}, {Hambly}, {Harrison}, {Hern{\'a}ndez},
  {Hestroffer}, {Hodgkin}, {Hutton}, {Jasniewicz}, {Jean-Antoine-Piccolo},
  {Jordan}, {Korn}, {Krone-Martins}, {Lanzafame}, {Lebzelter}, {L{\"o}ffler},
  {Manteiga}, {Marrese}, {Mart{\'\i}n-Fleitas}, {Moitinho}, {Mora}, {Muinonen},
  {Osinde}, {Pancino}, {Pauwels}, {Petit}, {Recio-Blanco}, {Richards},
  {Rimoldini}, {Robin}, {Sarro}, {Siopis}, {Smith}, {Sozzetti}, {S{\"u}veges},
  {Torra}, {van Reeven}, {Abbas}, {Abreu Aramburu}, {Accart}, {Aerts},
  {Altavilla}, {{\'A}lvarez}, {Alvarez}, {Alves}, {Anderson}, {Andrei},
  {Anglada Varela}, {Antiche}, {Antoja}, {Arcay}, {Astraatmadja}, {Bach},
  {Baker}, {Balaguer-N{\'u}{\~n}ez}, {Balm}, {Barache}, {Barata}, {Barbato},
  {Barblan}, {Barklem}, {Barrado}, {Barros}, {Barstow}, {Bartholom{\'e}
  Mu{\~n}oz}, {Bassilana}, {Becciani}, {Bellazzini}, {Berihuete}, {Bertone},
  {Bianchi}, {Bienaym{\'e}}, {Blanco-Cuaresma}, {Boch}, {Boeche}, {Bombrun},
  {Borrachero}, {Bossini}, {Bouquillon}, {Bourda}, {Bragaglia}, {Bramante},
  {Breddels}, {Bressan}, {Brouillet}, {Br{\"u}semeister}, {Brugaletta},
  {Bucciarelli}, {Burlacu}, {Busonero}, {Butkevich}, {Buzzi}, {Caffau},
  {Cancelliere}, {Cannizzaro}, {Cantat-Gaudin}, {Carballo}, {Carlucci},
  {Carrasco}, {Casamiquela}, {Castellani}, {Castro-Ginard}, {Charlot},
  {Chemin}, {Chiavassa}, {Cocozza}, {Costigan}, {Cowell}, {Crifo}, {Crosta},
  {Crowley}, {Cuypers}, {Dafonte}, {Damerdji}, {Dapergolas}, {David}, {David},
  {de Laverny}, {De Luise}, {De March}, {de Martino}, {de Souza}, {de Torres},
  {Debosscher}, {del Pozo}, {Delbo}, {Delgado}, {Delgado}, {Di Matteo},
  {Diakite}, {Diener}, {Distefano}, {Dolding}, {Drazinos}, {Dur{\'a}n},
  {Edvardsson}, {Enke}, {Eriksson}, {Esquej}, {Eynard Bontemps}, {Fabre},
  {Fabrizio}, {Faigler}, {Falc{\~a}o}, {Farr{\`a}s Casas}, {Federici},
  {Fedorets}, {Fernique}, {Figueras}, {Filippi}, {Findeisen}, {Fonti},
  {Fraile}, {Fraser}, {Fr{\'e}zouls}, {Gai}, {Galleti}, {Garabato},
  {Garc{\'\i}a-Sedano}, {Garofalo}, {Garralda}, {Gavel}, {Gavras}, {Gerssen},
  {Geyer}, {Giacobbe}, {Gilmore}, {Girona}, {Giuffrida}, {Glass}, {Gomes},
  {Granvik}, {Gueguen}, {Guerrier}, {Guiraud}, {Guti{\'e}rrez-S{\'a}nchez},
  {Haigron}, {Hatzidimitriou}, {Hauser}, {Haywood}, {Heiter}, {Helmi}, {Heu},
  {Hilger}, {Hobbs}, {Hofmann}, {Holland}, {Huckle}, {Hypki}, {Icardi},
  {Jan{\ss}en}, {Jevardat de Fombelle}, {Jonker}, {Juh{\'a}sz}, {Julbe},
  {Karampelas}, {Kewley}, {Klar}, {Kochoska}, {Kohley}, {Kolenberg},
  {Kontizas}, {Kontizas}, {Koposov}, {Kordopatis}, {Kostrzewa-Rutkowska},
  {Koubsky}, {Lambert}, {Lanza}, {Lasne}, {Lavigne}, {Le Fustec}, {Le
  Poncin-Lafitte}, {Lebreton}, {Leccia}, {Leclerc}, {Lecoeur-Taibi},
  {Lenhardt}, {Leroux}, {Liao}, {Licata}, {Lindstr{\o}m}, {Lister}, {Livanou},
  {Lobel}, {L{\'o}pez}, {Managau}, {Mann}, {Mantelet}, {Marchal}, {Marchant},
  {Marconi}, {Marinoni}, {Marschalk{\'o}}, {Marshall}, {Martino}, {Marton},
  {Mary}, {Massari}, {Matijevi{\v{c}}}, {Mazeh}, {McMillan}, {Messina},
  {Michalik}, {Millar}, {Molina}, {Molinaro}, {Moln{\'a}r}, {Montegriffo},
  {Mor}, {Morbidelli}, {Morel}, {Morris}, {Mulone}, {Muraveva}, {Musella},
  {Nelemans}, {Nicastro}, {Noval}, {O'Mullane}, {Ord{\'e}novic},
  {Ord{\'o}{\~n}ez-Blanco}, {Osborne}, {Pagani}, {Pagano}, {Pailler},
  {Palacin}, {Palaversa}, {Panahi}, {Pawlak}, {Piersimoni}, {Pineau}, {Plachy},
  {Plum}, {Poggio}, {Poujoulet}, {Pr{\v{s}}a}, {Pulone}, {Racero}, {Ragaini},
  {Rambaux}, {Ramos-Lerate}, {Regibo}, {Reyl{\'e}}, {Riclet}, {Ripepi}, {Riva},
  {Rivard}, {Rixon}, {Roegiers}, {Roelens}, {Romero-G{\'o}mez}, {Rowell},
  {Royer}, {Ruiz-Dern}, {Sadowski}, {Sagrist{\`a} Sell{\'e}s}, {Sahlmann},
  {Salgado}, {Salguero}, {Sanna}, {Santana-Ros}, {Sarasso}, {Savietto},
  {Schultheis}, {Sciacca}, {Segol}, {Segovia}, {S{\'e}gransan}, {Shih},
  {Siltala}, {Silva}, {Smart}, {Smith}, {Solano}, {Solitro}, {Sordo}, {Soria
  Nieto}, {Souchay}, {Spagna}, {Spoto}, {Stampa}, {Steele},
  {Steidelm{\"u}ller}, {Stephenson}, {Stoev}, {Suess}, {Surdej}, {Szabados},
  {Szegedi-Elek}, {Tapiador}, {Taris}, {Tauran}, {Taylor}, {Teixeira},
  {Terrett}, {Teyssand ier}, {Thuillot}, {Titarenko}, {Torra Clotet}, {Turon},
  {Ulla}, {Utrilla}, {Uzzi}, {Vaillant}, {Valentini}, {Valette}, {van Elteren},
  {Van Hemelryck}, {van Leeuwen}, {Vaschetto}, {Vecchiato}, {Veljanoski},
  {Viala}, {Vicente}, {Vogt}, {von Essen}, {Voss}, {Votruba}, {Voutsinas},
  {Walmsley}, {Weiler}, {Wertz}, {Wevers}, {Wyrzykowski}, {Yoldas},
  {{\v{Z}}erjal}, {Ziaeepour}, {Zorec}, {Zschocke}, {Zucker}, {Zurbach}, \&
  {Zwitter}}]{gbv18}
{Gaia Collaboration}, {Brown}, A.~G.~A., {Vallenari}, A., {et~al.} 2018,
  \href{http://dx.doi.org/10.1051/0004-6361/201833051}{\color{magenta}Astronomy
  \& Astrophysics},
  \href{https://ui.adsabs.harvard.edu/abs/2018A&A...616A...1G}{\color{blue}616},
  \href{https://ui.adsabs.harvard.edu/abs/2018A&A...616A...1G}{\color{blue}A1}

\bibitem[{{Galicher} {et~al.}(2011){Galicher}, {Marois}, {Macintosh}, {Barman},
  \& {Konopacky}}]{gmm11}
{Galicher}, R., {Marois}, C., {Macintosh}, B., {et~al.} 2011,
  \href{http://dx.doi.org/10.1088/2041-8205/739/2/L41}{\color{magenta}\apjl},
  \href{http://adsabs.harvard.edu/abs/2011ApJ...739L..41G}{\color{blue}739},
  \href{http://adsabs.harvard.edu/abs/2011ApJ...739L..41G}{\color{blue}L41}

\bibitem[{{Gardner} {et~al.}(2006){Gardner}, {Mather}, {Clampin}, {Doyon},
  {Greenhouse}, {Hammel}, {Hutchings}, {Jakobsen}, {Lilly}, {Long}, {Lunine},
  {McCaughrean}, {Mountain}, {Nella}, {Rieke}, {Rieke}, {Rix}, {Smith},
  {Sonneborn}, {Stiavelli}, {Stockman}, {Windhorst}, \& {Wright}}]{gmc06}
{Gardner}, J.~P., {Mather}, J.~C., {Clampin}, M., {et~al.} 2006,
  \href{http://dx.doi.org/10.1007/s11214-006-8315-7}{\color{magenta}\ssr},
  \href{https://ui.adsabs.harvard.edu/abs/2006SSRv..123..485G}{\color{blue}123},
  \href{https://ui.adsabs.harvard.edu/abs/2006SSRv..123..485G}{\color{blue}485}

\bibitem[{{Gauza} {et~al.}(2015){Gauza}, {B{\'e}jar}, {P{\'e}rez-Garrido},
  {Rosa Zapatero Osorio}, {Lodieu}, {Rebolo}, {Pall{\'e}}, \& {Nowak}}]{gbp15}
{Gauza}, B., {B{\'e}jar}, V.~J.~S., {P{\'e}rez-Garrido}, A., {et~al.} 2015,
  \href{http://dx.doi.org/10.1088/0004-637X/804/2/96}{\color{magenta}\apj},
  \href{http://adsabs.harvard.edu/abs/2015ApJ...804...96G}{\color{blue}804},
  \href{http://adsabs.harvard.edu/abs/2015ApJ...804...96G}{\color{blue}96}

\bibitem[{{Girard} {et~al.}(2018){Girard}, {Blair}, {Brooks}, {Brooks},
  {Brown}, {Bushouse}, {Canipe}, {Chen}, {Correnti}, {Hagan}, {Hilbert},
  {Hines}, {Leisenring}, {Long}, {Nickson}, {Perrin}, {Pontoppidan}, {Pueyo},
  {Rajan}, {Riedel}, {Soummer}, {Stansberry}, {Stark}, {Van Gorkom}, \&
  {York}}]{gbb18}
{Girard}, J.~H., {Blair}, W., {Brooks}, B., {et~al.} 2018,
  \href{http://dx.doi.org/10.1117/12.2314198}{\color{magenta}Proc.~SPIE},
  \href{https://ui.adsabs.harvard.edu/abs/2018SPIE10698E..3VG}{\color{blue}10698},
  \href{https://ui.adsabs.harvard.edu/abs/2018SPIE10698E..3VG}{\color{blue}106983V}

\bibitem[{{Gravity Collaboration} {et~al.}(2019){Gravity Collaboration},
  {Lacour}, {Nowak}, {Wang}, {Pfuhl}, {Eisenhauer}, {Abuter}, {Amorim},
  {Anugu}, {Benisty}, {Berger}, {Beust}, {Blind}, {Bonnefoy}, {Bonnet},
  {Bourget}, {Brandner}, {Buron}, {Collin}, {Charnay}, {Chapron}, {Cl{\'e}net},
  {Coud{\'e} Du Foresto}, {de Zeeuw}, {Deen}, {Dembet}, {Dexter}, {Duvert},
  {Eckart}, {F{\"o}rster Schreiber}, {F{\'e}dou}, {Garcia}, {Garcia Lopez},
  {Gao}, {Gendron}, {Genzel}, {Gillessen}, {Gordo}, {Greenbaum}, {Habibi},
  {Haubois}, {Hau{\ss}mann}, {Henning}, {Hippler}, {Horrobin}, {Hubert},
  {Jimenez Rosales}, {Jocou}, {Kendrew}, {Kervella}, {Kolb}, {Lagrange},
  {Lapeyr{\`e}re}, {Le Bouquin}, {L{\'e}na}, {Lippa}, {Lenzen}, {Maire},
  {Molli{\`e}re}, {Ott}, {Paumard}, {Perraut}, {Perrin}, {Pueyo}, {Rabien},
  {Ram{\'\i}rez}, {Rau}, {Rodr{\'\i}guez-Coira}, {Rousset}, {Sanchez-Bermudez},
  {Scheithauer}, {Schuhler}, {Straub}, {Straubmeier}, {Sturm}, {Tacconi},
  {Vincent}, {van Dishoeck}, {von Fellenberg}, {Wank}, {Waisberg}, {Widmann},
  {Wieprecht}, {Wiest}, {Wiezorrek}, {Woillez}, {Yazici}, {Ziegler}, \&
  {Zins}}]{gln19}
{Gravity Collaboration}, {Lacour}, S., {Nowak}, M., {et~al.} 2019,
  \href{http://dx.doi.org/10.1051/0004-6361/201935253}{\color{magenta}Astronomy
  \& Astrophysics},
  \href{https://ui.adsabs.harvard.edu/abs/2019A&A...623L..11G}{\color{blue}623},
  \href{https://ui.adsabs.harvard.edu/abs/2019A&A...623L..11G}{\color{blue}L11}

\bibitem[{{Green} {et~al.}(2005){Green}, {Beichman}, {Basinger}, {Horner},
  {Meyer}, {Redding}, {Rieke}, \& {Trauger}}]{gbb05}
{Green}, J.~J., {Beichman}, C., {Basinger}, S.~A., {et~al.} 2005,
  \href{http://dx.doi.org/10.1117/12.619343}{\color{magenta}Proc.~SPIE},
  \href{https://ui.adsabs.harvard.edu/abs/2005SPIE.5905..185G}{\color{blue}5905},
  \href{https://ui.adsabs.harvard.edu/abs/2005SPIE.5905..185G}{\color{blue}185}

\bibitem[{{Greenbaum} {et~al.}(2015){Greenbaum}, {Pueyo}, {Sivaramakrishnan},
  \& {Lacour}}]{gps15}
{Greenbaum}, A.~Z., {Pueyo}, L., {Sivaramakrishnan}, A., \& {Lacour}, S. 2015,
  \href{http://dx.doi.org/10.1088/0004-637X/798/2/68}{\color{magenta}\apj},
  \href{https://ui.adsabs.harvard.edu/abs/2015ApJ...798...68G}{\color{blue}798},
  \href{https://ui.adsabs.harvard.edu/abs/2015ApJ...798...68G}{\color{blue}68}

\bibitem[{{Grillmair} {et~al.}(2007){Grillmair}, {Charbonneau}, {Burrows},
  {Armus}, {Stauffer}, {Meadows}, {Van Cleve}, \& {Levine}}]{gcb07}
{Grillmair}, C.~J., {Charbonneau}, D., {Burrows}, A., {et~al.} 2007,
  \href{http://dx.doi.org/10.1086/513741}{\color{magenta}\apjl},
  \href{https://ui.adsabs.harvard.edu/abs/2007ApJ...658L.115G}{\color{blue}658},
  \href{https://ui.adsabs.harvard.edu/abs/2007ApJ...658L.115G}{\color{blue}L115}

\bibitem[{{Grillmair} {et~al.}(2008){Grillmair}, {Burrows}, {Charbonneau},
  {Armus}, {Stauffer}, {Meadows}, {van Cleve}, {von Braun}, \&
  {Levine}}]{gbc08}
{Grillmair}, C.~J., {Burrows}, A., {Charbonneau}, D., {et~al.} 2008,
  \href{http://dx.doi.org/10.1038/nature07574}{\color{magenta}\nat},
  \href{http://adsabs.harvard.edu/abs/2008Natur.456..767G}{\color{blue}456},
  \href{http://adsabs.harvard.edu/abs/2008Natur.456..767G}{\color{blue}767}

\bibitem[{{Hagan} {et~al.}(2018){Hagan}, {Choquet}, {Soummer}, \&
  {Vigan}}]{hcs18}
{Hagan}, J.~B., {Choquet}, {\'E}., {Soummer}, R., \& {Vigan}, A. 2018,
  \href{http://dx.doi.org/10.3847/1538-3881/aab14b}{\color{magenta}\aj},
  \href{https://ui.adsabs.harvard.edu/abs/2018AJ....155..179H}{\color{blue}155},
  \href{https://ui.adsabs.harvard.edu/abs/2018AJ....155..179H}{\color{blue}179}

\bibitem[{{Hinkley} {et~al.}(2011{\natexlab{a}}){Hinkley}, {Carpenter},
  {Ireland}, \& {Kraus}}]{hci11}
{Hinkley}, S., {Carpenter}, J.~M., {Ireland}, M.~J., \& {Kraus}, A.~L.
  2011{\natexlab{a}},
  \href{http://dx.doi.org/10.1088/2041-8205/730/2/L21}{\color{magenta}\apjl},
  \href{http://adsabs.harvard.edu/abs/2011ApJ...730L..21H}{\color{blue}730},
  \href{http://adsabs.harvard.edu/abs/2011ApJ...730L..21H}{\color{blue}L21+}

\bibitem[{{Hinkley} {et~al.}(2007){Hinkley}, {Oppenheimer}, {Soummer},
  {Sivaramakrishnan}, {Roberts}, {Kuhn}, {Makidon}, {Perrin}, {Lloyd},
  {Kratter}, \& {Brenner}}]{hos07}
{Hinkley}, S., {Oppenheimer}, B.~R., {Soummer}, R., {et~al.} 2007, \apj,
  \href{http://adsabs.harvard.edu/cgi-bin/nph-bib_query?bibcode=2007ApJ...654..633H&db_key=AST}{\color{blue}654},
  \href{http://adsabs.harvard.edu/cgi-bin/nph-bib_query?bibcode=2007ApJ...654..633H&db_key=AST}{\color{blue}633}

\bibitem[{{Hinkley} {et~al.}(2011{\natexlab{b}}){Hinkley}, {Oppenheimer},
  {Zimmerman}, {Brenner}, {Parry}, {Crepp}, {Vasisht}, {Ligon}, {King},
  {Soummer}, {Sivaramakrishnan}, {Beichman}, {Shao}, {Roberts}, {Bouchez},
  {Dekany}, {Pueyo}, {Roberts}, {Lockhart}, {Zhai}, {Shelton}, \&
  {Burruss}}]{hoz11}
{Hinkley}, S., {Oppenheimer}, B.~R., {Zimmerman}, N., {et~al.}
  2011{\natexlab{b}},
  \href{http://dx.doi.org/10.1086/658163}{\color{magenta}\pasp},
  \href{http://adsabs.harvard.edu/abs/2011PASP..123...74H}{\color{blue}123},
  \href{http://adsabs.harvard.edu/abs/2011PASP..123...74H}{\color{blue}74}

\bibitem[{{Hinkley} {et~al.}(2015{\natexlab{a}}){Hinkley}, {Bowler}, {Vigan},
  {Aller}, {Liu}, {Mawet}, {Matthews}, {Wahhaj}, {Kraus}, {Baraffe}, \&
  {Chabrier}}]{hbv15}
{Hinkley}, S., {Bowler}, B.~P., {Vigan}, A., {et~al.} 2015{\natexlab{a}},
  \href{http://dx.doi.org/10.1088/2041-8205/805/1/L10}{\color{magenta}\apjl},
  \href{http://adsabs.harvard.edu/abs/2015ApJ...805L..10H}{\color{blue}805},
  \href{http://adsabs.harvard.edu/abs/2015ApJ...805L..10H}{\color{blue}L10}

\bibitem[{{Hinkley} {et~al.}(2015{\natexlab{b}}){Hinkley}, {Kraus}, {Ireland},
  {Cheetham}, {Carpenter}, {Tuthill}, {Lacour}, {Evans}, \& {Haubois}}]{hki15}
{Hinkley}, S., {Kraus}, A.~L., {Ireland}, M.~J., {et~al.} 2015{\natexlab{b}},
  \href{http://dx.doi.org/10.1088/2041-8205/806/1/L9}{\color{magenta}\apjl},
  \href{http://adsabs.harvard.edu/abs/2015ApJ...806L...9H}{\color{blue}806},
  \href{http://adsabs.harvard.edu/abs/2015ApJ...806L...9H}{\color{blue}L9}

\bibitem[{{Hinkley} {et~al.}(2021){Hinkley}, {Matthews}, {Lefevre}, {Lestrade},
  {Kennedy}, {Mawet}, {Stapelfeldt}, {Ray}, {Mamajek}, {Bowler}, {Wilner},
  {Williams}, {Ansdell}, {Wyatt}, {Lau}, {Phillips}, {Fernandez}, {Gagn{\'e}},
  {Bubb}, {Sutlieff}, {Wilson}, {Matthews}, {Ngo}, {Piskorz}, {Crepp},
  {Gonzalez}, {Mann}, \& {Mace}}]{hml21}
{Hinkley}, S., {Matthews}, E.~C., {Lefevre}, C., {et~al.} 2021,
  \href{http://dx.doi.org/10.3847/1538-4357/abec6e}{\color{magenta}\apj},
  \href{https://ui.adsabs.harvard.edu/abs/2021ApJ...912..115H}{\color{blue}912},
  \href{https://ui.adsabs.harvard.edu/abs/2021ApJ...912..115H}{\color{blue}115}

\bibitem[{{Hubeny} \& {Burrows}(2007)}]{hb07}
{Hubeny}, I., \& {Burrows}, A. 2007,
  \href{http://dx.doi.org/10.1086/522107}{\color{magenta}\apj},
  \href{https://ui.adsabs.harvard.edu/abs/2007ApJ...669.1248H}{\color{blue}669},
  \href{https://ui.adsabs.harvard.edu/abs/2007ApJ...669.1248H}{\color{blue}1248}

\bibitem[{{Hughes} {et~al.}(2018){Hughes}, {Duch{\^e}ne}, \&
  {Matthews}}]{hdm18}
{Hughes}, A.~M., {Duch{\^e}ne}, G., \& {Matthews}, B.~C. 2018,
  \href{http://dx.doi.org/10.1146/annurev-astro-081817-052035}{\color{magenta}\araa},
  \href{https://ui.adsabs.harvard.edu/abs/2018ARA&A..56..541H}{\color{blue}56},
  \href{https://ui.adsabs.harvard.edu/abs/2018ARA&A..56..541H}{\color{blue}541}

\bibitem[{{Ireland}(2013)}]{i13}
{Ireland}, M.~J. 2013,
  \href{http://dx.doi.org/10.1093/mnras/stt859}{\color{magenta}\mnras},
  \href{http://adsabs.harvard.edu/abs/2013MNRAS.433.1718I}{\color{blue}433},
  \href{http://adsabs.harvard.edu/abs/2013MNRAS.433.1718I}{\color{blue}1718}

\bibitem[{{Jakobsen} {et~al.}(2022){Jakobsen}, {Ferruit}, {Alves de Oliveira},
  {Arribas}, {Bagnasco}, {Barho}, {Beck}, {Birkmann}, {B{\"o}ker}, {Bunker},
  {Charlot}, {de Jong}, {de Marchi}, {Ehrenwinkler}, {Falcolini}, {Fels},
  {Franx}, {Franz}, {Funke}, {Giardino}, {Gnata}, {Holota}, {Honnen}, {Jensen},
  {Jentsch}, {Johnson}, {Jollet}, {Karl}, {Kling}, {K{\"o}hler}, {Kolm},
  {Kumari}, {Lander}, {Lemke}, {L{\'o}pez-Caniego}, {L{\"u}tzgendorf},
  {Maiolino}, {Manjavacas}, {Marston}, {Maschmann}, {Maurer}, {Messerschmidt},
  {Moseley}, {Mosner}, {Mott}, {Muzerolle}, {Pirzkal}, {Pittet}, {Plitzke},
  {Posselt}, {Rapp}, {Rauscher}, {Rawle}, {Rix}, {R{\"o}del}, {Rumler},
  {Sabbi}, {Salvignol}, {Schmid}, {Sirianni}, {Smith}, {Strada}, {te Plate},
  {Valenti}, {Wettemann}, {Wiehe}, {Wiesmayer}, {Willott}, {Wright}, {Zeidler},
  \& {Zincke}}]{jfa22}
{Jakobsen}, P., {Ferruit}, P., {Alves de Oliveira}, C., {et~al.} 2022,
  \href{http://dx.doi.org/10.1051/0004-6361/202142663}{\color{magenta}\aap},
  \href{https://ui.adsabs.harvard.edu/abs/2022A&A...661A..80J}{\color{blue}661},
  \href{https://ui.adsabs.harvard.edu/abs/2022A&A...661A..80J}{\color{blue}A80}

\bibitem[{{Janson} {et~al.}(2015){Janson}, {Quanz}, {Carson}, {Thalmann},
  {Lafreni{\`e}re}, \& {Amara}}]{jqc15}
{Janson}, M., {Quanz}, S.~P., {Carson}, J.~C., {et~al.} 2015,
  \href{http://dx.doi.org/10.1051/0004-6361/201424944}{\color{magenta}Astronomy
  \& Astrophysics},
  \href{https://ui.adsabs.harvard.edu/abs/2015A&A...574A.120J}{\color{blue}574},
  \href{https://ui.adsabs.harvard.edu/abs/2015A&A...574A.120J}{\color{blue}A120}

\bibitem[{{Janson} {et~al.}(2021){Janson}, {Gratton}, {Rodet}, {Vigan},
  {Bonnefoy}, {Delorme}, {Mamajek}, {Reffert}, {Stock}, {Marleau}, {Langlois},
  {Chauvin}, {Desidera}, {Ringqvist}, {Mayer}, {Viswanath}, {Squicciarini},
  {Meyer}, {Samland}, {Petrus}, {Helled}, {Kenworthy}, {Quanz}, {Biller},
  {Henning}, {Mesa}, {Engler}, \& {Carson}}]{jgr21}
{Janson}, M., {Gratton}, R., {Rodet}, L., {et~al.} 2021,
  \href{http://dx.doi.org/10.1038/s41586-021-04124-8}{\color{magenta}\nat},
  \href{https://ui.adsabs.harvard.edu/abs/2021Natur.600..231J}{\color{blue}600},
  \href{https://ui.adsabs.harvard.edu/abs/2021Natur.600..231J}{\color{blue}231}

\bibitem[{{Kammerer} {et~al.}(2021){Kammerer}, {Lacour}, {Stolker},
  {Molli{\`e}re}, {Sing}, {Nasedkin}, {Kervella}, {Wang}, {Ward-Duong},
  {Nowak}, {Abuter}, {Amorim}, {Asensio-Torres}, {Baub{\"o}ck}, {Benisty},
  {Berger}, {Beust}, {Blunt}, {Boccaletti}, {Bohn}, {Bolzer}, {Bonnefoy},
  {Bonnet}, {Brandner}, {Cantalloube}, {Caselli}, {Charnay}, {Chauvin},
  {Choquet}, {Christiaens}, {Cl{\'e}net}, {Coud{\'e} du Foresto}, {Cridland},
  {Dembet}, {Dexter}, {de Zeeuw}, {Drescher}, {Duvert}, {Eckart}, {Eisenhauer},
  {Gao}, {Garcia}, {Garcia Lopez}, {Gendron}, {Genzel}, {Gillessen}, {Girard},
  {Haubois}, {Hei{\ss}el}, {Henning}, {Hinkley}, {Hippler}, {Horrobin},
  {Houll{\'e}}, {Hubert}, {Jocou}, {Keppler}, {Kreidberg}, {Lagrange},
  {Lapeyr{\`e}re}, {Le Bouquin}, {L{\'e}na}, {Lutz}, {Maire}, {M{\'e}rand},
  {Monnier}, {Mouillet}, {M{\"u}ller}, {Ott}, {Otten}, {Paladini}, {Paumard},
  {Perraut}, {Perrin}, {Pfuhl}, {Pueyo}, {Rameau}, {Rodet}, {Rousset},
  {Rustamkulov}, {Shangguan}, {Shimizu}, {Stadler}, {Straub}, {Straubmeier},
  {Sturm}, {Tacconi}, {van Dishoeck}, {Vigan}, {Vincent}, {von Fellenberg},
  {Widmann}, {Wieprecht}, {Wiezorrek}, {Woillez}, \& {Yazici}}]{kls21}
{Kammerer}, J., {Lacour}, S., {Stolker}, T., {et~al.} 2021,
  \href{http://dx.doi.org/10.1051/0004-6361/202140749}{\color{magenta}Astronomy
  \& Astrophysics},
  \href{https://ui.adsabs.harvard.edu/abs/2021A&A...652A..57K}{\color{blue}652},
  \href{https://ui.adsabs.harvard.edu/abs/2021A&A...652A..57K}{\color{blue}A57}

\bibitem[{{Kitzmann} \& {Stock}(2018)}]{ks18}
{Kitzmann}, D., \& {Stock}, J. 2018, {FastChem: An ultra-fast equilibrium
  chemistry}

\bibitem[{{Konishi} {et~al.}(2016){Konishi}, {Grady}, {Schneider}, {Shibai},
  {McElwain}, {Nesvold}, {Kuchner}, {Carson}, {Debes}, {Gaspar}, {Henning},
  {Hines}, {Hinz}, {Jang-Condell}, {Moro-Mart{\'\i}n}, {Perrin}, {Rodigas},
  {Serabyn}, {Silverstone}, {Stark}, {Tamura}, {Weinberger}, \&
  {Wisniewski}}]{kgs16}
{Konishi}, M., {Grady}, C.~A., {Schneider}, G., {et~al.} 2016,
  \href{http://dx.doi.org/10.3847/2041-8205/818/2/L23}{\color{magenta}\apjl},
  \href{https://ui.adsabs.harvard.edu/abs/2016ApJ...818L..23K}{\color{blue}818},
  \href{https://ui.adsabs.harvard.edu/abs/2016ApJ...818L..23K}{\color{blue}L23}

\bibitem[{{Konopacky} {et~al.}(2013){Konopacky}, {Barman}, {Macintosh}, \&
  {Marois}}]{kbm13}
{Konopacky}, Q.~M., {Barman}, T.~S., {Macintosh}, B.~A., \& {Marois}, C. 2013,
  \href{http://dx.doi.org/10.1126/science.1232003}{\color{magenta}Science},
  \href{http://adsabs.harvard.edu/abs/2013Sci...339.1398K}{\color{blue}339},
  \href{http://adsabs.harvard.edu/abs/2013Sci...339.1398K}{\color{blue}1398}

\bibitem[{{Kratter} {et~al.}(2010){Kratter}, {Murray-Clay}, \&
  {Youdin}}]{kmy10}
{Kratter}, K.~M., {Murray-Clay}, R.~A., \& {Youdin}, A.~N. 2010,
  \href{http://dx.doi.org/10.1088/0004-637X/710/2/1375}{\color{magenta}\apj},
  \href{http://adsabs.harvard.edu/abs/2010ApJ...710.1375K}{\color{blue}710},
  \href{http://adsabs.harvard.edu/abs/2010ApJ...710.1375K}{\color{blue}1375}

\bibitem[{{Kraus} \& {Ireland}(2012)}]{ki12}
{Kraus}, A.~L., \& {Ireland}, M.~J. 2012,
  \href{http://dx.doi.org/10.1088/0004-637X/745/1/5}{\color{magenta}\apj},
  \href{http://adsabs.harvard.edu/abs/2012ApJ...745....5K}{\color{blue}745},
  \href{http://adsabs.harvard.edu/abs/2012ApJ...745....5K}{\color{blue}5}

\bibitem[{{Kraus} {et~al.}(2011){Kraus}, {Ireland}, {Martinache}, \&
  {Hillenbrand}}]{kim11}
{Kraus}, A.~L., {Ireland}, M.~J., {Martinache}, F., \& {Hillenbrand}, L.~A.
  2011,
  \href{http://dx.doi.org/10.1088/0004-637X/731/1/8}{\color{magenta}\apj},
  \href{http://adsabs.harvard.edu/abs/2011ApJ...731....8K}{\color{blue}731},
  \href{http://adsabs.harvard.edu/abs/2011ApJ...731....8K}{\color{blue}8}

\bibitem[{{Krist} {et~al.}(2012){Krist}, {Stapelfeldt}, {Bryden}, \&
  {Plavchan}}]{ksb12}
{Krist}, J.~E., {Stapelfeldt}, K.~R., {Bryden}, G., \& {Plavchan}, P. 2012,
  \href{http://dx.doi.org/10.1088/0004-6256/144/2/45}{\color{magenta}\aj},
  \href{https://ui.adsabs.harvard.edu/abs/2012AJ....144...45K}{\color{blue}144},
  \href{https://ui.adsabs.harvard.edu/abs/2012AJ....144...45K}{\color{blue}45}

\bibitem[{{Krist} {et~al.}(2007){Krist}, {Beichman}, {Trauger}, {Rieke},
  {Somerstein}, {Green}, {Horner}, {Stansberry}, {Shi}, {Meyer}, {Stapelfeldt},
  \& {Roellig}}]{kbt07}
{Krist}, J.~E., {Beichman}, C.~A., {Trauger}, J.~T., {et~al.} 2007,
  \href{http://dx.doi.org/10.1117/12.734873}{\color{magenta}Proc.~SPIE},
  \href{https://ui.adsabs.harvard.edu/abs/2007SPIE.6693E..0HK}{\color{blue}6693},
  \href{https://ui.adsabs.harvard.edu/abs/2007SPIE.6693E..0HK}{\color{blue}66930H}

\bibitem[{{Kupka} {et~al.}(2018){Kupka}, {Zaussinger}, \& {Montgomery}}]{kzm18}
{Kupka}, F., {Zaussinger}, F., \& {Montgomery}, M.~H. 2018,
  \href{http://dx.doi.org/10.1093/mnras/stx3119}{\color{magenta}\mnras},
  \href{https://ui.adsabs.harvard.edu/abs/2018MNRAS.474.4660K}{\color{blue}474},
  \href{https://ui.adsabs.harvard.edu/abs/2018MNRAS.474.4660K}{\color{blue}4660}

\bibitem[{{Lacour} {et~al.}(2021){Lacour}, {Wang}, {Rodet}, {Nowak},
  {Shangguan}, {Beust}, {Lagrange}, {Abuter}, {Amorim}, {Asensio-Torres},
  {Benisty}, {Berger}, {Blunt}, {Boccaletti}, {Bohn}, {Bolzer}, {Bonnefoy},
  {Bonnet}, {Bourdarot}, {Brandner}, {Cantalloube}, {Caselli}, {Charnay},
  {Chauvin}, {Choquet}, {Christiaens}, {Cl{\'e}net}, {Coud{\'e} Du Foresto},
  {Cridland}, {Dembet}, {Dexter}, {de Zeeuw}, {Drescher}, {Duvert}, {Eckart},
  {Eisenhauer}, {Gao}, {Garcia}, {Garcia Lopez}, {Gendron}, {Genzel},
  {Gillessen}, {Girard}, {Haubois}, {Hei{\ss}el}, {Henning}, {Hinkley},
  {Hippler}, {Horrobin}, {Houll{\'e}}, {Hubert}, {Jocou}, {Kammerer},
  {Keppler}, {Kervella}, {Kreidberg}, {Lapeyr{\`e}re}, {Le Bouquin},
  {L{\'e}na}, {Lutz}, {Maire}, {M{\'e}rand}, {Molli{\`e}re}, {Monnier},
  {Mouillet}, {Nasedkin}, {Ott}, {Otten}, {Paladini}, {Paumard}, {Perraut},
  {Perrin}, {Pfuhl}, {Rickman}, {Pueyo}, {Rameau}, {Rousset}, {Rustamkulov},
  {Samland}, {Shimizu}, {Sing}, {Stadler}, {Stolker}, {Straub}, {Straubmeier},
  {Sturm}, {Tacconi}, {van Dishoeck}, {Vigan}, {Vincent}, {von Fellenberg},
  {Ward-Duong}, {Widmann}, {Wieprecht}, {Wiezorrek}, {Woillez}, {Yazici},
  {Young}, \& {Gravity Collaboration}}]{lwr21}
{Lacour}, S., {Wang}, J.~J., {Rodet}, L., {et~al.} 2021,
  \href{http://dx.doi.org/10.1051/0004-6361/202141889}{\color{magenta}Astronomy
  \& Astrophysics},
  \href{https://ui.adsabs.harvard.edu/abs/2021A&A...654L...2L}{\color{blue}654},
  \href{https://ui.adsabs.harvard.edu/abs/2021A&A...654L...2L}{\color{blue}L2}

\bibitem[{{Lafreni{\`e}re} {et~al.}(2009){Lafreni{\`e}re}, {Marois}, {Doyon},
  \& {Barman}}]{lmd09}
{Lafreni{\`e}re}, D., {Marois}, C., {Doyon}, R., \& {Barman}, T. 2009,
  \href{http://dx.doi.org/10.1088/0004-637X/694/2/L148}{\color{magenta}\apjl},
  \href{http://adsabs.harvard.edu/abs/2009ApJ...694L.148L}{\color{blue}694},
  \href{http://adsabs.harvard.edu/abs/2009ApJ...694L.148L}{\color{blue}L148}

\bibitem[{{Lafreni{\`e}re} {et~al.}(2007){Lafreni{\`e}re}, {Marois}, {Doyon},
  {Nadeau}, \& {Artigau}}]{lmd07}
{Lafreni{\`e}re}, D., {Marois}, C., {Doyon}, R., {et~al.} 2007, \apj,
  \href{http://adsabs.harvard.edu/abs/2007ApJ...660..770L}{\color{blue}660},
  \href{http://adsabs.harvard.edu/abs/2007ApJ...660..770L}{\color{blue}770}

\bibitem[{{Lagrange} {et~al.}(2010){Lagrange}, {Bonnefoy}, {Chauvin}, {Apai},
  {Ehrenreich}, {Boccaletti}, {Gratadour}, {Rouan}, {Mouillet}, {Lacour}, \&
  {Kasper}}]{lbc10}
{Lagrange}, A., {Bonnefoy}, M., {Chauvin}, G., {et~al.} 2010,
  \href{http://dx.doi.org/10.1126/science.1187187}{\color{magenta}Science},
  \href{http://adsabs.harvard.edu/abs/2010Sci...329...57L}{\color{blue}329},
  \href{http://adsabs.harvard.edu/abs/2010Sci...329...57L}{\color{blue}57}

\bibitem[{{Lagrange} {et~al.}(2020){Lagrange}, {Rubini}, {Nowak}, {Lacour},
  {Grandjean}, {Boccaletti}, {Langlois}, {Delorme}, {Gratton}, {Wang},
  {Flasseur}, {Galicher}, {Kral}, {Meunier}, {Beust}, {Babusiaux}, {Le
  Coroller}, {Thebault}, {Kervella}, {Zurlo}, {Maire}, {Wahhaj}, {Amorim},
  {Asensio-Torres}, {Benisty}, {Berger}, {Bonnefoy}, {Brandner}, {Cantalloube},
  {Charnay}, {Chauvin}, {Choquet}, {Cl{\'e}net}, {Christiaens}, {Coud{\'e} Du
  Foresto}, {de Zeeuw}, {Desidera}, {Duvert}, {Eckart}, {Eisenhauer},
  {Galland}, {Gao}, {Garcia}, {Garcia Lopez}, {Gendron}, {Genzel}, {Gillessen},
  {Girard}, {Hagelberg}, {Haubois}, {Henning}, {Heissel}, {Hippler},
  {Horrobin}, {Janson}, {Kammerer}, {Kenworthy}, {Keppler}, {Kreidberg},
  {Lapeyr{\`e}re}, {Le Bouquin}, {L{\'e}na}, {M{\'e}rand}, {Messina},
  {Molli{\`e}re}, {Monnier}, {Ott}, {Otten}, {Paumard}, {Paladini}, {Perraut},
  {Perrin}, {Pueyo}, {Pfuhl}, {Rodet}, {Rodriguez-Coira}, {Rousset}, {Samland},
  {Shangguan}, {Schmidt}, {Straub}, {Straubmeier}, {Stolker}, {Vigan},
  {Vincent}, {Widmann}, {Woillez}, \& {Gravity Collaboration}}]{lrn20}
{Lagrange}, A.~M., {Rubini}, P., {Nowak}, M., {et~al.} 2020,
  \href{http://dx.doi.org/10.1051/0004-6361/202038823}{\color{magenta}\aap},
  \href{https://ui.adsabs.harvard.edu/abs/2020A&A...642A..18L}{\color{blue}642},
  \href{https://ui.adsabs.harvard.edu/abs/2020A&A...642A..18L}{\color{blue}A18}

\bibitem[{{Lajoie} {et~al.}(2016){Lajoie}, {Soummer}, {Pueyo}, {Hines},
  {Nelan}, {Perrin}, {Clampin}, \& {Isaacs}}]{lsp16}
{Lajoie}, C.-P., {Soummer}, R., {Pueyo}, L., {et~al.} 2016,
  \href{http://dx.doi.org/10.1117/12.2233032}{\color{magenta}Proc.~SPIE},
  \href{https://ui.adsabs.harvard.edu/abs/2016SPIE.9904E..5KL}{\color{blue}9904},
  \href{https://ui.adsabs.harvard.edu/abs/2016SPIE.9904E..5KL}{\color{blue}99045K}

\bibitem[{{Langlois} {et~al.}(2021){Langlois}, {Gratton}, {Lagrange},
  {Delorme}, {Boccaletti}, {Bonnefoy}, {Maire}, {Mesa}, {Chauvin}, {Desidera},
  {Vigan}, {Cheetham}, {Hagelberg}, {Feldt}, {Meyer}, {Rubini}, {Le Coroller},
  {Cantalloube}, {Biller}, {Bonavita}, {Bhowmik}, {Brandner}, {Daemgen},
  {D'Orazi}, {Flasseur}, {Fontanive}, {Galicher}, {Girard}, {Janin-Potiron},
  {Janson}, {Keppler}, {Kopytova}, {Lagadec}, {Lannier}, {Lazzoni}, {Ligi},
  {Meunier}, {Perreti}, {Perrot}, {Rodet}, {Romero}, {Rouan}, {Samland},
  {Salter}, {Sissa}, {Schmidt}, {Zurlo}, {Mouillet}, {Denis}, {Thi{\'e}baut},
  {Milli}, {Wahhaj}, {Beuzit}, {Dominik}, {Henning}, {M{\'e}nard},
  {M{\"u}ller}, {Schmid}, {Turatto}, {Udry}, {Abe}, {Antichi}, {Allard},
  {Baruffolo}, {Baudoz}, {Baudrand}, {Bazzon}, {Blanchard}, {Carbillet},
  {Carle}, {Cascone}, {Charton}, {Claudi}, {Costille}, {De Caprio},
  {Delboulb{\'e}}, {Dohlen}, {Fantinel}, {Feautrier}, {Fusco}, {Gigan}, {Giro},
  {Gisler}, {Gluck}, {Gry}, {Hubin}, {Hugot}, {Jaquet}, {Kasper}, {Le Mignant},
  {Llored}, {Madec}, {Magnard}, {Martinez}, {Maurel}, {Messina},
  {M{\"o}ller-Nilsson}, {Mugnier}, {Moulin}, {Orign{\'e}}, {Pavlov}, {Perret},
  {Petit}, {Pragt}, {Puget}, {Rabou}, {Ramos}, {Rigal}, {Rochat}, {Roelfsema},
  {Rousset}, {Roux}, {Salasnich}, {Sauvage}, {Sevin}, {Soenke}, {Stadler},
  {Suarez}, {Weber}, {Wildi}, \& {Rickman}}]{lgl21}
{Langlois}, M., {Gratton}, R., {Lagrange}, A.~M., {et~al.} 2021,
  \href{http://dx.doi.org/10.1051/0004-6361/202039753}{\color{magenta}\aap},
  \href{https://ui.adsabs.harvard.edu/abs/2021A&A...651A..71L}{\color{blue}651},
  \href{https://ui.adsabs.harvard.edu/abs/2021A&A...651A..71L}{\color{blue}A71}

\bibitem[{{Liu} {et~al.}(2016){Liu}, {Dupuy}, \& {Allers}}]{lda16}
{Liu}, M.~C., {Dupuy}, T.~J., \& {Allers}, K.~N. 2016,
  \href{http://dx.doi.org/10.3847/1538-4357/833/1/96}{\color{magenta}\apj},
  \href{https://ui.adsabs.harvard.edu/abs/2016ApJ...833...96L}{\color{blue}833},
  \href{https://ui.adsabs.harvard.edu/abs/2016ApJ...833...96L}{\color{blue}96}

\bibitem[{{Looper} {et~al.}(2008){Looper}, {Kirkpatrick}, {Cutri}, {Barman},
  {Burgasser}, {Cushing}, {Roellig}, {McGovern}, {McLean}, {Rice}, {Swift}, \&
  {Schurr}}]{lkc08}
{Looper}, D.~L., {Kirkpatrick}, J.~D., {Cutri}, R.~M., {et~al.} 2008,
  \href{http://dx.doi.org/10.1086/591025}{\color{magenta}\apj},
  \href{https://ui.adsabs.harvard.edu/abs/2008ApJ...686..528L}{\color{blue}686},
  \href{https://ui.adsabs.harvard.edu/abs/2008ApJ...686..528L}{\color{blue}528}

\bibitem[{{Macintosh} {et~al.}(2015){Macintosh}, {Graham}, {Barman}, {De Rosa},
  {Konopacky}, {Marley}, {Marois}, {Nielsen}, {Pueyo}, {Rajan}, {Rameau},
  {Saumon}, {Wang}, {Ammons}, {Arriaga}, {Artigau}, {Beckwith}, {Brewster},
  {Bruzzone}, {Bulger}, {Burningham}, {Burrows}, {Chen}, {Duchene}, {Esposito},
  {Fabrycky}, {Fitzgerald}, {Follette}, {Fortney}, {Gerard}, {Goodsell},
  {Greenbaum}, {Hibon}, {Hinkley}, {Hufford}, {Hung}, {Ingraham},
  {Johnson-Groh}, {Kalas}, {Lafreniere}, {Larkin}, {Lee}, {Line}, {Long},
  {Maire}, {Marchis}, {Matthews}, {Max}, {Metchev}, {Millar-Blanchaer},
  {Mittal}, {Morley}, {Morzinski}, {Murray-Clay}, {Oppenheimer}, {Palmer},
  {Patel}, {Patience}, {Perrin}, {Poyneer}, {Rafikov}, {Rantakyro}, {Rice},
  {Rojo}, {Rudy}, {Ruffio}, {Ruiz}, {Sadakuni}, {Saddlemyer}, {Salama},
  {Savransky}, {Schneider}, {Sivaramakrishnan}, {Song}, {Soummer}, {Thomas},
  {Vasisht}, {Wallace}, {Ward-Duong}, {Wiktorowicz}, {Wolff}, \&
  {Zuckerman}}]{mgb15}
{Macintosh}, B., {Graham}, J.~R., {Barman}, T., {et~al.} 2015, Science,
  \href{http://adsabs.harvard.edu/abs/2015Sci...350...64M}{\color{blue}350},
  \href{http://adsabs.harvard.edu/abs/2015Sci...350...64M}{\color{blue}64}

\bibitem[{{Macintosh} {et~al.}(2018){Macintosh}, {Chilcote}, {Bailey}, {de
  Rosa}, {Nielsen}, {Norton}, {Poyneer}, {Wang}, {Ruffio}, {Graham}, {Marois},
  {Savransky}, \& {Veran}}]{mcb18}
{Macintosh}, B., {Chilcote}, J.~K., {Bailey}, V.~P., {et~al.} 2018,
  \href{http://dx.doi.org/10.1117/12.2314253}{\color{magenta}Proc.~SPIE},
  \href{https://ui.adsabs.harvard.edu/abs/2018SPIE10703E..0KM}{\color{blue}10703},
  \href{https://ui.adsabs.harvard.edu/abs/2018SPIE10703E..0KM}{\color{blue}107030K}

\bibitem[{{Madhusudhan} {et~al.}(2017){Madhusudhan}, {Bitsch}, {Johansen}, \&
  {Eriksson}}]{mbj17}
{Madhusudhan}, N., {Bitsch}, B., {Johansen}, A., \& {Eriksson}, L. 2017,
  \href{http://dx.doi.org/10.1093/mnras/stx1139}{\color{magenta}\mnras},
  \href{https://ui.adsabs.harvard.edu/abs/2017MNRAS.469.4102M}{\color{blue}469},
  \href{https://ui.adsabs.harvard.edu/abs/2017MNRAS.469.4102M}{\color{blue}4102}

\bibitem[{{Males} {et~al.}(2018){Males}, {Close}, {Miller}, {Schatz},
  {Doelman}, {Lumbres}, {Snik}, {Rodack}, {Knight}, {Van Gorkom}, {Long},
  {Hedglen}, {Kautz}, {Jovanovic}, {Morzinski}, {Guyon}, {Douglas}, {Follette},
  {Lozi}, {Bohlman}, {Durney}, {Gasho}, {Hinz}, {Ireland}, {Jean}, {Keller},
  {Kenworthy}, {Mazin}, {Noenickx}, {Alfred}, {Perez}, {Sanchez}, {Sauve},
  {Weinberger}, \& {Conrad}}]{mcm18}
{Males}, J.~R., {Close}, L.~M., {Miller}, K., {et~al.} 2018,
  \href{http://dx.doi.org/10.1117/12.2312992}{\color{magenta}Proc.~SPIE},
  \href{https://ui.adsabs.harvard.edu/abs/2018SPIE10703E..09M}{\color{blue}10703},
  \href{https://ui.adsabs.harvard.edu/abs/2018SPIE10703E..09M}{\color{blue}1070309}

\bibitem[{{Manjavacas} {et~al.}(2021){Manjavacas}, {Karalidi}, {Vos}, {Biller},
  \& {Lew}}]{mkv21}
{Manjavacas}, E., {Karalidi}, T., {Vos}, J.~M., {et~al.} 2021,
  \href{http://dx.doi.org/10.3847/1538-3881/ac174c}{\color{magenta}\aj},
  \href{https://ui.adsabs.harvard.edu/abs/2021AJ....162..179M}{\color{blue}162},
  \href{https://ui.adsabs.harvard.edu/abs/2021AJ....162..179M}{\color{blue}179}

\bibitem[{{Marley} {et~al.}(2012){Marley}, {Saumon}, {Cushing}, {Ackerman},
  {Fortney}, \& {Freedman}}]{msc12}
{Marley}, M.~S., {Saumon}, D., {Cushing}, M., {et~al.} 2012,
  \href{http://dx.doi.org/10.1088/0004-637X/754/2/135}{\color{magenta}\apj},
  \href{http://adsabs.harvard.edu/abs/2012ApJ...754..135M}{\color{blue}754},
  \href{http://adsabs.harvard.edu/abs/2012ApJ...754..135M}{\color{blue}135}

\bibitem[{{Marois} {et~al.}(2006){Marois}, {Lafreni{\`e}re}, {Doyon},
  {Macintosh}, \& {Nadeau}}]{mld06}
{Marois}, C., {Lafreni{\`e}re}, D., {Doyon}, R., {et~al.} 2006, \apj,
  \href{http://adsabs.harvard.edu/cgi-bin/nph-bib_query?bibcode=2006ApJ...641..556M&db_key=AST}{\color{blue}641},
  \href{http://adsabs.harvard.edu/cgi-bin/nph-bib_query?bibcode=2006ApJ...641..556M&db_key=AST}{\color{blue}556}

\bibitem[{{Marois} {et~al.}(2008){Marois}, {Macintosh}, {Barman}, {Zuckerman},
  {Song}, {Patience}, {Lafreni{\`e}re}, \& {Doyon}}]{mmb08}
{Marois}, C., {Macintosh}, B., {Barman}, T., {et~al.} 2008, Science,
  \href{http://adsabs.harvard.edu/abs/2008Sci...322.1348M}{\color{blue}322},
  \href{http://adsabs.harvard.edu/abs/2008Sci...322.1348M}{\color{blue}1348}

\bibitem[{{Marois} {et~al.}(2010){Marois}, {Zuckerman}, {Konopacky},
  {Macintosh}, \& {Barman}}]{mzk10}
{Marois}, C., {Zuckerman}, B., {Konopacky}, Q.~M., {et~al.} 2010,
  \href{http://dx.doi.org/10.1038/nature09684}{\color{magenta}\nat},
  \href{http://adsabs.harvard.edu/abs/2010Natur.468.1080M}{\color{blue}468},
  \href{http://adsabs.harvard.edu/abs/2010Natur.468.1080M}{\color{blue}1080}

\bibitem[{{Matthews} {et~al.}(2017){Matthews}, {Hinkley}, {Vigan}, {Kennedy},
  {Rizzuto}, {Stapelfeldt}, {Mawet}, {Booth}, {Chen}, \&
  {Jang-Condell}}]{mhv17}
{Matthews}, E., {Hinkley}, S., {Vigan}, A., {et~al.} 2017,
  \href{http://dx.doi.org/10.3847/2041-8213/aa7943}{\color{magenta}\apjl},
  \href{http://adsabs.harvard.edu/abs/2017ApJ...843L..12M}{\color{blue}843},
  \href{http://adsabs.harvard.edu/abs/2017ApJ...843L..12M}{\color{blue}L12}

\bibitem[{{Matthews} {et~al.}(2018){Matthews}, {Hinkley}, {Vigan}, {Kennedy},
  {Sutlieff}, {Wickenden}, {Treves}, {David}, {Meshkat}, {Mawet}, {Morales},
  {Shannon}, \& {Stapelfeldt}}]{mhv18}
---. 2018,
  \href{http://dx.doi.org/10.1093/mnras/sty1778}{\color{magenta}\mnras},
  \href{http://adsabs.harvard.edu/abs/2018MNRAS.480.2757M}{\color{blue}480},
  \href{http://adsabs.harvard.edu/abs/2018MNRAS.480.2757M}{\color{blue}2757}

\bibitem[{{Mawet} {et~al.}(2014){Mawet}, {Milli}, {Wahhaj}, {Pelat}, {Absil},
  {Delacroix}, {Boccaletti}, {Kasper}, {Kenworthy}, {Marois}, {Mennesson}, \&
  {Pueyo}}]{mmw14}
{Mawet}, D., {Milli}, J., {Wahhaj}, Z., {et~al.} 2014,
  \href{http://dx.doi.org/10.1088/0004-637X/792/2/97}{\color{magenta}\apj},
  \href{https://ui.adsabs.harvard.edu/abs/2014ApJ...792...97M}{\color{blue}792},
  \href{https://ui.adsabs.harvard.edu/abs/2014ApJ...792...97M}{\color{blue}97}

\bibitem[{{Mawet} {et~al.}(2017){Mawet}, {Choquet}, {Absil}, {Huby}, {Bottom},
  {Serabyn}, {Femenia}, {Lebreton}, {Matthews}, {Gomez Gonzalez}, {Wertz},
  {Carlomagno}, {Christiaens}, {Defr{\`e}re}, {Delacroix}, {Forsberg},
  {Habraken}, {Jolivet}, {Karlsson}, {Milli}, {Pinte}, {Piron}, {Reggiani},
  {Surdej}, \& {Vargas Catalan}}]{Mawet2017}
{Mawet}, D., {Choquet}, {\'E}., {Absil}, O., {et~al.} 2017,
  \href{http://dx.doi.org/10.3847/1538-3881/153/1/44}{\color{magenta}\aj},
  \href{https://ui.adsabs.harvard.edu/abs/2017AJ....153...44M}{\color{blue}153},
  \href{https://ui.adsabs.harvard.edu/abs/2017AJ....153...44M}{\color{blue}44}

\bibitem[{{Mawet} {et~al.}(2018){Mawet}, {Bond}, {Delorme}, {Jovanovic},
  {Cetre}, {Chun}, {Echeverri}, {Hall}, {Lilley}, {Wallace}, \&
  {Wizinowich}}]{mbd18}
{Mawet}, D., {Bond}, C.~Z., {Delorme}, J.~R., {et~al.} 2018,
  \href{http://dx.doi.org/10.1117/12.2314037}{\color{magenta}Proc.~SPIE},
  \href{https://ui.adsabs.harvard.edu/abs/2018SPIE10703E..06M}{\color{blue}10703},
  \href{https://ui.adsabs.harvard.edu/abs/2018SPIE10703E..06M}{\color{blue}1070306}

\bibitem[{{Metchev} \& {Hillenbrand}(2006)}]{mh06}
{Metchev}, S.~A., \& {Hillenbrand}, L.~A. 2006,
  \href{http://dx.doi.org/10.1086/507836}{\color{magenta}\apj},
  \href{http://adsabs.harvard.edu/abs/2006ApJ...651.1166M}{\color{blue}651},
  \href{http://adsabs.harvard.edu/abs/2006ApJ...651.1166M}{\color{blue}1166}

\bibitem[{{Metchev} {et~al.}(2015){Metchev}, {Heinze}, {Apai}, {Flateau},
  {Radigan}, {Burgasser}, {Marley}, {Artigau}, {Plavchan}, \&
  {Goldman}}]{mha15}
{Metchev}, S.~A., {Heinze}, A., {Apai}, D., {et~al.} 2015,
  \href{http://dx.doi.org/10.1088/0004-637X/799/2/154}{\color{magenta}\apj},
  \href{https://ui.adsabs.harvard.edu/abs/2015ApJ...799..154M}{\color{blue}799},
  \href{https://ui.adsabs.harvard.edu/abs/2015ApJ...799..154M}{\color{blue}154}

\bibitem[{{Miles} {et~al.}(2018){Miles}, {Skemer}, {Barman}, {Allers}, \&
  {Stone}}]{msb18}
{Miles}, B.~E., {Skemer}, A.~J., {Barman}, T.~S., {et~al.} 2018,
  \href{http://dx.doi.org/10.3847/1538-4357/aae6cd}{\color{magenta}\apj},
  \href{https://ui.adsabs.harvard.edu/abs/2018ApJ...869...18M}{\color{blue}869},
  \href{https://ui.adsabs.harvard.edu/abs/2018ApJ...869...18M}{\color{blue}18}

\bibitem[{{Miles} {et~al.}(2020){Miles}, {Skemer}, {Morley}, {Marley},
  {Fortney}, {Allers}, {Faherty}, {Geballe}, {Visscher}, {Schneider}, {Lupu},
  {Freedman}, \& {Bjoraker}}]{msm20}
{Miles}, B.~E., {Skemer}, A. J.~I., {Morley}, C.~V., {et~al.} 2020,
  \href{http://dx.doi.org/10.3847/1538-3881/ab9114}{\color{magenta}\aj},
  \href{https://ui.adsabs.harvard.edu/abs/2020AJ....160...63M}{\color{blue}160},
  \href{https://ui.adsabs.harvard.edu/abs/2020AJ....160...63M}{\color{blue}63}

\bibitem[{{Millar-Blanchaer} {et~al.}(2015){Millar-Blanchaer}, {Graham},
  {Pueyo}, {Kalas}, {Dawson}, {Wang}, {Perrin}, {moon}, {Macintosh}, {Ammons},
  {Barman}, {Cardwell}, {Chen}, {Chiang}, {Chilcote}, {Cotten}, {De Rosa},
  {Draper}, {Dunn}, {Duch{\^e}ne}, {Esposito}, {Fitzgerald}, {Follette},
  {Goodsell}, {Greenbaum}, {Hartung}, {Hibon}, {Hinkley}, {Ingraham},
  {Jensen-Clem}, {Konopacky}, {Larkin}, {Long}, {Maire}, {Marchis}, {Marley},
  {Marois}, {Morzinski}, {Nielsen}, {Palmer}, {Oppenheimer}, {Poyneer},
  {Rajan}, {Rantakyr{\"o}}, {Ruffio}, {Sadakuni}, {Saddlemyer}, {Schneider},
  {Sivaramakrishnan}, {Soummer}, {Thomas}, {Vasisht}, {Vega}, {Wallace},
  {Ward-Duong}, {Wiktorowicz}, \& {Wolff}}]{mgp15}
{Millar-Blanchaer}, M.~A., {Graham}, J.~R., {Pueyo}, L., {et~al.} 2015,
  \href{http://dx.doi.org/10.1088/0004-637X/811/1/18}{\color{magenta}\apj},
  \href{https://ui.adsabs.harvard.edu/abs/2015ApJ...811...18M}{\color{blue}811},
  \href{https://ui.adsabs.harvard.edu/abs/2015ApJ...811...18M}{\color{blue}18}

\bibitem[{{Moerchen} {et~al.}(2010){Moerchen}, {Telesco}, \& {Packham}}]{mtp10}
{Moerchen}, M.~M., {Telesco}, C.~M., \& {Packham}, C. 2010,
  \href{http://dx.doi.org/10.1088/0004-637X/723/2/1418}{\color{magenta}\apj},
  \href{https://ui.adsabs.harvard.edu/abs/2010ApJ...723.1418M}{\color{blue}723},
  \href{https://ui.adsabs.harvard.edu/abs/2010ApJ...723.1418M}{\color{blue}1418}

\bibitem[{{Molli{\`e}re} {et~al.}(2020){Molli{\`e}re}, {Stolker}, {Lacour},
  {Otten}, {Shangguan}, {Charnay}, {Molyarova}, {Nowak}, {Henning}, {Marleau},
  {Semenov}, {van Dishoeck}, {Eisenhauer}, {Garcia}, {Garcia Lopez}, {Girard},
  {Greenbaum}, {Hinkley}, {Kervella}, {Kreidberg}, {Maire}, {Nasedkin},
  {Pueyo}, {Snellen}, {Vigan}, {Wang}, {de Zeeuw}, \& {Zurlo}}]{msl20}
{Molli{\`e}re}, P., {Stolker}, T., {Lacour}, S., {et~al.} 2020,
  \href{http://dx.doi.org/10.1051/0004-6361/202038325}{\color{magenta}\aap},
  \href{https://ui.adsabs.harvard.edu/abs/2020A&A...640A.131M}{\color{blue}640},
  \href{https://ui.adsabs.harvard.edu/abs/2020A&A...640A.131M}{\color{blue}A131}

\bibitem[{{Molli{\`e}re} {et~al.}(2022){Molli{\`e}re}, {Molyarova}, {Bitsch},
  {Henning}, {Schneider}, {Kreidberg}, {Eistrup}, {Burn}, {Nasedkin},
  {Semenov}, {Mordasini}, {Schlecker}, {Schwarz}, {Lacour}, {Nowak}, \&
  {Schulik}}]{mmb22}
{Molli{\`e}re}, P., {Molyarova}, T., {Bitsch}, B., {et~al.} 2022,
  \href{https://arxiv.org/abs/2204.13714}{\color{magenta}arXiv},
  \href{https://ui.adsabs.harvard.edu/abs/2022arXiv220413714M}{\color{blue}arXiv:2204.13714}

\bibitem[{{Mordasini} {et~al.}(2016){Mordasini}, {van Boekel}, {Molli{\`e}re},
  {Henning}, \& {Benneke}}]{mvm16}
{Mordasini}, C., {van Boekel}, R., {Molli{\`e}re}, P., {et~al.} 2016,
  \href{http://dx.doi.org/10.3847/0004-637X/832/1/41}{\color{magenta}\apj},
  \href{https://ui.adsabs.harvard.edu/abs/2016ApJ...832...41M}{\color{blue}832},
  \href{https://ui.adsabs.harvard.edu/abs/2016ApJ...832...41M}{\color{blue}41}

\bibitem[{{Morley} {et~al.}(2012){Morley}, {Fortney}, {Marley}, {Visscher},
  {Saumon}, \& {Leggett}}]{mfm12}
{Morley}, C.~V., {Fortney}, J.~J., {Marley}, M.~S., {et~al.} 2012,
  \href{http://dx.doi.org/10.1088/0004-637X/756/2/172}{\color{magenta}\apj},
  \href{https://ui.adsabs.harvard.edu/abs/2012ApJ...756..172M}{\color{blue}756},
  \href{https://ui.adsabs.harvard.edu/abs/2012ApJ...756..172M}{\color{blue}172}

\bibitem[{{Morley} {et~al.}(2017){Morley}, {Kreidberg}, {Rustamkulov},
  {Robinson}, \& {Fortney}}]{mkr17}
{Morley}, C.~V., {Kreidberg}, L., {Rustamkulov}, Z., {et~al.} 2017,
  \href{http://dx.doi.org/10.3847/1538-4357/aa927b}{\color{magenta}\apj},
  \href{https://ui.adsabs.harvard.edu/abs/2017ApJ...850..121M}{\color{blue}850},
  \href{https://ui.adsabs.harvard.edu/abs/2017ApJ...850..121M}{\color{blue}121}

\bibitem[{{Morrison} \& {Malhotra}(2015)}]{mm15}
{Morrison}, S., \& {Malhotra}, R. 2015,
  \href{http://dx.doi.org/10.1088/0004-637X/799/1/41}{\color{magenta}\apj},
  \href{https://ui.adsabs.harvard.edu/abs/2015ApJ...799...41M}{\color{blue}799},
  \href{https://ui.adsabs.harvard.edu/abs/2015ApJ...799...41M}{\color{blue}41}

\bibitem[{{Mouillet} {et~al.}(2001){Mouillet}, {Lagrange}, {Augereau}, \&
  {M{\'e}nard}}]{mla01}
{Mouillet}, D., {Lagrange}, A.~M., {Augereau}, J.~C., \& {M{\'e}nard}, F. 2001,
  \href{http://dx.doi.org/10.1051/0004-6361:20010660}{\color{magenta}\aap},
  \href{https://ui.adsabs.harvard.edu/abs/2001A&A...372L..61M}{\color{blue}372},
  \href{https://ui.adsabs.harvard.edu/abs/2001A&A...372L..61M}{\color{blue}L61}

\bibitem[{{Nielsen} {et~al.}(2019){Nielsen}, {De Rosa}, {Macintosh}, {Wang},
  {Ruffio}, {Chiang}, {Marley}, {Saumon}, {Savransky}, {Ammons}, {Bailey},
  {Barman}, {Blain}, {Bulger}, {Burrows}, {Chilcote}, {Cotten}, {Czekala},
  {Doyon}, {Duch{\^e}ne}, {Esposito}, {Fabrycky}, {Fitzgerald}, {Follette},
  {Fortney}, {Gerard}, {Goodsell}, {Graham}, {Greenbaum}, {Hibon}, {Hinkley},
  {Hirsch}, {Hom}, {Hung}, {Dawson}, {Ingraham}, {Kalas}, {Konopacky},
  {Larkin}, {Lee}, {Lin}, {Maire}, {Marchis}, {Marois}, {Metchev},
  {Millar-Blanchaer}, {Morzinski}, {Oppenheimer}, {Palmer}, {Patience},
  {Perrin}, {Poyneer}, {Pueyo}, {Rafikov}, {Rajan}, {Rameau}, {Rantakyr{\"o}},
  {Ren}, {Schneider}, {Sivaramakrishnan}, {Song}, {Soummer}, {Tallis},
  {Thomas}, {Ward-Duong}, \& {Wolff}}]{ndm19}
{Nielsen}, E.~L., {De Rosa}, R.~J., {Macintosh}, B., {et~al.} 2019,
  \href{http://dx.doi.org/10.3847/1538-3881/ab16e9}{\color{magenta}AJ},
  \href{https://ui.adsabs.harvard.edu/abs/2019AJ....158...13N}{\color{blue}158},
  \href{https://ui.adsabs.harvard.edu/abs/2019AJ....158...13N}{\color{blue}13}

\bibitem[{{Nowak} {et~al.}(2020){Nowak}, {Lacour}, {Lagrange}, {Rubini},
  {Wang}, {Stolker}, {Abuter}, {Amorim}, {Asensio-Torres}, {Baub{\"o}ck},
  {Benisty}, {Berger}, {Beust}, {Blunt}, {Boccaletti}, {Bonnefoy}, {Bonnet},
  {Brandner}, {Cantalloube}, {Charnay}, {Choquet}, {Christiaens}, {Cl{\'e}net},
  {Coud{\'e} Du Foresto}, {Cridland}, {de Zeeuw}, {Dembet}, {Dexter},
  {Drescher}, {Duvert}, {Eckart}, {Eisenhauer}, {Gao}, {Garcia}, {Garcia
  Lopez}, {Gardner}, {Gendron}, {Genzel}, {Gillessen}, {Girard}, {Grandjean},
  {Haubois}, {Hei{\ss}el}, {Henning}, {Hinkley}, {Hippler}, {Horrobin},
  {Houll{\'e}}, {Hubert}, {Jim{\'e}nez-Rosales}, {Jocou}, {Kammerer},
  {Kervella}, {Keppler}, {Kreidberg}, {Kulikauskas}, {Lapeyr{\`e}re}, {Le
  Bouquin}, {L{\'e}na}, {M{\'e}rand}, {Maire}, {Molli{\`e}re}, {Monnier},
  {Mouillet}, {M{\"u}ller}, {Nasedkin}, {Ott}, {Otten}, {Paumard}, {Paladini},
  {Perraut}, {Perrin}, {Pueyo}, {Pfuhl}, {Rameau}, {Rodet},
  {Rodr{\'\i}guez-Coira}, {Rousset}, {Scheithauer}, {Shangguan}, {Stadler},
  {Straub}, {Straubmeier}, {Sturm}, {Tacconi}, {van Dishoeck}, {Vigan},
  {Vincent}, {von Fellenberg}, {Ward-Duong}, {Widmann}, {Wieprecht},
  {Wiezorrek}, {Woillez}, \& {Gravity Collaboration}}]{nll20}
{Nowak}, M., {Lacour}, S., {Lagrange}, A.~M., {et~al.} 2020,
  \href{http://dx.doi.org/10.1051/0004-6361/202039039}{\color{magenta}\aap},
  \href{https://ui.adsabs.harvard.edu/abs/2020A&A...642L...2N}{\color{blue}642},
  \href{https://ui.adsabs.harvard.edu/abs/2020A&A...642L...2N}{\color{blue}L2}

\bibitem[{{{\"O}berg} {et~al.}(2011){{\"O}berg}, {Murray-Clay}, \&
  {Bergin}}]{omb11}
{{\"O}berg}, K.~I., {Murray-Clay}, R., \& {Bergin}, E.~A. 2011,
  \href{http://dx.doi.org/10.1088/2041-8205/743/1/L16}{\color{magenta}\apjl},
  \href{http://adsabs.harvard.edu/abs/2011ApJ...743L..16O}{\color{blue}743},
  \href{http://adsabs.harvard.edu/abs/2011ApJ...743L..16O}{\color{blue}L16}

\bibitem[{{Otten} {et~al.}(2021){Otten}, {Vigan}, {Muslimov}, {N'Diaye},
  {Choquet}, {Seemann}, {Dohlen}, {Houll{\'e}}, {Cristofari}, {Phillips},
  {Charles}, {Baraffe}, {Beuzit}, {Costille}, {Dorn}, {El Morsy}, {Kasper},
  {Lopez}, {Mordasini}, {Pourcelot}, {Reiners}, \& {Sauvage}}]{ovm21}
{Otten}, G.~P.~P.~L., {Vigan}, A., {Muslimov}, E., {et~al.} 2021,
  \href{http://dx.doi.org/10.1051/0004-6361/202038517}{\color{magenta}\aap},
  \href{https://ui.adsabs.harvard.edu/abs/2021A&A...646A.150O}{\color{blue}646},
  \href{https://ui.adsabs.harvard.edu/abs/2021A&A...646A.150O}{\color{blue}A150}

\bibitem[{{Perrin} {et~al.}(2018){Perrin}, {Pueyo}, {Van Gorkom}, {Brooks},
  {Rajan}, {Girard}, \& {Lajoie}}]{ppv18}
{Perrin}, M.~D., {Pueyo}, L., {Van Gorkom}, K., {et~al.} 2018,
  \href{http://dx.doi.org/10.1117/12.2313552}{\color{magenta}Proc.~SPIE},
  \href{https://ui.adsabs.harvard.edu/abs/2018SPIE10698E..09P}{\color{blue}10698},
  \href{https://ui.adsabs.harvard.edu/abs/2018SPIE10698E..09P}{\color{blue}1069809}

\bibitem[{{Perrot} {et~al.}(2016){Perrot}, {Boccaletti}, {Pantin}, {Augereau},
  {Lagrange}, {Galicher}, {Maire}, {Mazoyer}, {Milli}, {Rousset}, {Gratton},
  {Bonnefoy}, {Brandner}, {Buenzli}, {Langlois}, {Lannier}, {Mesa}, {Peretti},
  {Salter}, {Sissa}, {Chauvin}, {Desidera}, {Feldt}, {Vigan}, {Di Folco},
  {Dutrey}, {P{\'e}ricaud}, {Baudoz}, {Benisty}, {De Boer}, {Garufi}, {Girard},
  {Menard}, {Olofsson}, {Quanz}, {Mouillet}, {Christiaens}, {Casassus},
  {Beuzit}, {Blanchard}, {Carle}, {Fusco}, {Giro}, {Hubin}, {Maurel},
  {Moeller-Nilsson}, {Sevin}, \& {Weber}}]{pbp16}
{Perrot}, C., {Boccaletti}, A., {Pantin}, E., {et~al.} 2016,
  \href{http://dx.doi.org/10.1051/0004-6361/201628396}{\color{magenta}\aap},
  \href{https://ui.adsabs.harvard.edu/abs/2016A&A...590L...7P}{\color{blue}590},
  \href{https://ui.adsabs.harvard.edu/abs/2016A&A...590L...7P}{\color{blue}L7}

\bibitem[{{Petrus} {et~al.}(2021){Petrus}, {Bonnefoy}, {Chauvin}, {Charnay},
  {Marleau}, {Gratton}, {Lagrange}, {Rameau}, {Mordasini}, {Nowak}, {Delorme},
  {Boccaletti}, {Carlotti}, {Houll{\'e}}, {Vigan}, {Allard}, {Desidera},
  {D'Orazi}, {Hoeijmakers}, {Wyttenbach}, \& {Lavie}}]{pbc21}
{Petrus}, S., {Bonnefoy}, M., {Chauvin}, G., {et~al.} 2021,
  \href{http://dx.doi.org/10.1051/0004-6361/202038914}{\color{magenta}\aap},
  \href{https://ui.adsabs.harvard.edu/abs/2021A&A...648A..59P}{\color{blue}648},
  \href{https://ui.adsabs.harvard.edu/abs/2021A&A...648A..59P}{\color{blue}A59}

\bibitem[{{Phillips} {et~al.}(2020){Phillips}, {Tremblin}, {Baraffe},
  {Chabrier}, {Allard}, {Spiegelman}, {Goyal}, {Drummond}, \&
  {H{\'e}brard}}]{ptb20}
{Phillips}, M.~W., {Tremblin}, P., {Baraffe}, I., {et~al.} 2020,
  \href{http://dx.doi.org/10.1051/0004-6361/201937381}{\color{magenta}\aap},
  \href{https://ui.adsabs.harvard.edu/abs/2020A&A...637A..38P}{\color{blue}637},
  \href{https://ui.adsabs.harvard.edu/abs/2020A&A...637A..38P}{\color{blue}A38}

\bibitem[{{Pinte} {et~al.}(2006){Pinte}, {M{\'e}nard}, {Duch{\^e}ne}, \&
  {Bastien}}]{Pinte2006}
{Pinte}, C., {M{\'e}nard}, F., {Duch{\^e}ne}, G., \& {Bastien}, P. 2006,
  \href{http://dx.doi.org/10.1051/0004-6361:20053275}{\color{magenta}\aap},
  \href{https://ui.adsabs.harvard.edu/abs/2006A&A...459..797P}{\color{blue}459},
  \href{https://ui.adsabs.harvard.edu/abs/2006A&A...459..797P}{\color{blue}797}

\bibitem[{{Poleski} {et~al.}(2021){Poleski}, {Skowron}, {Mr{\'o}z}, {Udalski},
  {Szyma{\'n}ski}, {Pietrukowicz}, {Ulaczyk}, {Rybicki}, {Iwanek}, {Wrona}, \&
  {Gromadzki}}]{psm21}
{Poleski}, R., {Skowron}, J., {Mr{\'o}z}, P., {et~al.} 2021,
  \href{http://dx.doi.org/10.32023/0001-5237/71.1.1}{\color{magenta}\actaa},
  \href{https://ui.adsabs.harvard.edu/abs/2021AcA....71....1P}{\color{blue}71},
  \href{https://ui.adsabs.harvard.edu/abs/2021AcA....71....1P}{\color{blue}1}

\bibitem[{{Pollack} {et~al.}(1996){Pollack}, {Hubickyj}, {Bodenheimer},
  {Lissauer}, {Podolak}, \& {Greenzweig}}]{phb96}
{Pollack}, J.~B., {Hubickyj}, O., {Bodenheimer}, P., {et~al.} 1996, Icarus,
  \href{http://adsabs.harvard.edu/abs/1996Icar..124...62P}{\color{blue}124},
  \href{http://adsabs.harvard.edu/abs/1996Icar..124...62P}{\color{blue}62}

\bibitem[{{Pontoppidan} {et~al.}(2016){Pontoppidan}, {Pickering}, {Laidler},
  {Gilbert}, {Sontag}, {Slocum}, {Sienkiewicz}, {Hanley}, {Earl}, {Pueyo},
  {Ravindranath}, {Karakla}, {Robberto}, {Noriega-Crespo}, \& {Barker}}]{ppl16}
{Pontoppidan}, K.~M., {Pickering}, T.~E., {Laidler}, V.~G., {et~al.} 2016,
  \href{http://dx.doi.org/10.1117/12.2231768}{\color{magenta}Proc.~SPIE},
  \href{https://ui.adsabs.harvard.edu/abs/2016SPIE.9910E..16P}{\color{blue}9910},
  \href{https://ui.adsabs.harvard.edu/abs/2016SPIE.9910E..16P}{\color{blue}991016}

\bibitem[{{Preibisch} \& {Mamajek}(2008)}]{pm08}
{Preibisch}, T., \& {Mamajek}, E. 2008, {The Nearest OB Association:
  Scorpius-Centaurus (Sco OB2)}, ed. B.~{Reipurth},
  \href{http://adsabs.harvard.edu/abs/2008hsf2.book..235P}{\color{blue}235}

\bibitem[{{Pueyo}(2016)}]{p16}
{Pueyo}, L. 2016,
  \href{http://dx.doi.org/10.3847/0004-637X/824/2/117}{\color{magenta}\apj},
  \href{https://ui.adsabs.harvard.edu/abs/2016ApJ...824..117P}{\color{blue}824},
  \href{https://ui.adsabs.harvard.edu/abs/2016ApJ...824..117P}{\color{blue}117}

\bibitem[{{Quanz} {et~al.}(2021){Quanz}, {Absil}, {Benz}, {Bonfils}, {Berger},
  {Defr{\`e}re}, {van Dishoeck}, {Ehrenreich}, {Fortney}, {Glauser},
  {Grenfell}, {Janson}, {Kraus}, {Krause}, {Labadie}, {Lacour}, {Line}, {Linz},
  {Loicq}, {Miguel}, {Pall{\'e}}, {Queloz}, {Rauer}, {Ribas}, {Rugheimer},
  {Selsis}, {Snellen}, {Sozzetti}, {Stapelfeldt}, {Udry}, \& {Wyatt}}]{qab21}
{Quanz}, S.~P., {Absil}, O., {Benz}, W., {et~al.} 2021,
  \href{http://dx.doi.org/10.1007/s10686-021-09791-z}{\color{magenta}Experimental
  Astronomy}

\bibitem[{{Radigan} {et~al.}(2014){Radigan}, {Lafreni{\`e}re}, {Jayawardhana},
  \& {Artigau}}]{rlj14}
{Radigan}, J., {Lafreni{\`e}re}, D., {Jayawardhana}, R., \& {Artigau}, E. 2014,
  \href{http://dx.doi.org/10.1088/0004-637X/793/2/75}{\color{magenta}\apj},
  \href{https://ui.adsabs.harvard.edu/abs/2014ApJ...793...75R}{\color{blue}793},
  \href{https://ui.adsabs.harvard.edu/abs/2014ApJ...793...75R}{\color{blue}75}

\bibitem[{{Rameau} {et~al.}(2013){Rameau}, {Chauvin}, {Lagrange}, {Boccaletti},
  {Quanz}, {Bonnefoy}, {Girard}, {Delorme}, {Desidera}, {Klahr}, {Mordasini},
  {Dumas}, \& {Bonavita}}]{rameau95086}
{Rameau}, J., {Chauvin}, G., {Lagrange}, A.-M., {et~al.} 2013,
  \href{http://dx.doi.org/10.1088/2041-8205/772/2/L15}{\color{magenta}\apjl},
  \href{http://adsabs.harvard.edu/abs/2013ApJ...772L..15R}{\color{blue}772},
  \href{http://adsabs.harvard.edu/abs/2013ApJ...772L..15R}{\color{blue}L15}

\bibitem[{{Ren} {et~al.}(2020){Ren}, {Pueyo}, {Chen}, {Choquet}, {Debes},
  {Duch{\^e}ne}, {M{\'e}nard}, \& {Perrin}}]{Ren2020}
{Ren}, B., {Pueyo}, L., {Chen}, C., {et~al.} 2020,
  \href{http://dx.doi.org/10.3847/1538-4357/ab7024}{\color{magenta}\apj},
  \href{https://ui.adsabs.harvard.edu/abs/2020ApJ...892...74R}{\color{blue}892},
  \href{https://ui.adsabs.harvard.edu/abs/2020ApJ...892...74R}{\color{blue}74}

\bibitem[{{Ren} {et~al.}(2018){Ren}, {Pueyo}, {Zhu}, {Debes}, \&
  {Duch{\^e}ne}}]{Ren2018}
{Ren}, B., {Pueyo}, L., {Zhu}, G.~B., {et~al.} 2018,
  \href{http://dx.doi.org/10.3847/1538-4357/aaa1f2}{\color{magenta}\apj},
  \href{https://ui.adsabs.harvard.edu/abs/2018ApJ...852..104R}{\color{blue}852},
  \href{https://ui.adsabs.harvard.edu/abs/2018ApJ...852..104R}{\color{blue}104}

\bibitem[{{Rich} {et~al.}(2016){Rich}, {Currie}, {Wisniewski}, {Hashimoto},
  {Brandt}, {Carson}, {Kuzuhara}, \& {Uyama}}]{rcw16}
{Rich}, E.~A., {Currie}, T., {Wisniewski}, J.~P., {et~al.} 2016,
  \href{http://dx.doi.org/10.3847/0004-637X/830/2/114}{\color{magenta}\apj},
  \href{http://adsabs.harvard.edu/abs/2016ApJ...830..114R}{\color{blue}830},
  \href{http://adsabs.harvard.edu/abs/2016ApJ...830..114R}{\color{blue}114}

\bibitem[{{Rieke} {et~al.}(2015){Rieke}, {Wright}, {B{\"o}ker}, {Bouwman},
  {Colina}, {Glasse}, {Gordon}, {Greene}, {G{\"u}del}, {Henning}, {Justtanont},
  {Lagage}, {Meixner}, {N{\o}rgaard-Nielsen}, {Ray}, {Ressler}, {van Dishoeck},
  \& {Waelkens}}]{rwb15}
{Rieke}, G.~H., {Wright}, G.~S., {B{\"o}ker}, T., {et~al.} 2015,
  \href{http://dx.doi.org/10.1086/682252}{\color{magenta}\pasp},
  \href{https://ui.adsabs.harvard.edu/abs/2015PASP..127..584R}{\color{blue}127},
  \href{https://ui.adsabs.harvard.edu/abs/2015PASP..127..584R}{\color{blue}584}

\bibitem[{{Rieke} {et~al.}(2005){Rieke}, {Kelly}, \& {Horner}}]{rkh05}
{Rieke}, M.~J., {Kelly}, D., \& {Horner}, S. 2005,
  \href{http://dx.doi.org/10.1117/12.615554}{\color{magenta}Proc.~SPIE},
  \href{https://ui.adsabs.harvard.edu/abs/2005SPIE.5904....1R}{\color{blue}5904},
  \href{https://ui.adsabs.harvard.edu/abs/2005SPIE.5904....1R}{\color{blue}1}

\bibitem[{{Rizzuto} {et~al.}(2011){Rizzuto}, {Ireland}, \& {Robertson}}]{rir11}
{Rizzuto}, A.~C., {Ireland}, M.~J., \& {Robertson}, J.~G. 2011,
  \href{http://dx.doi.org/10.1111/j.1365-2966.2011.19256.x}{\color{magenta}\mnras},
  \href{http://adsabs.harvard.edu/abs/2011MNRAS.416.3108R}{\color{blue}416},
  \href{http://adsabs.harvard.edu/abs/2011MNRAS.416.3108R}{\color{blue}3108}

\bibitem[{{Rouan} {et~al.}(2000){Rouan}, {Riaud}, {Boccaletti}, {Cl{\'e}net},
  \& {Labeyrie}}]{rrb00}
{Rouan}, D., {Riaud}, P., {Boccaletti}, A., {et~al.} 2000,
  \href{http://dx.doi.org/10.1086/317707}{\color{magenta}\pasp},
  \href{https://ui.adsabs.harvard.edu/abs/2000PASP..112.1479R}{\color{blue}112},
  \href{https://ui.adsabs.harvard.edu/abs/2000PASP..112.1479R}{\color{blue}1479}

\bibitem[{{Ruane} {et~al.}(2019){Ruane}, {Ngo}, {Mawet}, {Absil}, {Choquet},
  {Cook}, {Gomez Gonzalez}, {Huby}, {Matthews}, {Meshkat}, {Reggiani},
  {Serabyn}, {Wallack}, \& {Xuan}}]{rnm19}
{Ruane}, G., {Ngo}, H., {Mawet}, D., {et~al.} 2019,
  \href{http://dx.doi.org/10.3847/1538-3881/aafee2}{\color{magenta}\aj},
  \href{https://ui.adsabs.harvard.edu/abs/2019AJ....157..118R}{\color{blue}157},
  \href{https://ui.adsabs.harvard.edu/abs/2019AJ....157..118R}{\color{blue}118}

\bibitem[{{Sallum} \& {Skemer}(2019)}]{ss19}
{Sallum}, S., \& {Skemer}, A. 2019,
  \href{http://dx.doi.org/10.1117/1.JATIS.5.1.018001}{\color{magenta}JATIS},
  \href{https://ui.adsabs.harvard.edu/abs/2019JATIS...5a8001S}{\color{blue}5},
  \href{https://ui.adsabs.harvard.edu/abs/2019JATIS...5a8001S}{\color{blue}018001}

\bibitem[{{Sallum} {et~al.}(2015){Sallum}, {Follette}, {Eisner}, {Close},
  {Hinz}, {Kratter}, {Males}, {Skemer}, {Macintosh}, {Tuthill}, {Bailey},
  {Defr{\`e}re}, {Morzinski}, {Rodigas}, {Spalding}, {Vaz}, \&
  {Weinberger}}]{sfe15}
{Sallum}, S., {Follette}, K.~B., {Eisner}, J.~A., {et~al.} 2015,
  \href{http://dx.doi.org/10.1038/nature15761}{\color{magenta}\nat},
  \href{https://ui.adsabs.harvard.edu/abs/2015Natur.527..342S}{\color{blue}527},
  \href{https://ui.adsabs.harvard.edu/abs/2015Natur.527..342S}{\color{blue}342}

\bibitem[{{Sanghi} {et~al.}(2022){Sanghi}, {Zhou}, \& {Bowler}}]{szb22}
{Sanghi}, A., {Zhou}, Y., \& {Bowler}, B.~P. 2022,
  \href{http://dx.doi.org/10.3847/1538-3881/ac477e}{\color{magenta}\aj},
  \href{https://ui.adsabs.harvard.edu/abs/2022AJ....163..119S}{\color{blue}163},
  \href{https://ui.adsabs.harvard.edu/abs/2022AJ....163..119S}{\color{blue}119}

\bibitem[{{Schlieder} {et~al.}(2016){Schlieder}, {Beichman}, {Meyer}, \&
  {Greene}}]{sbm16}
{Schlieder}, J.~E., {Beichman}, C.~A., {Meyer}, M.~R., \& {Greene}, T. 2016,
  \href{http://dx.doi.org/10.1017/S1743921315006407}{\color{magenta}IAU
  Symposium},
  \href{https://ui.adsabs.harvard.edu/abs/2016IAUS..314..288S}{\color{blue}314},
  \href{https://ui.adsabs.harvard.edu/abs/2016IAUS..314..288S}{\color{blue}288}

\bibitem[{{Schneider} {et~al.}(2017){Schneider}, {Gaspar}, {Debes}, {Gull},
  {Hines}, {Apai}, \& {Rieke}}]{sgd17}
{Schneider}, G., {Gaspar}, A., {Debes}, J., {et~al.} 2017, {Enabling
  Narrow(est) IWA Coronagraphy with STIS BAR5 and BAR10 Occulters}, Tech. rep.,
  STScI,
  \href{https://ui.adsabs.harvard.edu/abs/2017stis.rept....3S}{\color{blue}3}

\bibitem[{{Schneider} {et~al.}(2014){Schneider}, {Grady}, {Hines}, {Stark},
  {Debes}, {Carson}, {Kuchner}, {Perrin}, {Weinberger}, {Wisniewski},
  {Silverstone}, {Jang-Condell}, {Henning}, {Woodgate}, {Serabyn},
  {Moro-Martin}, {Tamura}, {Hinz}, \& {Rodigas}}]{sgh14}
{Schneider}, G., {Grady}, C.~A., {Hines}, D.~C., {et~al.} 2014,
  \href{http://dx.doi.org/10.1088/0004-6256/148/4/59}{\color{magenta}\aj},
  \href{https://ui.adsabs.harvard.edu/abs/2014AJ....148...59S}{\color{blue}148},
  \href{https://ui.adsabs.harvard.edu/abs/2014AJ....148...59S}{\color{blue}59}

\bibitem[{{Sing} {et~al.}(2016){Sing}, {Fortney}, {Nikolov}, {Wakeford},
  {Kataria}, {Evans}, {Aigrain}, {Ballester}, {Burrows}, {Deming},
  {D{\'e}sert}, {Gibson}, {Henry}, {Huitson}, {Knutson}, {Lecavelier Des
  Etangs}, {Pont}, {Showman}, {Vidal-Madjar}, {Williamson}, \&
  {Wilson}}]{sfn16}
{Sing}, D.~K., {Fortney}, J.~J., {Nikolov}, N., {et~al.} 2016,
  \href{http://dx.doi.org/10.1038/nature16068}{\color{magenta}\nat},
  \href{https://ui.adsabs.harvard.edu/abs/2016Natur.529...59S}{\color{blue}529},
  \href{https://ui.adsabs.harvard.edu/abs/2016Natur.529...59S}{\color{blue}59}

\bibitem[{{Singh} {et~al.}(2021){Singh}, {Bhowmik}, {Boccaletti},
  {Th{\'e}bault}, {Kral}, {Milli}, {Mazoyer}, {Pantin}, {van Holstein},
  {Olofsson}, {Boukrouche}, {Di Folco}, {Janson}, {Langlois}, {Maire}, {Vigan},
  {Benisty}, {Augereau}, {Perrot}, {Gratton}, {Henning}, {M{\'e}nard},
  {Rickman}, {Wahhaj}, {Zurlo}, {Biller}, {Bonnefoy}, {Chauvin}, {Delorme},
  {Desidera}, {D'Orazi}, {Feldt}, {Hagelberg}, {Keppler}, {Kopytova},
  {Lagadec}, {Lagrange}, {Mesa}, {Meyer}, {Rouan}, {Sissa}, {Schmidt},
  {Jaquet}, {Fusco}, {Pavlov}, \& {Rabou}}]{sbb21}
{Singh}, G., {Bhowmik}, T., {Boccaletti}, A., {et~al.} 2021,
  \href{http://dx.doi.org/10.1051/0004-6361/202140319}{\color{magenta}\aap},
  \href{https://ui.adsabs.harvard.edu/abs/2021A&A...653A..79S}{\color{blue}653},
  \href{https://ui.adsabs.harvard.edu/abs/2021A&A...653A..79S}{\color{blue}A79}

\bibitem[{{Sivaramakrishnan} {et~al.}(2010){Sivaramakrishnan},
  {Lafreni{\`e}re}, {Tuthill}, {Ireland}, {Lloyd}, {Martinache}, {Makidon},
  {Soummer}, {Doyon}, {Beaulieu}, {Parmentier}, \& {Beichman}}]{slt10}
{Sivaramakrishnan}, A., {Lafreni{\`e}re}, D., {Tuthill}, P.~G., {et~al.} 2010,
  \href{http://dx.doi.org/10.1117/12.858161}{\color{magenta}Proc.~SPIE},
  \href{https://ui.adsabs.harvard.edu/abs/2010SPIE.7731E..3WS}{\color{blue}7731},
  \href{https://ui.adsabs.harvard.edu/abs/2010SPIE.7731E..3WS}{\color{blue}77313W}

\bibitem[{{Sivaramakrishnan} {et~al.}(2012){Sivaramakrishnan},
  {Lafreni{\`e}re}, {Ford}, {McKernan}, {Cheetham}, {Greenbaum}, {Tuthill},
  {Lloyd}, {Ireland}, {Doyon}, {Beaulieu}, {Martel}, {Koekemoer}, {Martinache},
  \& {Teuben}}]{slf12}
{Sivaramakrishnan}, A., {Lafreni{\`e}re}, D., {Ford}, K.~E.~S., {et~al.} 2012,
  \href{http://dx.doi.org/10.1117/12.925565}{\color{magenta}Proc.~SPIE},
  \href{https://ui.adsabs.harvard.edu/abs/2012SPIE.8442E..2SS}{\color{blue}8442},
  \href{https://ui.adsabs.harvard.edu/abs/2012SPIE.8442E..2SS}{\color{blue}84422S}

\bibitem[{{Skemer} {et~al.}(2012){Skemer}, {Hinz}, {Esposito}, {Burrows},
  {Leisenring}, {Skrutskie}, {Desidera}, {Mesa}, {Arcidiacono}, {Mannucci},
  {Rodigas}, {Close}, {McCarthy}, {Kulesa}, {Agapito}, {Apai}, {Argomedo},
  {Bailey}, {Boutsia}, {Briguglio}, {Brusa}, {Busoni}, {Claudi}, {Eisner},
  {Fini}, {Follette}, {Garnavich}, {Gratton}, {Guerra}, {Hill}, {Hoffmann},
  {Jones}, {Krejny}, {Males}, {Masciadri}, {Meyer}, {Miller}, {Morzinski},
  {Nelson}, {Pinna}, {Puglisi}, {Quanz}, {Quiros-Pacheco}, {Riccardi},
  {Stefanini}, {Vaitheeswaran}, {Wilson}, \& {Xompero}}]{she12}
{Skemer}, A.~J., {Hinz}, P.~M., {Esposito}, S., {et~al.} 2012,
  \href{http://dx.doi.org/10.1088/0004-637X/753/1/14}{\color{magenta}\apj},
  \href{http://adsabs.harvard.edu/abs/2012ApJ...753...14S}{\color{blue}753},
  \href{http://adsabs.harvard.edu/abs/2012ApJ...753...14S}{\color{blue}14}

\bibitem[{{Skemer} {et~al.}(2014){Skemer}, {Marley}, {Hinz}, {Morzinski},
  {Skrutskie}, {Leisenring}, {Close}, {Saumon}, {Bailey}, {Briguglio},
  {Defrere}, {Esposito}, {Follette}, {Hill}, {Males}, {Puglisi}, {Rodigas}, \&
  {Xompero}}]{smh14}
{Skemer}, A.~J., {Marley}, M.~S., {Hinz}, P.~M., {et~al.} 2014,
  \href{http://dx.doi.org/10.1088/0004-637X/792/1/17}{\color{magenta}\apj},
  \href{http://adsabs.harvard.edu/abs/2014ApJ...792...17S}{\color{blue}792},
  \href{http://adsabs.harvard.edu/abs/2014ApJ...792...17S}{\color{blue}17}

\bibitem[{{Skemer} {et~al.}(2016){Skemer}, {Morley}, {Zimmerman}, {Skrutskie},
  {Leisenring}, {Buenzli}, {Bonnefoy}, {Bailey}, {Hinz}, {Defr{\'e}re},
  {Esposito}, {Apai}, {Biller}, {Brandner}, {Close}, {Crepp}, {De Rosa},
  {Desidera}, {Eisner}, {Fortney}, {Freedman}, {Henning}, {Hofmann},
  {Kopytova}, {Lupu}, {Maire}, {Males}, {Marley}, {Morzinski}, {Oza},
  {Patience}, {Rajan}, {Rieke}, {Schertl}, {Schlieder}, {Stone}, {Su}, {Vaz},
  {Visscher}, {Ward-Duong}, {Weigelt}, \& {Woodward}}]{smz16}
{Skemer}, A.~J., {Morley}, C.~V., {Zimmerman}, N.~T., {et~al.} 2016,
  \href{http://dx.doi.org/10.3847/0004-637X/817/2/166}{\color{magenta}\apj},
  \href{https://ui.adsabs.harvard.edu/abs/2016ApJ...817..166S}{\color{blue}817},
  \href{https://ui.adsabs.harvard.edu/abs/2016ApJ...817..166S}{\color{blue}166}

\bibitem[{{Sloan} {et~al.}(2005){Sloan}, {Keller}, {Forrest}, {Leibensperger},
  {Sargent}, {Li}, {Najita}, {Watson}, {Brandl}, {Chen}, {Green},
  {Markwick-Kemper}, {Herter}, {D'Alessio}, {Morris}, {Barry}, {Hall}, {Myers},
  \& {Houck}}]{skf05}
{Sloan}, G.~C., {Keller}, L.~D., {Forrest}, W.~J., {et~al.} 2005,
  \href{http://dx.doi.org/10.1086/444371}{\color{magenta}\apj},
  \href{https://ui.adsabs.harvard.edu/abs/2005ApJ...632..956S}{\color{blue}632},
  \href{https://ui.adsabs.harvard.edu/abs/2005ApJ...632..956S}{\color{blue}956}

\bibitem[{{Snellen} {et~al.}(2015){Snellen}, {de Kok}, {Birkby}, {Brandl},
  {Brogi}, {Keller}, {Kenworthy}, {Schwarz}, \& {Stuik}}]{sdb15}
{Snellen}, I., {de Kok}, R., {Birkby}, J.~L., {et~al.} 2015,
  \href{http://dx.doi.org/10.1051/0004-6361/201425018}{\color{magenta}\aap},
  \href{https://ui.adsabs.harvard.edu/abs/2015A&A...576A..59S}{\color{blue}576},
  \href{https://ui.adsabs.harvard.edu/abs/2015A&A...576A..59S}{\color{blue}A59}

\bibitem[{{Sorahana} \& {Yamamura}(2012)}]{sy12}
{Sorahana}, S., \& {Yamamura}, I. 2012,
  \href{http://dx.doi.org/10.1088/0004-637X/760/2/151}{\color{magenta}\apj},
  \href{https://ui.adsabs.harvard.edu/abs/2012ApJ...760..151S}{\color{blue}760},
  \href{https://ui.adsabs.harvard.edu/abs/2012ApJ...760..151S}{\color{blue}151}

\bibitem[{{Soulain} {et~al.}(2020){Soulain}, {Sivaramakrishnan}, {Tuthill},
  {Thatte}, {Volk}, {Cooper}, {Albert}, {Artigau}, {Cook}, {Doyon},
  {Johnstone}, {Lafreni{\`e}re}, \& {Martel}}]{sst20}
{Soulain}, A., {Sivaramakrishnan}, A., {Tuthill}, P., {et~al.} 2020,
  \href{http://dx.doi.org/10.1117/12.2560804}{\color{magenta}Proc.~SPIE},
  \href{https://ui.adsabs.harvard.edu/abs/2020SPIE11446E..11S}{\color{blue}11446},
  \href{https://ui.adsabs.harvard.edu/abs/2020SPIE11446E..11S}{\color{blue}1144611}

\bibitem[{{Soummer} {et~al.}(2011){Soummer}, {Hagan}, {Pueyo}, {Thormann},
  {Rajan}, \& {Marois}}]{shp11}
{Soummer}, R., {Hagan}, J.~B., {Pueyo}, L., {et~al.} 2011,
  \href{http://dx.doi.org/10.1088/0004-637X/741/1/55}{\color{magenta}\apj},
  \href{https://ui.adsabs.harvard.edu/abs/2011ApJ...741...55S}{\color{blue}741},
  \href{https://ui.adsabs.harvard.edu/abs/2011ApJ...741...55S}{\color{blue}55}

\bibitem[{{Soummer} {et~al.}(2012){Soummer}, {Pueyo}, \& {Larkin}}]{spl12}
{Soummer}, R., {Pueyo}, L., \& {Larkin}, J. 2012,
  \href{http://dx.doi.org/10.1088/2041-8205/755/2/L28}{\color{magenta}\apjl},
  \href{http://adsabs.harvard.edu/abs/2012ApJ...755L..28S}{\color{blue}755},
  \href{http://adsabs.harvard.edu/abs/2012ApJ...755L..28S}{\color{blue}L28}

\bibitem[{{Soummer} {et~al.}(2014){Soummer}, {Perrin}, {Pueyo}, {Choquet},
  {Chen}, {Golimowski}, {Hagan}, {Mittal}, {Moerchen}, {N'Diaye}, {Rajan},
  {Wolff}, {Debes}, {Hines}, \& {Schneider}}]{spp14}
{Soummer}, R., {Perrin}, M.~D., {Pueyo}, L., {et~al.} 2014,
  \href{http://dx.doi.org/10.1088/2041-8205/786/2/L23}{\color{magenta}\apjl},
  \href{https://ui.adsabs.harvard.edu/abs/2014ApJ...786L..23S}{\color{blue}786},
  \href{https://ui.adsabs.harvard.edu/abs/2014ApJ...786L..23S}{\color{blue}L23}

\bibitem[{{Stolker} {et~al.}(2020){Stolker}, {Quanz}, {Todorov}, {K{\"u}hn},
  {Molli{\`e}re}, {Meyer}, {Currie}, {Daemgen}, \& {Lavie}}]{sqt20}
{Stolker}, T., {Quanz}, S.~P., {Todorov}, K.~O., {et~al.} 2020,
  \href{http://dx.doi.org/10.1051/0004-6361/201937159}{\color{magenta}Astronomy
  \& Astrophysics},
  \href{https://ui.adsabs.harvard.edu/abs/2020A&A...635A.182S}{\color{blue}635},
  \href{https://ui.adsabs.harvard.edu/abs/2020A&A...635A.182S}{\color{blue}A182}

\bibitem[{{Stone} {et~al.}(2016){Stone}, {Skemer}, {Kratter}, {Dupuy}, {Close},
  {Eisner}, {Fortney}, {Hinz}, {Males}, {Morley}, {Morzinski}, \&
  {Ward-Duong}}]{ssk16}
{Stone}, J.~M., {Skemer}, A.~J., {Kratter}, K.~M., {et~al.} 2016,
  \href{http://dx.doi.org/10.3847/2041-8205/818/1/L12}{\color{magenta}\apjl},
  \href{http://adsabs.harvard.edu/abs/2016ApJ...818L..12S}{\color{blue}818},
  \href{http://adsabs.harvard.edu/abs/2016ApJ...818L..12S}{\color{blue}L12}

\bibitem[{{Takeuchi} \& {Artymowicz}(2001)}]{ta01}
{Takeuchi}, T., \& {Artymowicz}, P. 2001,
  \href{http://dx.doi.org/10.1086/322252}{\color{magenta}\apj},
  \href{https://ui.adsabs.harvard.edu/abs/2001ApJ...557..990T}{\color{blue}557},
  \href{https://ui.adsabs.harvard.edu/abs/2001ApJ...557..990T}{\color{blue}990}

\bibitem[{{The LUVOIR Team}(2019)}]{t19}
{The LUVOIR Team}. 2019,
  \href{https://arxiv.org/abs/1912.06219}{\color{magenta}arXiv},
  \href{https://ui.adsabs.harvard.edu/abs/2019arXiv191206219T}{\color{blue}arXiv:1912.06219}

\bibitem[{{Th{\'e}bault}(2009)}]{t09}
{Th{\'e}bault}, P. 2009,
  \href{http://dx.doi.org/10.1051/0004-6361/200912396}{\color{magenta}\aap},
  \href{https://ui.adsabs.harvard.edu/abs/2009A&A...505.1269T}{\color{blue}505},
  \href{https://ui.adsabs.harvard.edu/abs/2009A&A...505.1269T}{\color{blue}1269}

\bibitem[{{Thi} {et~al.}(2014){Thi}, {Pinte}, {Pantin}, {Augereau}, {Meeus},
  {M{\'e}nard}, {Martin-Za{\"\i}di}, {Woitke}, {Riviere-Marichalar}, {Kamp},
  {Carmona}, {Sandell}, {Eiroa}, {Dent}, {Montesinos}, {Aresu}, {Meijerink},
  {Spaans}, {White}, {Ardila}, {Lebreton}, {Mendigut{\'\i}a}, \&
  {Brittain}}]{Thi2014}
{Thi}, W.~F., {Pinte}, C., {Pantin}, E., {et~al.} 2014,
  \href{http://dx.doi.org/10.1051/0004-6361/201322150}{\color{magenta}\aap},
  \href{https://ui.adsabs.harvard.edu/abs/2014A&A...561A..50T}{\color{blue}561},
  \href{https://ui.adsabs.harvard.edu/abs/2014A&A...561A..50T}{\color{blue}A50}

\bibitem[{{Tremblin} {et~al.}(2016){Tremblin}, {Amundsen}, {Chabrier},
  {Baraffe}, {Drummond}, {Hinkley}, {Mourier}, \& {Venot}}]{tac16}
{Tremblin}, P., {Amundsen}, D.~S., {Chabrier}, G., {et~al.} 2016,
  \href{http://dx.doi.org/10.3847/2041-8205/817/2/L19}{\color{magenta}\apjl},
  \href{https://ui.adsabs.harvard.edu/abs/2016ApJ...817L..19T}{\color{blue}817},
  \href{https://ui.adsabs.harvard.edu/abs/2016ApJ...817L..19T}{\color{blue}L19}

\bibitem[{{Tremblin} {et~al.}(2015){Tremblin}, {Amundsen}, {Mourier},
  {Baraffe}, {Chabrier}, {Drummond}, {Homeier}, \& {Venot}}]{tam15}
{Tremblin}, P., {Amundsen}, D.~S., {Mourier}, P., {et~al.} 2015,
  \href{http://dx.doi.org/10.1088/2041-8205/804/1/L17}{\color{magenta}\apjl},
  \href{https://ui.adsabs.harvard.edu/abs/2015ApJ...804L..17T}{\color{blue}804},
  \href{https://ui.adsabs.harvard.edu/abs/2015ApJ...804L..17T}{\color{blue}L17}

\bibitem[{{Tuthill} {et~al.}(2000){Tuthill}, {Monnier}, {Danchi}, {Wishnow}, \&
  {Haniff}}]{tmd00}
{Tuthill}, P.~G., {Monnier}, J.~D., {Danchi}, W.~C., {et~al.} 2000, \pasp,
  \href{http://adsabs.harvard.edu/abs/2000PASP..112..555T}{\color{blue}112},
  \href{http://adsabs.harvard.edu/abs/2000PASP..112..555T}{\color{blue}555}

\bibitem[{{Vigan} {et~al.}(2021){Vigan}, {Fontanive}, {Meyer}, {Biller},
  {Bonavita}, {Feldt}, {Desidera}, {Marleau}, {Emsenhuber}, {Galicher}, {Rice},
  {Forgan}, {Mordasini}, {Gratton}, {Le Coroller}, {Maire}, {Cantalloube},
  {Chauvin}, {Cheetham}, {Hagelberg}, {Lagrange}, {Langlois}, {Bonnefoy},
  {Beuzit}, {Boccaletti}, {D'Orazi}, {Delorme}, {Dominik}, {Henning}, {Janson},
  {Lagadec}, {Lazzoni}, {Ligi}, {Menard}, {Mesa}, {Messina}, {Moutou},
  {M{\"u}ller}, {Perrot}, {Samland}, {Schmid}, {Schmidt}, {Sissa}, {Turatto},
  {Udry}, {Zurlo}, {Abe}, {Antichi}, {Asensio-Torres}, {Baruffolo}, {Baudoz},
  {Baudrand}, {Bazzon}, {Blanchard}, {Bohn}, {Brown Sevilla}, {Carbillet},
  {Carle}, {Cascone}, {Charton}, {Claudi}, {Costille}, {De Caprio},
  {Delboulb{\'e}}, {Dohlen}, {Engler}, {Fantinel}, {Feautrier}, {Fusco},
  {Gigan}, {Girard}, {Giro}, {Gisler}, {Gluck}, {Gry}, {Hubin}, {Hugot},
  {Jaquet}, {Kasper}, {Le Mignant}, {Llored}, {Madec}, {Magnard}, {Martinez},
  {Maurel}, {M{\"o}ller-Nilsson}, {Mouillet}, {Moulin}, {Orign{\'e}}, {Pavlov},
  {Perret}, {Petit}, {Pragt}, {Puget}, {Rabou}, {Ramos}, {Rickman}, {Rigal},
  {Rochat}, {Roelfsema}, {Rousset}, {Roux}, {Salasnich}, {Sauvage}, {Sevin},
  {Soenke}, {Stadler}, {Suarez}, {Wahhaj}, {Weber}, \& {Wildi}}]{vfm21}
{Vigan}, A., {Fontanive}, C., {Meyer}, M., {et~al.} 2021,
  \href{http://dx.doi.org/10.1051/0004-6361/202038107}{\color{magenta}\aap},
  \href{https://ui.adsabs.harvard.edu/abs/2021A&A...651A..72V}{\color{blue}651},
  \href{https://ui.adsabs.harvard.edu/abs/2021A&A...651A..72V}{\color{blue}A72}

\bibitem[{{Vos} {et~al.}(2017){Vos}, {Allers}, \& {Biller}}]{vab17}
{Vos}, J.~M., {Allers}, K.~N., \& {Biller}, B.~A. 2017,
  \href{http://dx.doi.org/10.3847/1538-4357/aa73cf}{\color{magenta}\apj},
  \href{https://ui.adsabs.harvard.edu/abs/2017ApJ...842...78V}{\color{blue}842},
  \href{https://ui.adsabs.harvard.edu/abs/2017ApJ...842...78V}{\color{blue}78}

\bibitem[{{Vos} {et~al.}(2020){Vos}, {Biller}, {Allers}, {Faherty}, {Liu},
  {Metchev}, {Eriksson}, {Manjavacas}, {Dupuy}, {Janson}, {Radigan-Hoffman},
  {Crossfield}, {Bonnefoy}, {Best}, {Homeier}, {Schlieder}, {Brandner},
  {Henning}, {Bonavita}, \& {Buenzli}}]{vba20}
{Vos}, J.~M., {Biller}, B.~A., {Allers}, K.~N., {et~al.} 2020,
  \href{http://dx.doi.org/10.3847/1538-3881/ab9642}{\color{magenta}\aj},
  \href{https://ui.adsabs.harvard.edu/abs/2020AJ....160...38V}{\color{blue}160},
  \href{https://ui.adsabs.harvard.edu/abs/2020AJ....160...38V}{\color{blue}38}

\bibitem[{{Wagner} {et~al.}(2019){Wagner}, {Apai}, \& {Kratter}}]{wak19}
{Wagner}, K., {Apai}, D., \& {Kratter}, K.~M. 2019,
  \href{http://dx.doi.org/10.3847/1538-4357/ab1904}{\color{magenta}\apj},
  \href{https://ui.adsabs.harvard.edu/abs/2019ApJ...877...46W}{\color{blue}877},
  \href{https://ui.adsabs.harvard.edu/abs/2019ApJ...877...46W}{\color{blue}46}

\bibitem[{{Wagner} {et~al.}(2021){Wagner}, {Boehle}, {Pathak}, {Kasper},
  {Arsenault}, {Jakob}, {K{\"a}ufl}, {Leveratto}, {Maire}, {Pantin},
  {Siebenmorgen}, {Zins}, {Absil}, {Ageorges}, {Apai}, {Carlotti}, {Choquet},
  {Delacroix}, {Dohlen}, {Duhoux}, {Forsberg}, {Fuenteseca}, {Gutruf}, {Guyon},
  {Huby}, {Kampf}, {Karlsson}, {Kervella}, {Kirchbauer}, {Klupar}, {Kolb},
  {Mawet}, {N'Diaye}, {Orban de Xivry}, {Quanz}, {Reutlinger}, {Ruane},
  {Riquelme}, {Soenke}, {Sterzik}, {Vigan}, \& {de Zeeuw}}]{wbp21}
{Wagner}, K., {Boehle}, A., {Pathak}, P., {et~al.} 2021,
  \href{http://dx.doi.org/10.1038/s41467-021-21176-6}{\color{magenta}Nature
  Communications},
  \href{https://ui.adsabs.harvard.edu/abs/2021NatCo..12..922W}{\color{blue}12},
  \href{https://ui.adsabs.harvard.edu/abs/2021NatCo..12..922W}{\color{blue}922}

\bibitem[{{Wahhaj} {et~al.}(2021){Wahhaj}, {Milli}, {Romero}, {Cieza}, {Zurlo},
  {Vigan}, {Pe{\~n}a}, {Valdes}, {Cantalloube}, {Girard}, \& {Pantoja}}]{wmr21}
{Wahhaj}, Z., {Milli}, J., {Romero}, C., {et~al.} 2021,
  \href{http://dx.doi.org/10.1051/0004-6361/202038794}{\color{magenta}\aap},
  \href{https://ui.adsabs.harvard.edu/abs/2021A&A...648A..26W}{\color{blue}648},
  \href{https://ui.adsabs.harvard.edu/abs/2021A&A...648A..26W}{\color{blue}A26}

\bibitem[{{Waldmann} {et~al.}(2015{\natexlab{a}}){Waldmann}, {Rocchetto},
  {Tinetti}, {Barton}, {Yurchenko}, \& {Tennyson}}]{wrt15}
{Waldmann}, I.~P., {Rocchetto}, M., {Tinetti}, G., {et~al.} 2015{\natexlab{a}},
  \href{http://dx.doi.org/10.1088/0004-637X/813/1/13}{\color{magenta}\apj},
  \href{https://ui.adsabs.harvard.edu/abs/2015ApJ...813...13W}{\color{blue}813},
  \href{https://ui.adsabs.harvard.edu/abs/2015ApJ...813...13W}{\color{blue}13}

\bibitem[{{Waldmann} {et~al.}(2015{\natexlab{b}}){Waldmann}, {Tinetti},
  {Rocchetto}, {Barton}, {Yurchenko}, \& {Tennyson}}]{wtr15}
{Waldmann}, I.~P., {Tinetti}, G., {Rocchetto}, M., {et~al.} 2015{\natexlab{b}},
  \href{http://dx.doi.org/10.1088/0004-637X/802/2/107}{\color{magenta}\apj},
  \href{https://ui.adsabs.harvard.edu/abs/2015ApJ...802..107W}{\color{blue}802},
  \href{https://ui.adsabs.harvard.edu/abs/2015ApJ...802..107W}{\color{blue}107}

\bibitem[{{Wang} {et~al.}(2015){Wang}, {Ruffio}, {De Rosa}, {Aguilar}, {Wolff},
  \& {Pueyo}}]{wrd15}
{Wang}, J.~J., {Ruffio}, J.-B., {De Rosa}, R.~J., {et~al.} 2015, {pyKLIP: PSF
  Subtraction for Exoplanets and Disks}

\bibitem[{{Wang} {et~al.}(2021{\natexlab{a}}){Wang}, {Vigan}, {Lacour},
  {Nowak}, {Stolker}, {De Rosa}, {Ginzburg}, {Gao}, {Abuter}, {Amorim},
  {Asensio-Torres}, {Baub{\"o}ck}, {Benisty}, {Berger}, {Beust}, {Beuzit},
  {Blunt}, {Boccaletti}, {Bohn}, {Bonnefoy}, {Bonnet}, {Brandner},
  {Cantalloube}, {Caselli}, {Charnay}, {Chauvin}, {Choquet}, {Christiaens},
  {Cl{\'e}net}, {Coud{\'e} Du Foresto}, {Cridland}, {de Zeeuw}, {Dembet},
  {Dexter}, {Drescher}, {Duvert}, {Eckart}, {Eisenhauer}, {Facchini}, {Gao},
  {Garcia}, {Garcia Lopez}, {Gardner}, {Gendron}, {Genzel}, {Gillessen},
  {Girard}, {Haubois}, {Hei{\ss}el}, {Henning}, {Hinkley}, {Hippler},
  {Horrobin}, {Houll{\'e}}, {Hubert}, {Jim{\'e}nez-Rosales}, {Jocou},
  {Kammerer}, {Keppler}, {Kervella}, {Meyer}, {Kreidberg}, {Lagrange},
  {Lapeyr{\`e}re}, {Le Bouquin}, {L{\'e}na}, {Lutz}, {Maire}, {M{\'e}nard},
  {M{\'e}rand}, {Molli{\`e}re}, {Monnier}, {Mouillet}, {M{\"u}ller},
  {Nasedkin}, {Ott}, {Otten}, {Paladini}, {Paumard}, {Perraut}, {Perrin},
  {Pfuhl}, {Pueyo}, {Rameau}, {Rodet}, {Rodr{\'\i}guez-Coira}, {Rousset},
  {Scheithauer}, {Shangguan}, {Shimizu}, {Stadler}, {Straub}, {Straubmeier},
  {Sturm}, {Tacconi}, {van Dishoeck}, {Vincent}, {von Fellenberg},
  {Ward-Duong}, {Widmann}, {Wieprecht}, {Wiezorrek}, {Woillez}, \& {Gravity
  Collaboration}}]{wvl21}
{Wang}, J.~J., {Vigan}, A., {Lacour}, S., {et~al.} 2021{\natexlab{a}},
  \href{http://dx.doi.org/10.3847/1538-3881/abdb2d}{\color{magenta}Astronomical
  Journal},
  \href{https://ui.adsabs.harvard.edu/abs/2021AJ....161..148W}{\color{blue}161},
  \href{https://ui.adsabs.harvard.edu/abs/2021AJ....161..148W}{\color{blue}148}

\bibitem[{{Wang} {et~al.}(2021{\natexlab{b}}){Wang}, {Ruffio}, {Morris},
  {Delorme}, {Jovanovic}, {Pezzato}, {Echeverri}, {Finnerty}, {Hood},
  {Zanazzi}, {Bryan}, {Bond}, {Cetre}, {Martin}, {Mawet}, {Skemer}, {Baker},
  {Xuan}, {Wallace}, {Wang}, {Bartos}, {Blake}, {Boden}, {Buzard}, {Calvin},
  {Chun}, {Doppmann}, {Dupuy}, {Duch{\^e}ne}, {Feng}, {Fitzgerald}, {Fortney},
  {Freedman}, {Knutson}, {Konopacky}, {Lilley}, {Liu}, {Lopez}, {Lupu},
  {Marley}, {Meshkat}, {Miles}, {Millar-Blanchaer}, {Ragland}, {Roy}, {Ruane},
  {Sappey}, {Schofield}, {Weiss}, {Wetherell}, {Wizinowich}, \&
  {Ygouf}}]{wrm21}
{Wang}, J.~J., {Ruffio}, J.-B., {Morris}, E., {et~al.} 2021{\natexlab{b}},
  \href{http://dx.doi.org/10.3847/1538-3881/ac1349}{\color{magenta}\aj},
  \href{https://ui.adsabs.harvard.edu/abs/2021AJ....162..148W}{\color{blue}162},
  \href{https://ui.adsabs.harvard.edu/abs/2021AJ....162..148W}{\color{blue}148}

\bibitem[{{Weinberger} {et~al.}(1999){Weinberger}, {Becklin}, {Schneider},
  {Smith}, {Lowrance}, {Silverstone}, {Zuckerman}, \& {Terrile}}]{wbs99}
{Weinberger}, A.~J., {Becklin}, E.~E., {Schneider}, G., {et~al.} 1999,
  \href{http://dx.doi.org/10.1086/312334}{\color{magenta}\apjl},
  \href{https://ui.adsabs.harvard.edu/abs/1999ApJ...525L..53W}{\color{blue}525},
  \href{https://ui.adsabs.harvard.edu/abs/1999ApJ...525L..53W}{\color{blue}L53}

\bibitem[{{Weinberger} {et~al.}(2000){Weinberger}, {Rich}, {Becklin},
  {Zuckerman}, \& {Matthews}}]{wrb00}
{Weinberger}, A.~J., {Rich}, R.~M., {Becklin}, E.~E., {et~al.} 2000,
  \href{http://dx.doi.org/10.1086/317243}{\color{magenta}\apj},
  \href{https://ui.adsabs.harvard.edu/abs/2000ApJ...544..937W}{\color{blue}544},
  \href{https://ui.adsabs.harvard.edu/abs/2000ApJ...544..937W}{\color{blue}937}

\bibitem[{{Wright} {et~al.}(2015){Wright}, {Wright}, {Goodson}, {Rieke},
  {Aitink-Kroes}, {Amiaux}, {Aricha-Yanguas}, {Azzollini}, {Banks},
  {Barrado-Navascues}, {Belenguer-Davila}, {Bloemmart}, {Bouchet}, {Brandl},
  {Colina}, {Detre}, {Diaz-Catala}, {Eccleston}, {Friedman},
  {Garc{\'\i}a-Mar{\'\i}n}, {G{\"u}del}, {Glasse}, {Glauser}, {Greene},
  {Groezinger}, {Grundy}, {Hastings}, {Henning}, {Hofferbert}, {Hunter},
  {Jessen}, {Justtanont}, {Karnik}, {Khorrami}, {Krause}, {Labiano}, {Lagage},
  {Langer}, {Lemke}, {Lim}, {Lorenzo-Alvarez}, {Mazy}, {McGowan}, {Meixner},
  {Morris}, {Morrison}, {M{\"u}ller}, {rgaard-Nielson}, {Olofsson},
  {O'Sullivan}, {Pel}, {Penanen}, {Petach}, {Pye}, {Ray}, {Renotte}, {Renouf},
  {Ressler}, {Samara-Ratna}, {Scheithauer}, {Schneider}, {Shaughnessy},
  {Stevenson}, {Sukhatme}, {Swinyard}, {Sykes}, {Thatcher}, {Tikkanen}, {van
  Dishoeck}, {Waelkens}, {Walker}, {Wells}, \& {Zhender}}]{wwg15}
{Wright}, G.~S., {Wright}, D., {Goodson}, G.~B., {et~al.} 2015,
  \href{http://dx.doi.org/10.1086/682253}{\color{magenta}\pasp},
  \href{https://ui.adsabs.harvard.edu/abs/2015PASP..127..595W}{\color{blue}127},
  \href{https://ui.adsabs.harvard.edu/abs/2015PASP..127..595W}{\color{blue}595}

\bibitem[{{Wyatt}(2003)}]{w03}
{Wyatt}, M.~C. 2003,
  \href{http://dx.doi.org/10.1086/379064}{\color{magenta}\apj},
  \href{https://ui.adsabs.harvard.edu/abs/2003ApJ...598.1321W}{\color{blue}598},
  \href{https://ui.adsabs.harvard.edu/abs/2003ApJ...598.1321W}{\color{blue}1321}

\bibitem[{{Wyatt}(2005)}]{w05}
---. 2005,
  \href{http://dx.doi.org/10.1051/0004-6361:20053391}{\color{magenta}\aap},
  \href{https://ui.adsabs.harvard.edu/abs/2005A&A...440..937W}{\color{blue}440},
  \href{https://ui.adsabs.harvard.edu/abs/2005A&A...440..937W}{\color{blue}937}

\bibitem[{{Wyatt}(2008)}]{w08}
---. 2008, \araa,
  \href{http://adsabs.harvard.edu/abs/2008ARA%26A..46..339W}{\color{blue}46},
  \href{http://adsabs.harvard.edu/abs/2008ARA%26A..46..339W}{\color{blue}339}

\bibitem[{{Wyatt} {et~al.}(2015){Wyatt}, {Pani{\'c}}, {Kennedy}, \&
  {Matr{\`a}}}]{wpk15}
{Wyatt}, M.~C., {Pani{\'c}}, O., {Kennedy}, G.~M., \& {Matr{\`a}}, L. 2015,
  \href{http://dx.doi.org/10.1007/s10509-015-2315-6}{\color{magenta}\apss},
  \href{https://ui.adsabs.harvard.edu/abs/2015Ap&SS.357..103W}{\color{blue}357},
  \href{https://ui.adsabs.harvard.edu/abs/2015Ap&SS.357..103W}{\color{blue}103}

\bibitem[{{Xuan} {et~al.}(2018){Xuan}, {Mawet}, {Ngo}, {Ruane}, {Bailey},
  {Choquet}, {Absil}, {Alvarez}, {Bryan}, {Cook}, {Femen{\'\i}a Castell{\'a}},
  {Gomez Gonzalez}, {Huby}, {Knutson}, {Matthews}, {Ragland}, {Serabyn}, \&
  {Zawol}}]{xmn18}
{Xuan}, W.~J., {Mawet}, D., {Ngo}, H., {et~al.} 2018,
  \href{http://dx.doi.org/10.3847/1538-3881/aadae6}{\color{magenta}\aj},
  \href{https://ui.adsabs.harvard.edu/abs/2018AJ....156..156X}{\color{blue}156},
  \href{https://ui.adsabs.harvard.edu/abs/2018AJ....156..156X}{\color{blue}156}

\bibitem[{{Zahnle} {et~al.}(2016){Zahnle}, {Marley}, {Morley}, \&
  {Moses}}]{zmm16}
{Zahnle}, K., {Marley}, M.~S., {Morley}, C.~V., \& {Moses}, J.~I. 2016,
  \href{http://dx.doi.org/10.3847/0004-637X/824/2/137}{\color{magenta}\apj},
  \href{https://ui.adsabs.harvard.edu/abs/2016ApJ...824..137Z}{\color{blue}824},
  \href{https://ui.adsabs.harvard.edu/abs/2016ApJ...824..137Z}{\color{blue}137}

\bibitem[{{Zhou} {et~al.}(2020){Zhou}, {Bowler}, {Morley}, {Apai}, {Kataria},
  {Bryan}, \& {Benneke}}]{zbm20}
{Zhou}, Y., {Bowler}, B.~P., {Morley}, C.~V., {et~al.} 2020,
  \href{http://dx.doi.org/10.3847/1538-3881/ab9e04}{\color{magenta}\aj},
  \href{https://ui.adsabs.harvard.edu/abs/2020AJ....160...77Z}{\color{blue}160},
  \href{https://ui.adsabs.harvard.edu/abs/2020AJ....160...77Z}{\color{blue}77}

\bibitem[{{Zhou} {et~al.}(2021){Zhou}, {Bowler}, {Wagner}, {Schneider}, {Apai},
  {Kraus}, {Close}, {Herczeg}, \& {Fang}}]{zbw21}
{Zhou}, Y., {Bowler}, B.~P., {Wagner}, K.~R., {et~al.} 2021,
  \href{http://dx.doi.org/10.3847/1538-3881/abeb7a}{\color{magenta}\aj},
  \href{https://ui.adsabs.harvard.edu/abs/2021AJ....161..244Z}{\color{blue}161},
  \href{https://ui.adsabs.harvard.edu/abs/2021AJ....161..244Z}{\color{blue}244}

\end{thebibliography}

\end{document}